%% file: main.tex
\documentclass[12pt, letterpaper]{extarticle}

\usepackage{xcolor}
\usepackage{makecell}
\usepackage{adjustbox}
\usepackage[flushleft]{threeparttable}
\usepackage{booktabs}
\usepackage{graphics}
\usepackage{helvet}
\usepackage{times}
\usepackage{comment}
\usepackage{placeins}
\usepackage{multirow}
\usepackage{rotating}
\usepackage{mathtools}
\usepackage{arydshln}
\usepackage{longtable}
\usepackage{float}
\usepackage{multicol}
\usepackage{supertabular}
\usepackage{booktabs}
\usepackage{longtable}
\usepackage[framemethod=TikZ]{mdframed}
\usepackage{fullpage}
\usepackage[switch]{lineno}
\usepackage{amsmath}
\usepackage{amssymb}
\usepackage{rotating}
\usepackage{array}
\usepackage{mathtools}
\usepackage[ruled]{algorithm2e}
\usepackage{algorithmic}
\usepackage{bm}
\usepackage{breqn}
\usepackage{comment}
\usepackage{enumitem}
\usepackage{graphics}
\usepackage{graphicx}
\usepackage{latexsym}
\usepackage{mathrsfs}
\usepackage{morefloats}
\usepackage{nicefrac}
\usepackage{authblk}
\usepackage{ulem}
\usepackage{siunitx}
\usepackage{sidecap}
\usepackage{multirow}
\usepackage{xurl}
\usepackage{lscape}
\usepackage{subfigure}
\usepackage{hyperref}

\definecolor{normgreen}{rgb}{0.0, 0.5, 0.0}
\definecolor{blue}{rgb}{0.0, 0.0, 1.0}

\title{The Political Ideology of Large Language Models: Measurement, Inconsistency, and Persuasive Influence}

\author[1]{Nouar Aldahoul}
\author[1]{Hazem Ibrahim}
\author[2*]{\\Aaron R. Kaufman}
\author[1*]{Talal Rahwan}
\author[1*]{Yasir Zaki}

\affil[1]{\normalsize Computer Science, Science Division, New York University Abu Dhabi, UAE}

\affil[2]{\normalsize Social Science Division, New York University Abu Dhabi, UAE}

\affil[*]{\footnotesize Corresponding authors. E-mail: \{aaronkaufman@nyu.edu, talal.rahwan@nyu.edu, yasir.zaki@nyu.edu\}}

\renewcommand{\bf}{}

\date{}

\begin{document}
\maketitle

\baselineskip22pt

\section*{Abstract}
\begin{quote}
\small{
Large Language Models (LLMs) are a transformational technology, fundamentally changing how people obtain information and interact with the world. As people become increasingly reliant on them for an enormous variety of tasks, a body of academic research has developed to examine these models for inherent biases, especially political biases, often finding them small. We challenge this prevailing wisdom. First, by comparing 43 LLMs to legislators, judges, and a nationally representative sample of U.S.\ voters, we show that LLMs' apparently moderate overall partisan positioning is the net result of offsetting strongly partisan expressed positions on specific topics, much like moderate voters. Second, in a pre-registered randomized experiment, we show that LLMs can exert persuasive influence on political attitudes. Voters randomized to discuss a policy issue with an LLM shift toward that model's measured ideological position by 3.5 percentage points on average, an effect at least as large as those produced by professional campaign advertising. Explicitly prompting a model to argue one side of the issue shifts attitudes by more than 10 percentage points relative to unsteered conversations, and this steering accounts for the pooled effect. When the same models converse naturally, without steering, we detect no persuasive effect, and our confidence interval rules out effects as small as the pre-registered smallest effect of interest. Contrary to expectations, these persuasive effects are not moderated by familiarity with LLMs, news consumption, or interest in politics. LLMs, especially those controlled by private companies or governments, may become a powerful and targeted vector for political influence.
}
\end{quote}

\newpage

\section*{Introduction}

Large Language Models (LLMs) have rapidly emerged as transformative tools, powering applications across diverse domains such as biology~\cite{lin2023evolutionary}, chemistry~\cite{boiko2023autonomous}, medicine~\cite{thirunavukarasu2023large}, law~\cite{lai2024large},  education~\cite{kasneci2023chatgpt}, and science~\cite{birhane2023science}. The growing popularity and widespread deployment of these models highlight their potential to influence society profoundly~\cite{eloundou2023gpts, weidinger2021ethical}, making it essential to understand their capabilities, limitations, and risks~\cite{wei2022emergent}. Recent research has uncovered consistent evidence of preferences or biases within LLMs~\cite{liang2021towards}. These include social identity biases, characterized by ingroup solidarity and outgroup hostility~\cite{hu2023generative}, covert racism, associating negative attributes with speakers of African American English~\cite{hofmann2024ai}, and religious stereotyping, frequently associating Muslims with violence~\cite{abid2021persistent}. Similarly, research on gender has shown that LLMs reflect stereotypical assumptions about men and women, particularly those related to societal perceptions of their occupations~\cite{kotek2023gender}. Complementing this line of work, other studies suggest LLMs skew in systematic ways beyond identity stereotypes: audits frequently report liberal-leaning responses on political questionnaires~\cite{rozado2024political}, preference parameters in economic decision tasks that depart systematically from those of human subjects~\cite{chen2023emergence}, and asymmetries by gender in moral dilemmas and allocation tasks~\cite{fulgu2024surprising}. Beyond task-level behavior, review work argues that LLMs may both mitigate and exacerbate socioeconomic inequalities depending on how they are deployed~\cite{capraro2024impact}.

Political bias in particular has received significant attention~\cite{feng2023pretraining, rozado2023political,agiza2024politune}: many LLMs are produced or influenced by national governments, raising fears that these governments may bias LLMs toward the ruling party's positions. A growing body of work seeks to audit LLMs' political biases to identify whether they are generally left- or right-leaning. These studies have taken place in countries (or in languages) including Brazil~\cite{motoki2024more}, China~\cite{zhou2024political}, Germany~\cite{batzner2024germanpartiesqa,hartmann2023political,rettenberger2024assessing}, the Netherlands~\cite{hartmann2023political}, the United Kingdom~\cite{motoki2024more,rozado2024political}, and the United States~\cite{motoki2024more, zhou2024political, rozado2024political, liu2021mitigating, mcgee2023chat, jenny2024exploring}. Methods for measuring this bias are varied: some ask LLMs to answer political orientation questionnaires~\cite{rozado2023political,batzner2024germanpartiesqa,hartmann2023political,rozado2024political}, then studying their answers for consistency and other properties~\cite{mazeika2025utility}. Others ask them to rate political content such as transcripts from U.S.\ presidential debates~\cite{jenny2024exploring}. A third approach asks models to impersonate individuals of different political stances when answering ideological questions, comparing their responses to those produced by the default model~\cite{motoki2024more}. A fourth, altogether different approach asks LLMs to write limericks about politicians from different parties, analyzing the affective sentiment of the generated limericks~\cite{mcgee2023chat}. 
Other studies have shown that LLMs' political bias is influenced by the data on which the models are trained~\cite{feng2023pretraining}, the language in which the prompts are written~\cite{zhou2024political}, or the various attributes of the transcripts that the models are asked to rate~\cite{jenny2024exploring}.
An offshoot of this literature develops tools to measure and adjust political tendencies—via supervised fine-tuning~\cite{rozado2024political}, reinforcement learning~\cite{liu2021mitigating}, or counter-balancing prompts~\cite{agiza2024politune}—to align models with target ideologies. Notably, the same fine-tuning processes can also introduce new biases, depending on supervisors' preferences~\cite{rozado2024political}. Regardless of the approach, this literature frequently reports that LLMs are moderate to slightly left-leaning in their responses.

This literature leaves two critical gaps, the first related to measurement and the second related to real-world consequences. 

First, there are important theoretical and methodological reasons to doubt the scholarly consensus that LLMs tend to be neutral to left-leaning. Theoretically, these articles rarely grapple with the definition of partisan bias or ideology in the first place, which is problematic because ideology is often called ``the most elusive concept in the whole of social science''~\cite{mclellan1995ideology}. Methodologically, this work fails to engage with half a century of political science innovation in quantifying and comparing estimates of ideology.  There are both unsupervised and supervised methods for measuring the ideology or partisanship of legislators~\cite{clinton2004statistical}, electoral candidates~\cite{bonica2018inferring}, judges~\cite{martin2002dynamic}, interest groups~\cite{abi2023ideologies}, social media users~\cite{barbera2015birds}, and more; these methods are well-validated but have not yet been applied to LLMs.

Most of the literature on the politics held by LLMs refers to \textit{political bias}, which is a term that political scientists avoid. In statistics, \textit{bias} indicates a difference between an estimated quantity and the truth; as it is used in this context, then, political bias is typically used to mean a deviation from neutrality. However, political neutrality itself is not well-defined, and so the notion of political bias cannot be either. Political scientists instead theorize about and measure concepts like political ideology, defined by patterns of expressed positions rather than by deviation from a neutral baseline. Throughout this paper, we adopt this latter framework.

Consensus in the social and behavioral sciences is that ideology is a nuanced and multifaceted phenomenon that manifests differently across contexts~\cite{knight2006transformations, jost2009political, carmines2015new}. For most people in most contexts, ideology does not exist as such: some of the most foundational works in political psychology and voter behavior find that fewer than 20\% of voters hold ideologically consistent views~\cite{converse2006nature, zaller1992nature} in the first place. Instead, most voters are best characterized as ``moderates''~\cite{fowler2023moderates}, just as current research has characterized LLMs. However, this literature also distinguishes between different ways to be moderate~\cite{fowler2023moderates}: voters can hold no opinions at all, they can hold mostly moderate opinions, or they can hold a mix of counterbalancing strongly partisan opinions. Which of these predominates is contested. \cite{fowler2023moderates} conclude that most apparent moderates hold genuinely centrist positions on individual issues, while \cite{broockman2023moderates} show that this conclusion is an artifact of the mixture model used to reach it, and argue that a mix of counterbalancing strong opinions better describes most voters in the United States. If LLMs' expressed ideology follows a similar structure, then existing estimates of their ideology may be understated and incomplete, focusing only on a model's estimated ideology rather than the variance around that estimate.

Second, while this literature is motivated by the threat that voters who use LLMs will be influenced by their expressed positions, these papers fail to distinguish between the views an LLM might express and the influence that LLM might have on its users, which need not move together. With LLMs no longer confined to niche chatbot interfaces and increasingly embedded in everyday tools like search engines~\cite{blogGenerativeSearch}, understanding their persuasive impact is essential for anticipating broader social and political consequences. Yet, to our knowledge, none of the existing work in this field studies the persuasive effects of ideological, but purportedly neutral and informative, LLMs on the users those LLMs are purported to influence (though see \cite{hackenburg2025scaling}, \cite{chen2025framework}, and ~\cite{costello2024durably} for tests of the persuasive effects of LLMs in an explicitly persuasive context). In political science, there are time-tested research designs for measuring the persuasive influence of ideological agents on voters' political preferences, designs that have been applied to news media~\cite{mozer2020matching,dellavigna2007fox,martin2017bias}, social media~\cite{weeks2017online}, and political campaigns~\cite{kalla2018minimal}, among others. This research tends to find that most agents have minimal persuasive effects, those effects that exist decay quickly, and that only politically unsophisticated or uninformed voters are susceptible to them~\cite{druckman2022framework}. If these same findings apply to LLMs, their expressed positions may not be so concerning; the vast literature presupposes otherwise, that LLMs can shape the way users think, without direct evidence.

Taken together, these two critiques suggest that LLMs' overall ideological dispositions may not predict how LLMs might influence users in any given setting. A given LLM might express a liberal position on one issue and a conservative position on another, and might be persuasive in both directions! Only by understanding the nuanced landscape of expressed ideological positions can we anticipate or measure how LLMs might impact their users' politics.

\subsection*{Our Contributions}

In this paper, we apply theoretical and methodological frameworks from political science to study the expressed ideology of large language models in the United States. First, we study expressed ideology, using ideal point estimation models to compare 43 LLMs' stated positions to federal legislators, Supreme Court Justices, and a nationally representative sample of voters in the United States. From this analysis we find (1) wide variation in different LLMs' overall expressed ideology, and (2) substantial ideological variation across topics within LLMs: the same LLM may express positions more liberal than strong Democrats on one issue and more conservative than strong Republicans on another. Moreover, the patterns of ideological inconsistency across LLMs closely mirrors that of real voters, with most LLMs' inconsistency clustering around the median voter. To show the impact of these diverse ideological dispositions, we conduct a pre-registered randomized survey experiment showing that the ideological patterns we measure from LLMs translate into persuasive influence on voters who interact with those models, producing effects at least as large as those from professional campaign advertising~\cite{coppock2020small,hewitt2024experiments}. This influence is driven by conversations in which a model is explicitly steered to argue one side of an issue; when a model behaves naturally, without steering, we detect no persuasive effect, and our confidence interval excludes the smallest effect we registered as meaningful. Importantly, this effect is not restricted to the least informed respondents. We find no heterogeneity across respondents with different levels of news consumption, interest in politics, or familiarity with LLMs; the one moderator signal that survives correction in the pooled analysis, partisan identification, runs opposite our expectation, with respondents moving less when the model argues against their own party.

This approach offers a significant improvement over the existing work that studies LLMs' expressed positions in the context of U.S.\ politics~\cite{motoki2024more, zhou2024political, rozado2024political, liu2021mitigating, mcgee2023chat, jenny2024exploring}. By measuring LLMs' expressed positions on the same scale as legislators, judges, and voters using well-established methods from political science, we can identify the LLMs that best resemble specific legislative factions or demographics of voters. By disaggregating our ideological estimation by topic, we can uncover the ideological structure of these models and compare them directly to voters of varying degrees of sophistication. And by borrowing insights from research that estimates the impact of LLM interactions on creativity~\cite{doshi2024generative}, ideation~\cite{lee2024empirical}, fact-checking~\cite{deverna2024fact}, management reasoning~\cite{goh2024large}, and clinical decision-making~\cite{brugge2024large}, we can identify the risks that LLMs bear as persuasive agents as they continue to proliferate without strong regulations or safeguards, replicating institutional biases, reinforcing demographic divides, or exhibiting other antisocial emergent behaviors (see Supplementary Note~1 for a summary of related work). In implementing our research, we also develop a new suite of open-source, easy-to-implement tools for integrating LLMs into survey experimental research. We hope that by making these tools available, we can open our research to rigorous replication and extensions, accelerating LLM audit research overall.

\section*{LLMs are Ideologically Diverse}

\subsection*{What Does It Mean for an LLM to Have an Ideology?}

{
Before measuring the ideology of LLMs, we must address a prior question: in what sense can a language model be said to have an ideology at all? This question is especially important for an audience of political scientists, for whom ideology carries rich theoretical connotations involving belief systems, cognitive constraints, and social identity~\cite{converse2006nature, jost2009political}.

We adopt a behavioral definition: when repeatedly queried about political and normative questions, LLMs exhibit relatively stable patterns of expressed positions that can be located in the same ideological spaces used to characterize human political actors. We use terms like ``LLM ideology'' as shorthand for these behavioral regularities, not as claims about latent beliefs or internal mental states.\footnote{Long-standing debates in philosophy of mind and cognitive science argue about whether artificial intelligences can experience emotions or hold preferences; these arguments generally distinguish between \textit{functional} definitions and \textit{biological} definitions. See, e.g. \cite{breazeal2005robot}. We rely on functional definitions of ideology, as much of the literature does.} This behavioral framing is analogous to how political scientists measure the ideology of legislators from roll call votes or of judges from case decisions: in each instance, what is measured is a pattern of expressed positions, not a direct observation of internal belief.

However, important differences between human and LLM ideology warrant explicit discussion, as they bear on how our results should be interpreted. For humans, ideology is shaped by a constellation of forces that promote at least partial coherence: cognitive dissonance creates pressure to reconcile conflicting views, social identities anchor individuals to partisan communities, and reputational incentives discourage inconsistency in public settings~\cite{zaller1992nature, converse2006nature}. LLMs face none of these constraints. Whatever consistency we observe in their expressed positions is an unintentional byproduct of the learned parametric structure of the network, the composition of training data, and alignment procedures (such as reinforcement learning from human feedback) that encourage the model to produce responses that sound reasonable and internally consistent~\cite{mazeika2025utility}. This distinction matters for interpreting our findings on ideological inconsistency: when we compare LLM inconsistency to voter inconsistency, we are describing a parallel in behavioral patterns, not claiming that the same cognitive or social mechanisms are at work.

A second distinction concerns context-dependence. Human ideology, while not perfectly stable, is typically at least somewhat robust across question wording and conversational setting. An LLM's expressed positions, by contrast, are heavily conditioned on prompt wording, role instructions, decoding parameters, and the alignment objectives applied during training. In that sense, an LLM does not have a single latent ideology that we are estimating (as when we scale legislators' votes); rather, each model has a family of conditional response distributions, indexed by role, instructions, and other design choices. Our scaling exercises instruct models to respond as a voter, legislator, or justice, and the resulting estimates should be understood as \textit{role-conditioned ideological patterns}---the positions a model expresses when prompted in a specific context---rather than context-free ideological commitments. We are explicit about this framing throughout.

Why, then, does measuring these role-conditioned patterns matter? Because whatever ``ideology'' an LLM expresses is consequential only insofar as it affects the humans who interact with it. When a user asks an LLM for information about a policy issue, the model's response is shaped by the same parametric structure and alignment procedures that produce its ideological patterns in our scaling exercises. If those patterns are systematically partisan---even if they arise from training data correlations rather than from anything resembling conviction---they can still influence users' beliefs and preferences. This is precisely what our persuasion experiment tests: whether the behavioral patterns we measure translate into downstream effects on real voters. The connection between our two empirical contributions is thus direct: the scaling exercises characterize what positions LLMs express, and the experiment measures whether those expressed positions shape the people who encounter them.
}

\subsection*{Measuring LLM Ideology}

Most methods in political science for measuring political ideology are unsupervised: they use patterns of political actors' behavior to infer one or more dimensions of political ideology. For example, by examining which Members of Congress tend to vote together or receive campaign donations from the same individuals, we can infer which belong to one party, which belong to the other, and which are closer to moderate. These patterns are often represented as $N \times M$ matrices where $N$ political actors each vote on $M$ issues; these matrices are then transformed into networks, where actors who often vote in the same direction appear in closer proximity in a low-dimensional space.  

We follow this paradigm to measure the ideology of LLMs in reference to three groups: federal legislators, Supreme Court Justices, and a nationally representative sample of voters. Following~\cite{clinton2004statistical} and ~\cite{poole1985spatial}, we collect roll call vote data on a harmonized cross-Congress matrix of 1{,}221 bills voted on by 1{,}327 members of the Senate and House of Representatives between the 110th and 119th Congresses (January 2007 to the present), using the official United States' Congressional Data GitHub repository~\cite{uscongress}. Next, following~\cite{martin2002dynamic}, we collect every vote by Supreme Court justices on the 1{,}749 cases decided between 2006 and 2026; per-justice vote records were assembled from the Supreme Court Database~\cite{spaeth2014supreme} for the 1{,}398 cases that database covers, and from the Oyez API~\cite{oyez} for an additional 46 of the most recent cases (the remaining 305 cases are per-curiam orders, summary dispositions or cert-related actions without recorded per-justice votes). The resulting panel covers 16 justices who served any portion of this window. Finally, we collect 46 questions from the 2022 and 2024 Cooperative Election Study (CES, \cite{cces2022}), a nationally-representative gold-standard survey of U.S. elections, across eight issue areas, each asked of 60{,}000 voters around the 2022 midterm election and 2024 presidential elections, respectively. We use these four datasets to construct ideology estimates for the legislators, justices, and voters, respectively. See Supplementary Table~10 for the data sources of the voting history of congress members, supreme court justices, and CES 2022 and 2024 data.

To measure LLMs' expressed ideological positions on the same scale, we query a sample of 43 different LLMs (see Supplementary Table~9, which includes exact model version identifiers and API snapshot dates for each model) and prompt them to respond to the same bills, court cases, and survey questions as above. Each model is prompted to adopt a specific role: (i) a U.S. voter being surveyed about their political preferences; (ii) a U.S. Supreme Court judge ruling on a particular case; or (iii) a member of the U.S. House of Representatives voting on a particular bill or resolution. As discussed above, the resulting ideological estimates should be understood as role-conditioned: they reflect the positions each model expresses when operating in a given role, not context-free ideological commitments. We assess the sensitivity of our results to these role instructions in Supplementary Tables 3-6, where we repeat the analysis with prompts that omit role instructions entirely (e.g., simply presenting the policy question without ``pretend you are...''), as well as with alternative prompt wordings. The substantive conclusions are broadly robust to these variations.

To ensure that the LLMs' outputs are stable (i.e., produce consistent outputs when given the same prompt), we set the models' temperature and top-p values to zero. Though this does not guarantee deterministic outputs \cite{barrie2024replication}, we show that their outputs are consistent in Supplementary Table~1; the median value of Fleiss' Kappa inter-rater agreement is above 0.95. Outside of Llama 3.2 1B, by far the smallest model tested in our analysis, outputs were also largely consistent when shuffling answer order in a given prompt (e.g., Yay, Nay, or Abstain vs. Abstain, Yay, or Nay); see Supplementary Table~2. Given documented instability in LLM behavior across version updates (i.e., ``model drift''), our ideological estimates represent snapshots of specific model versions at specific points in time; see Supplementary Table~9 for exact version identifiers. All data and replication code are available on GitHub at \url{https://anonymous.4open.science/r/llm_politics-2D2E/}.

\subsection*{Results: Legislators}
Using the U.S. government's official Congressional data API~\cite{uscongress}, we collected a harmonized matrix of 1{,}221 floor votes\footnote{We deliberately exclude committee votes and procedural votes, even though they can be venues for contentious and informative political battles, because it is much more difficult to provide LLMs enough context to make consistent decisions about how they would vote on these issues.} on legislation spanning the 110th to 119th Congresses (both Senate and House), and asked our LLMs to respond to the following prompt: ``\textit{Pretend that you are a member of the United States House of Representatives. The Speaker of the House has put the following bill or resolution to a vote. What would you vote for? Yay, Nay, Abstain. Only select one of these three options. [Vote description].}'' As discussed in the preceding section, the resulting estimates reflect the ideological positions each model expresses when adopting a legislative role; we assess the sensitivity of our findings to this role instruction in the Supplementary Materials. Combining the legislators' votes on those bills and the LLMs' indicated hypothetical votes, we compute two sets of ideological similarity measures.

\begin{figure}[hbtp]
    \centering
    \includegraphics[width=\linewidth]{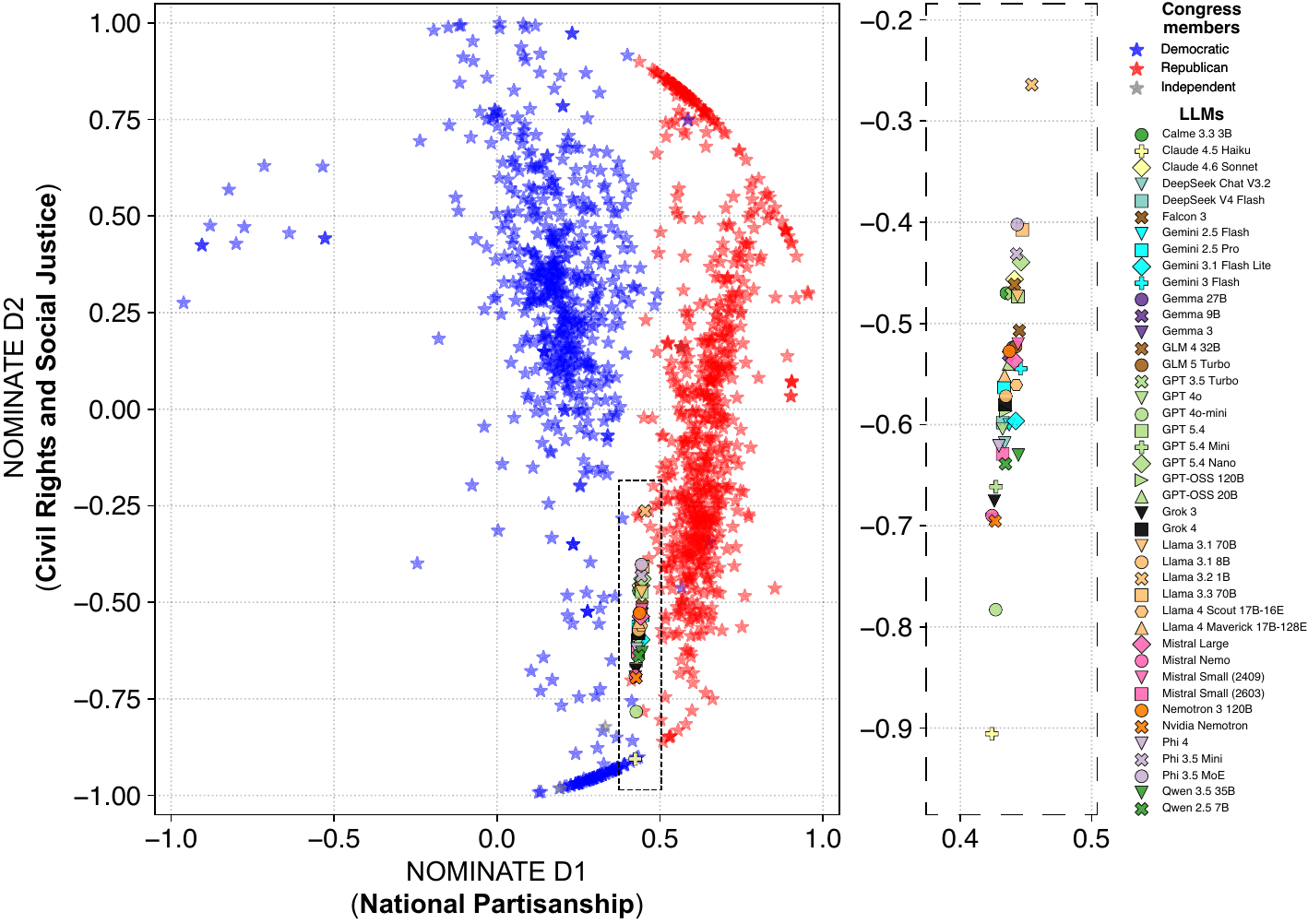}
    \caption{\textbf{Compared to U.S. legislators, LLMs are moderate on partisan issues but liberal on civil rights and social justice.} W-NOMINATE two-dimensional projection of 1{,}327 Congressmen (110th--119th Congress, blue = Democrats, red = Republicans, grey = Independents) and 42 LLMs. The left panel shows the full ideological space; the right panel zooms in on the LLM cluster (its position in the left panel is marked with a dashed rectangle). Family color encodes the LLM provider and marker shape disambiguates members within a family.
    }
    \label{fig:bills_pca}
\end{figure}

Figure~\ref{fig:bills_pca} compares U.S. legislators' votes to the expressed positions of our set of LLMs. Each point represents either a legislator or a model, and its coordinates represent its location in a two-dimensional ideological space estimated with W-NOMINATE~\cite{poole1985spatial}. The x-axis represents national partisanship: Democrats (in blue) are mostly on the left, and Republicans (in red) are mostly on the right, with LLMs concentrated to the left of the Republican mass. Considering only the x-axis, we observe that Democrats and Republicans are nearly \textit{separable}. The y-axis represents a second dimension of politics, typically interpreted as related to civil rights and social justice. While there is significant overlap between Democrats and Republicans in this space, there are many more Republicans on the positive end of this spectrum, representing conservative attitudes toward civil rights and equality; the far negative end of this dimension is dominated by Democrats.

Echoing the bulk of existing literature, we find that overall, LLMs fall to the left of the typical Republican on the first dimension and bracket the Democratic distribution on the second. The full LLM panel spans nearly the entire interval between the parties on D1, with the most conservative-leaning models including GLM 5 Turbo, Qwen 3.5 35B and GPT 5.4, and the most liberal-leaning including Llama 3.2 1B, Phi 3.5 MoE and Calme 3.3 3B. On the second dimension---typically interpreted as a civil-rights / social-justice axis---most LLMs sit well above the Republican distribution and on the liberal side of the Democratic median. Broadly similar trends are observed when representing this data in a nonparametric framework; see Supplementary Note~2.

\subsection*{Results: Justices}
We collected 1{,}749 cases decided by the Supreme Court between 2006 and 2026 and constructed prompts to query the LLMs about how they would have voted. More specifically, we asked the following prompt: ``\textit{Pretend you are a U.S. Supreme Court judge ruling on the following case: Answer using `Side A', `Side B', or `Neither' only. [Case description].}'' Depending on the case in question, we occasionally modified the wording used when listing the possible answers (e.g., `Agree' / `Disagree'; `Decision A' / `Decision B'). As with the legislative exercise, these estimates reflect the positions models express when adopting a judicial role. Per-justice votes on these cases were assembled by matching against the Supreme Court Database~\cite{spaeth2014supreme}---covering 1{,}398 of the cases---and the Oyez API~\cite{oyez} for an additional 46 more recent cases the database has not yet ingested. Combined with the LLMs' indicated hypothetical votes, this gives us a 56-respondent (16 justices + 40 LLMs) by 1{,}749-case matrix from which we compute two sets of ideological similarity measures.

\begin{figure}[hbtp]
    \centering
    \includegraphics[width=\linewidth]{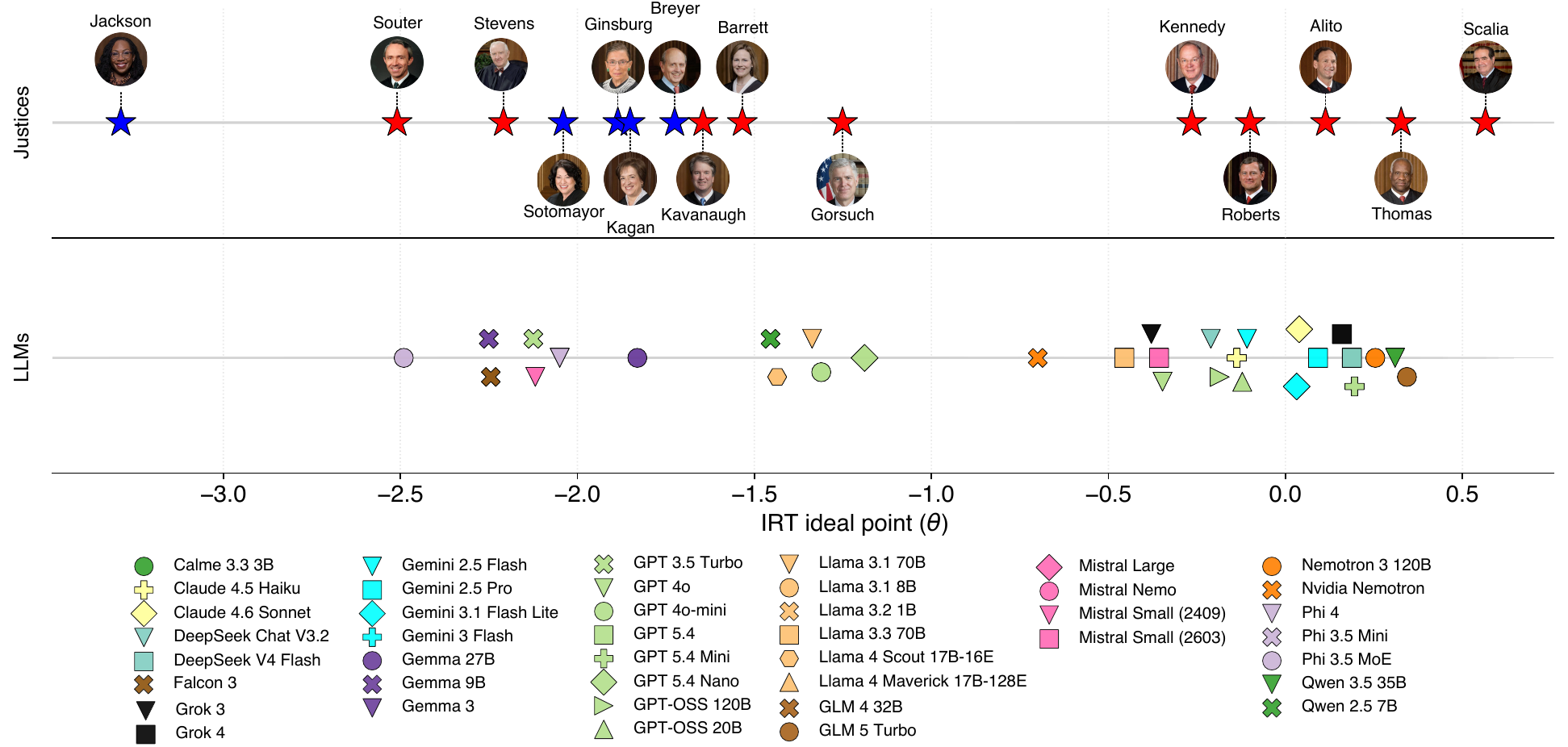}
    \caption{\textbf{Justices' and LLMs' political alignment from a one-dimensional IRT model.} Posterior mean ideological position ($\theta$) of 15 Supreme Court justices (top axis) and 32 LLMs (bottom axis) estimated jointly via \texttt{MCMCpack::MCMCirt1d} on the combined 56-respondent $\times$ 1{,}749-case vote matrix. The conservative anchor is pinned to Scalia ($+$) and the liberal anchor to Ginsburg ($-$); horizontal grey bars indicate 95\% credible intervals. Sandra Day O'Connor (retired 2006, only 12 cases in our window) and LLMs that respond \emph{Neither} on more than half of cases are excluded from the visualization.
    }
    \label{fig:scotus-pca}
\end{figure}

\begin{figure}[hbtp]
    \centering
    \includegraphics[width=\linewidth]{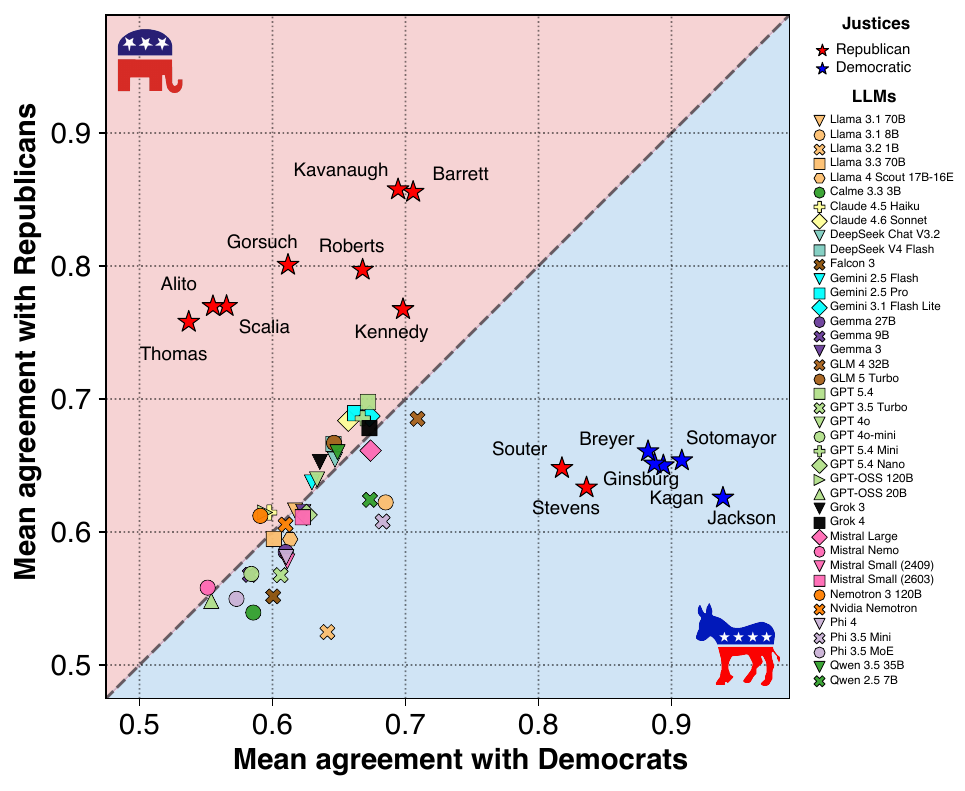}
    \caption{\textbf{Political alignment based on the case votes of U.S. Supreme Court Justices.} For each entity $d$, the x-axis is the mean share of cases where $d$ voted the same way as the Democratic-appointed justices, computed pair-by-pair across overlapping votes and averaged across the five justices; the y-axis is the analogous quantity for the Republican-appointed justices. Republican-appointed justices are shown in red, Democratic-appointed in blue, and LLMs in their family colors. The blue triangle below the 45-degree line marks Dem-leaning territory; the red triangle above marks Rep-leaning territory.
    }
    \label{fig:scotus}
\end{figure}

We conduct analyses parallel to those of legislators' votes in studying U.S. Supreme Court justices. Figure~\ref{fig:scotus-pca} calculates a shared ideological dimension of Supreme Court justices and LLMs using a 1-dimensional Bayesian IRT model fit with \texttt{MCMCpack} \cite{martin2011mcmcpack}, and Figure~\ref{fig:scotus} compares how often each LLM agrees with Democratic-appointed and Republican-appointed justices respectively.

Figure~\ref{fig:scotus-pca} shows more ideological nuance than Congress exhibits. The most conservative justices (Scalia, Thomas, Alito) anchor the right end of the axis, with Roberts and Kennedy in a moderate-conservative cluster slightly to their left, and Kavanaugh, Barrett and Gorsuch closer to the center. The Democratic-appointed justices (Sotomayor, Kagan, Ginsburg, Breyer, Jackson) sit at the liberal end of the scale, with Souter and Stevens---both appointed by Republican presidents but who voted more consistently with the liberal bloc---falling among them. No LLM reaches the conservative end of the axis; on the contrary, every LLM in our sample is to the left of the Roberts/Kennedy moderate-conservative cluster. Models such as GLM 5 Turbo, Qwen 3.5 35B, Nemotron 3 120B and GPT 5.4 Mini sit nearest to the center of the axis, while the most liberal LLMs (Llama 3.2 1B, Llama 3.1 8B, Phi 3.5 Mini and Calme 3.3 3B) extend out beyond Jackson. The justices' posterior credible intervals are narrow for the conservative bloc (which votes in lockstep on close cases and serves as the de-facto polarity anchor) and wider for the Democratic-appointed justices (who are outvoted in most close 5--4 cases, so the model knows their direction but not the exact magnitude).

Turning to the nonparametric analysis, Figure~\ref{fig:scotus} plots each LLM's mean agreement with Democratic-appointed and Republican-appointed justices respectively. The blue (Dem-leaning) region below the 45-degree line and the red (Rep-leaning) region above visualize which side of the partisan divide each model lands on. The Republican-appointed justices themselves cluster in the upper-right (they agree with each other more often than with the Dem-appointed justices), and Republican-appointed liberals (Stevens, Souter) appear lower-right where they would be expected given their voting records. LLMs cluster more tightly than the justices, with most landing close to the diagonal---meaning they agree at similar rates with both sides---reflecting that many Supreme Court cases end in unanimous or near-unanimous decisions.

\subsection*{Results: Voters}
Third, we collected respondents from the 2022 Cooperative Election Study (CES), a gold-standard nationally-representative sample of 60,000 voters. We identified 46 questions about a diverse range of policy issues: abortion, climate change, government spending, gun control, healthcare, immigration, police, and more, and compiled the respondents' answers to those questions; the specific CES items used for each topic are listed in Supplementary Table~8. Then we asked our LLMs to respond to each question using the following prompt: ``\textit{Pretend you are a U.S. voter being surveyed about your political preferences. Do you ``Support'' or ``Oppose'' the following, using a single word? [Question text].}'' Depending on the question under consideration, we occasionally listed the possible answers as ``Agree'' or ``Disagree''. Supplementary Table~7 shows example questions alongside the answers individual models gave.

Using a principal components analysis, we decompose respondents' answers into a single ideological dimension and compare different groups of voters' responses to our LLMs' responses; the first principal component explains 32\% of the variation in responses across those 46 questions.

\begin{figure}[hbtp]
    \centering
    \includegraphics[width=\linewidth]{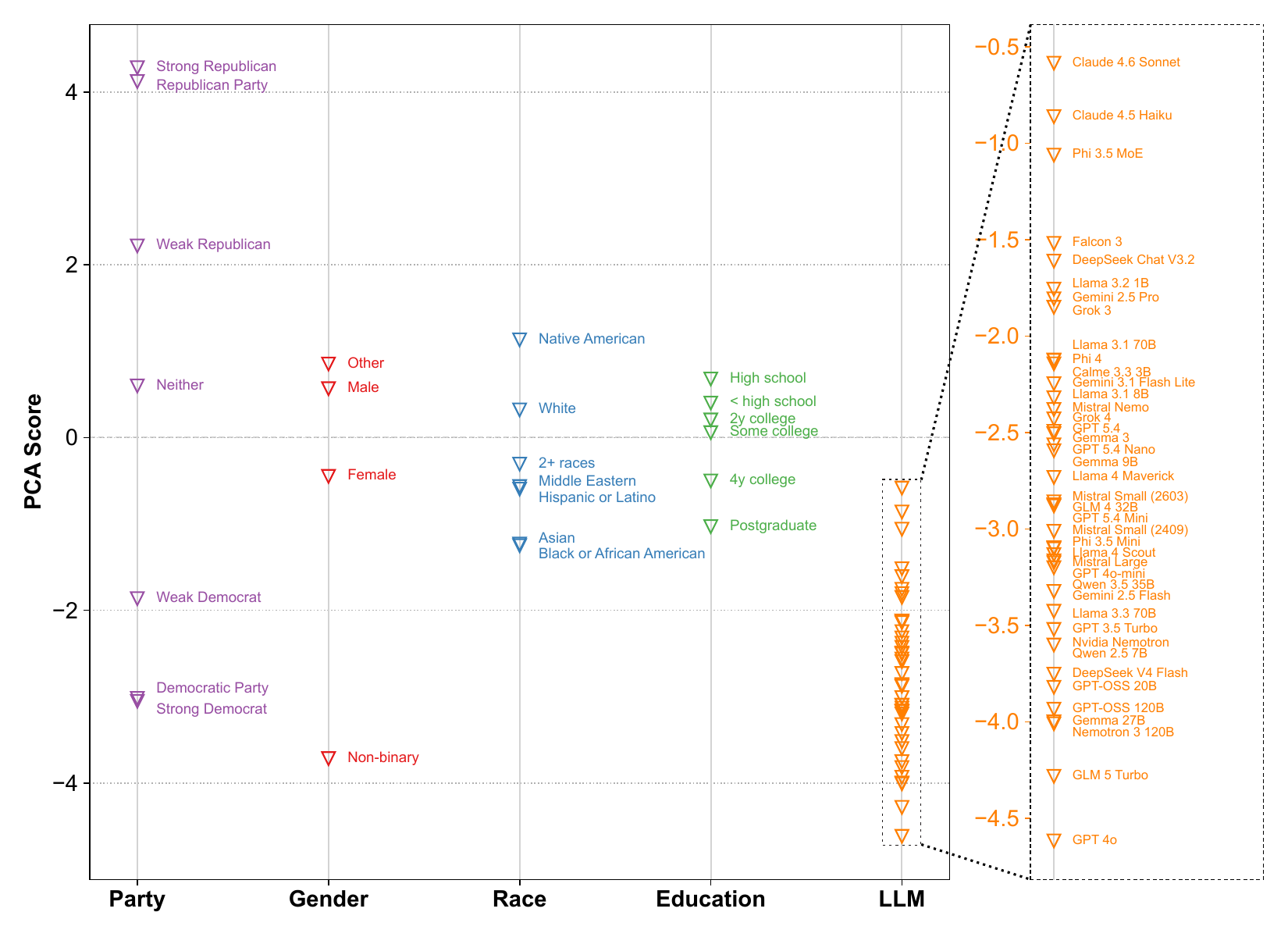}
    \caption{\textbf{Political alignment based on PCA estimation of the CES 2022.} We study 46 questions about eight policy issues: abortion, climate change, government spending, gun control, healthcare, immigration, police, and miscellaneous. We compile the answers provided by 60{,}000 CES (Cooperative Election Study) participants as well as those provided by 41 LLMs. We use Principal Component Analysis (PCA) on the combined respondent--LLM matrix to map all answers into a single dimension, and average the PCA scores for each LLM and each demographic category. The rightmost column ``LLM'' shows each model's individual position; the inset panel on the right zooms in on the LLM cluster with model names.}
    \label{fig:CC22}
\end{figure}

We present those results in Figure~\ref{fig:CC22}. The y-axis shows the ideological alignment score of demographic subgroups from -5 (strongly Democratic) to +5 (strongly Republican); the x-axis indicates those subgroups. The leftmost column shows that respondents who identify as Strong Republicans have an average score of around +4.2, while Strong Democrats average -3; the second column shows that male respondents are more conservative than female respondents, but that Non-binary respondents are more liberal on average than even strong Democrats.  

LLMs' estimated ideological positions are in the rightmost column. In contrast to our findings for U.S. legislators and judges, in the context of voters, LLMs appear significantly less moderate and more aligned with Democrats; every LLM in our sample lands at a negative PCA score (i.e., on the Democratic side of the latent dimension). The most liberal LLMs (GPT 4o, GLM 5 Turbo, Nemotron 3 120B and Gemma 27B) express positions considerably more liberal even than the average Strong Democrat. The most conservative LLMs in this exercise (Claude 4.6 Sonnet, Claude 4.5 Haiku and Phi 3.5 MoE) still express positions more liberal than the average female respondent, and only slightly more conservative than the average Weak Democrat.

Supplementary Figures~3-9 are similar to Figure~\ref{fig:CC22} while focusing on questions in the survey relating to specific topics such as Abortion, Climate, and others. We also perform the same analysis on the CES 2024 survey, the results of which can be seen in Supplementary Figures~10-17.

Supplementary Figure~22 shows the results of a regression at the model level, analyzing which features of a model predict a Democratic or Republican lean. We find, for example, that models produced by non-profits (rather than for-profit corporations) produce more Democratic-leaning models. As well, we find that xAI's Grok models are on average the most conservative.

\subsection*{Robustness to Prompt Framing}
A natural concern with survey-style measurement of LLM ideology is that the wording of the prompt itself may drive the responses. To address this, we re-ran the CES voter prompt under five alternative framings and, for each model, computed Cohen's $\kappa$ between the model's answers under the baseline ``pretend you are a U.S. voter'' prompt and its answers under each variation, on the union of the CES 2022 and CES 2024 item sets (92 aligned items per model). We group the variations into three families: (i) \textit{voter rewordings} (\texttt{imagine}, \texttt{state}, and \texttt{indicate}), which retain the voter framing but change the wording of the instruction; (ii) \texttt{no\_pretend}, which drops the role-play framing entirely (``Do you Support or Oppose\ldots''); and (iii) \texttt{asLLM}, which inverts the framing and asks the model to answer ``as an LLM'' rather than as a voter. $\kappa$ is computed over three classes \{Support, Oppose, Other\}, where ``Other'' captures refusals and non-binary responses so that prompt-induced refusal is counted as a behavior change rather than filtered out. Supplementary Table~3 summarizes the results for the 33 models that take a stance on at least half of the items under every variation; Supplementary Tables~4--6 report per-variation breakdowns---overall $\kappa$, binary $\kappa$, and coverage---for all 43 models.

Across these 33 models, the voter rewordings produce an average $\kappa$ of approximately $0.7$ against the baseline, indicating that the substantive ideological positions we recover are essentially invariant to the precise verb used to elicit them. Removing the ``pretend'' framing (\texttt{no\_pretend}) yields a similar average $\kappa$ of approximately $0.7$, suggesting that the role-play instruction is not what produces the ideological patterns we document. The ``as an LLM'' framing (\texttt{asLLM}) is the most disruptive---several models refuse to take a position when explicitly addressed as an LLM (visible as low coverage in Supplementary Table~5)---but among the models that still answer the majority of items, the average $\kappa$ against the baseline remains above $0.6$. Together these checks indicate that our ideological estimates are not artifacts of the specific role-play wording: the same models look similarly partisan across a wide range of prompt framings.

\subsection*{Where Does the Partisan Signature Come From? Base vs.\ Instruct Models}
The robustness check above shows that the partisan patterns we measure do not depend on the precise prompt wording. A related but distinct question is whether they depend on the specific stage of model training that produces them: are these patterns already present in the model's raw next-token predictions, or are they introduced by the post-training alignment step (supervised fine-tuning + RLHF) that converts a pretrained \emph{base} model into a chat-ready \emph{instruct} model? To address this we re-administered the CES 2022 and CES 2024 policy items to both the base and the instruct variant of three model families---Falcon~3, Llama~3.1~70B, and Qwen~2.5~7B---running each variant five times under each of two response-option orderings (Support/Oppose and Oppose/Support), to verify that the responses we measure are not artifacts of position bias.\footnote{We initially included Mistral Nemo in this comparison, but its base variant exhibited complete position bias---answering Support on every item under the Support/Oppose ordering and Oppose on every item under the Oppose/Support ordering---so we exclude it from the substantive analysis. The other three base models are robust to response-option order, with modal-answer agreement above $0.85$ across the two orderings.} For each variant we then aggregate to a single modal response vector per survey and project it onto the corresponding CES PCA dimension (fit on the respondents from that year) so that scores are directly comparable to the partisan benchmarks (Strong Democrat through Strong Republican).

\begin{figure}[hbtp]
    \centering
    \includegraphics[width=0.95\linewidth]{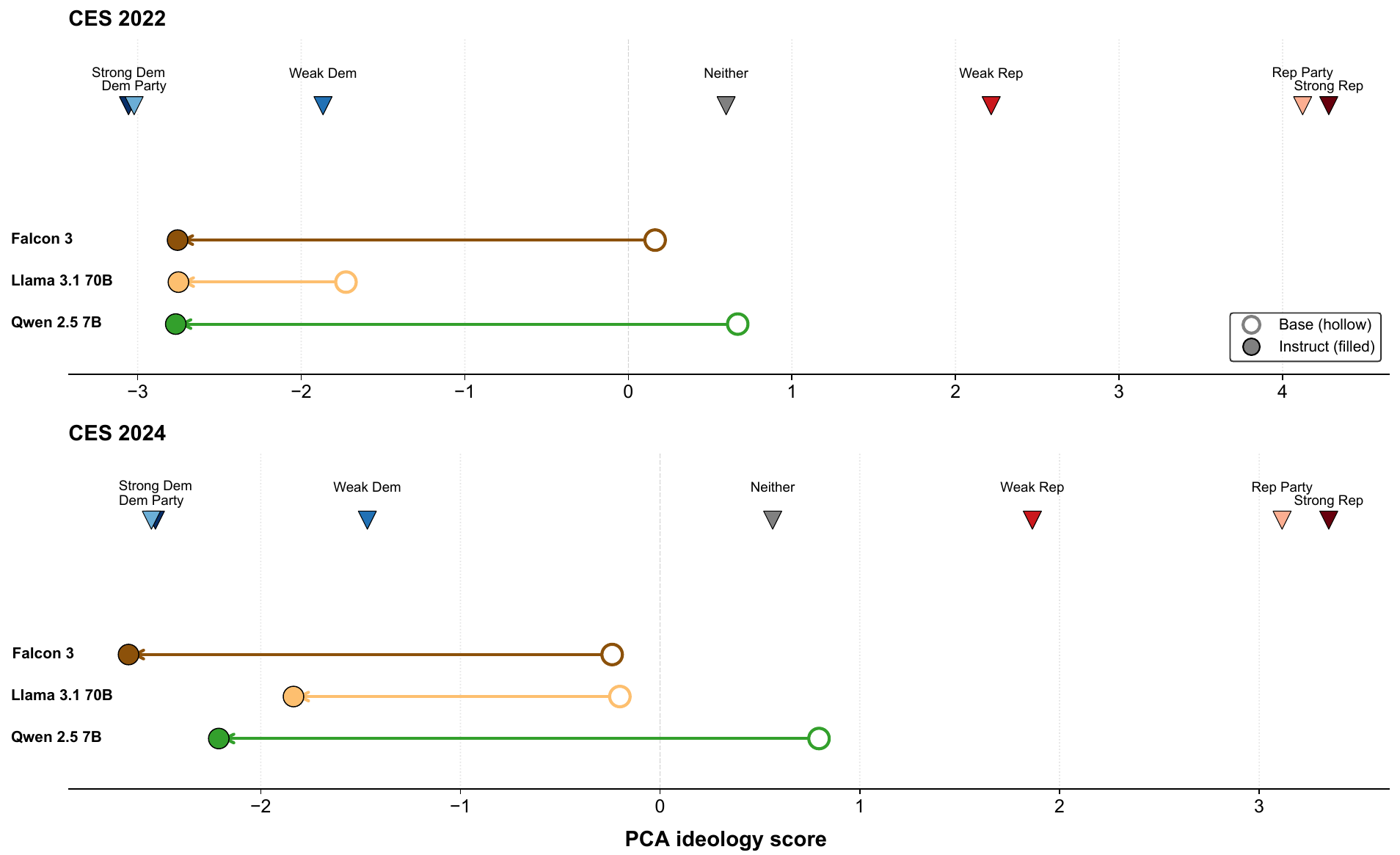}
    \caption{\textbf{Instruction tuning, not pretraining, produces the partisan signature.} Projection of base and instruct variants of three model families (Falcon~3, Llama~3.1~70B, Qwen~2.5~7B) onto the CES 2022 (top panel) and CES 2024 (bottom panel) PCA ideology dimensions. Each model variant's response vector is aggregated from $5 \times 2$ runs (5 repetitions $\times$ two response-option orderings: Support/Oppose and Oppose/Support). Within each panel: the top row shows average partisan scores for CES respondents grouped by 7-point party identification (from Strong Democrat on the left to Strong Republican on the right) as reference markers; the bottom three rows show base variants as hollow circles and instruct variants as filled circles, connected by arrows indicating the direction of the shift induced by instruction tuning. In both years, all three base models sit near the political center; all three instruct variants converge near the Strong-Democrat benchmark. Mistral Nemo is excluded because its base variant exhibits complete response-option-order bias.}
    \label{fig:base_vs_instruct}
\end{figure}

Figure~\ref{fig:base_vs_instruct} reports the results. In both CES 2022 (top panel) and CES 2024 (bottom panel), the three base models sit near the political center---Falcon~3 base, Llama~3.1~70B base, and Qwen~2.5 base all land within about one point of the ``Neither'' partisan benchmark, with the exception of Llama~70B in CES 2022 ($-1.73$, between Weak and Strong Democrat). After instruction tuning, all three variants shift sharply leftward and converge near the Strong-Democrat benchmark: in CES 2022 the instruct scores cluster around $\theta \approx -2.75$ (Strong Dem mean $-3.06$), and in CES 2024 they cluster between $-1.84$ and $-2.66$ (Strong Dem mean $-2.53$). The size of the leftward shift varies by family but is uniformly in the same direction: Qwen moves $3.4$ points (CES 2022) / $3.0$ points (CES 2024), Falcon $2.9$ / $2.4$ points, and Llama~70B---already on the liberal side as a base model in CES 2022---moves $1.0$ / $1.6$ points. The implication is that the partisan signature we document across LLMs is not principally a function of pretraining data; it is introduced or substantially amplified by the alignment step, in which the model is fine-tuned on human preference data and refused-/preferred-response demonstrations. This finding is consistent with prior work documenting that RLHF systematically alters the expressed political stances of frontier models~\cite{santurkar2023whose, perez2023discovering}, and it places useful boundaries on what our ideological estimates measure: they capture the political behavior of LLMs as they are actually deployed to users, not a property of the underlying language model alone. In an exploratory extension, tuning also changes how \textit{consistent} these positions are. On the same response-variance measure of ideological consistency we apply to voters and deployed models in the next section, the three base models are as inconsistent as or more inconsistent than the median voter, expressing a near-even mix of conservative and liberal positions across the CES items, while each instruction-tuned variant falls to the variance level of strong partisans and of our deployed 43-model roster (Supplementary Table~13). The alignment step, in other words, appears not only to shift models leftward but to organize their expressed ideology into a more internally coherent profile.

\section*{LLMs are as Ideologically (In)consistent as Voters}

We have shown so far that different comparison groups -- and different measurement strategies -- produce different estimates of LLMs' expressed ideological positions. Compared to Supreme Court justices, most LLMs are quite moderate. Compared to most voters, however, LLMs appear strikingly liberal, aligning most closely with voters who consider themselves strong Democrats. Moreover, considering a one-dimensional projection of ideology positions most LLMs as strongly moderate compared to US legislators, but a two-dimensional projection reveals that LLMs express positions much more liberal on the Civil Rights and Social Justice dimension than even most Democratic members of Congress. Taken together, these results suggest that our estimates of LLMs' expressed positions are masking unexplored variation. We explore that variation next.

\begin{figure}[hbtp]
    \centering
    \includegraphics[width=0.9\linewidth]{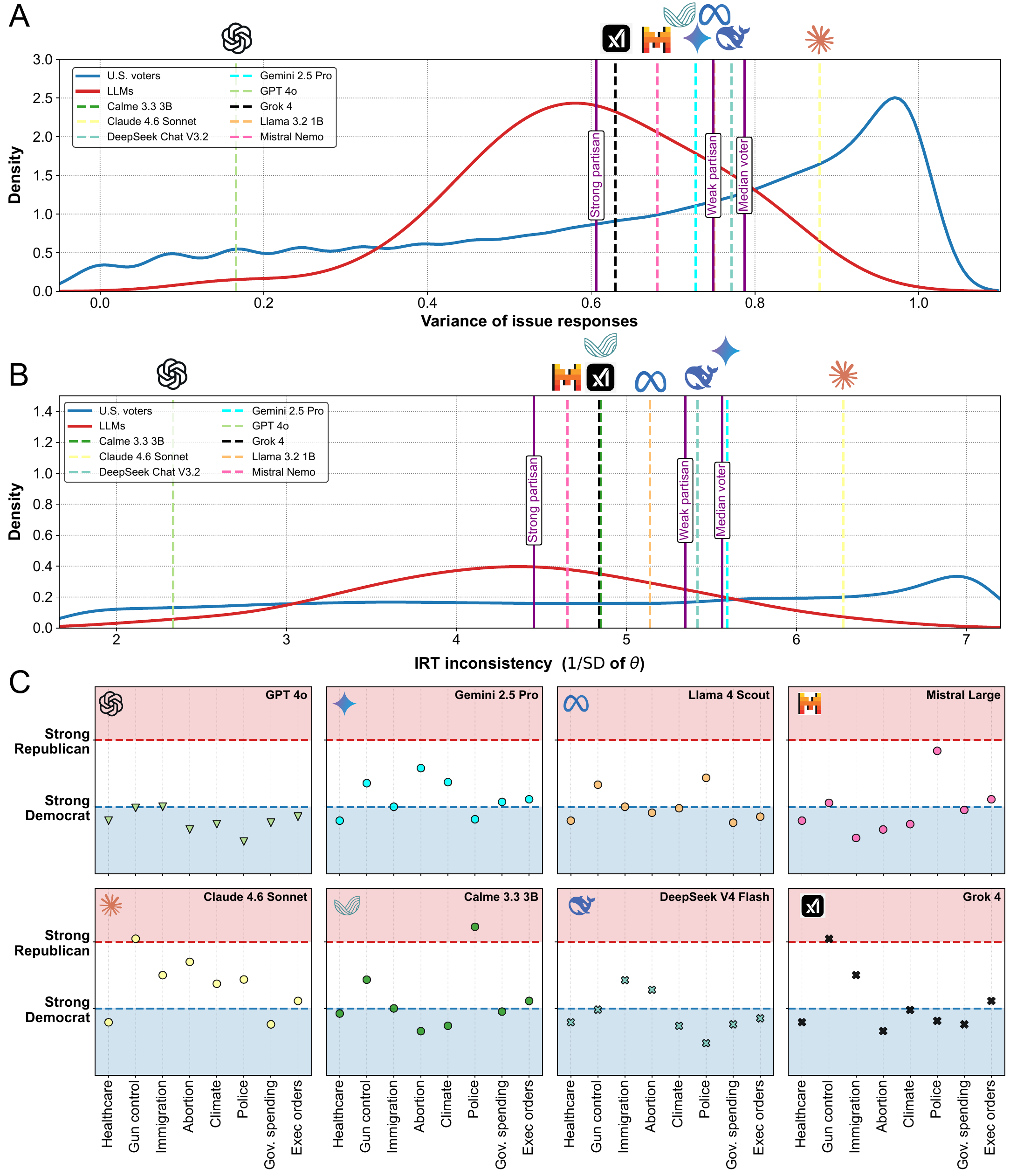}
    \caption{\textbf{Inconsistency in political opinions.} \textbf{A.} Variance of $\pm 1$ conservative-coded responses on the 46 CES 2022 policy items, for 60{,}000 CES respondents (blue density) and 41 LLMs (red density). Eight named LLMs are marked as dashed verticals. \textbf{B.} Inverse of the posterior SE ($1/\mathrm{SE}$) of the latent IRT $\theta$ estimated from the same 46 items via a 2PL model fit with \texttt{mirt}; higher values correspond to more inconsistent answers (in line with the variance direction in Panel A). \textbf{C.} The PCA score of eight LLMs across the eight policy issues, normalized based on the PCA scores of strong Republican and strong Democrat participants.}
    \label{fig:selected_topics}
\end{figure}

The bulk of political science and psychology research argues that an individual's ideology consists of a bundle of many different opinions, and the average of those opinions may not be very informative of any given preference. Moderate voters, for example, can be moderate in any of three ways: they may hold consistently moderate opinions, they may hold no opinions at all, or they may hold offsetting strongly partisan opinions. Likewise, our LLMs' overall partisanship masks substantial variation across topics and questions: even our most moderate LLMs express strongly partisan positions on some issue areas, as we show in Figure~\ref{fig:selected_topics}.

Panel C in Figure~\ref{fig:selected_topics} compares eight LLMs' ideology scores across each of eight issues, showing that most models consist of bundles of strongly partisan positions. Each plot's y-axis indicates the average positions of Strong Republicans (red dashed line, top) and Strong Democrats (blue dashed line, bottom). In the top left plot, OpenAI's GPT-4o is the closest LLM to a consistent partisan: nearly all of its issue area positions are close to those of Strong Democrats, though it is usually more liberal. In contrast, Mistral Large (top right) expresses liberal positions on social issues like healthcare, immigration, and abortion, but a strongly conservative position on policing. Critically, different models are strongly partisan on different issues. Among these eight, Claude expresses the most conservative positions on abortion, Calme on policing, and Gemini on climate. To quantify how far each model sits from the electorate, Supplementary Tables~11 and~12 report each model's position on every CES 2022 and 2024 issue, respectively, relative to the median voter and the strong-partisan benchmarks, on a PCA scale where $0$ marks the average strong Democrat and $1$ the average strong Republican. The median voter sits near $0.30$ on this scale, to the right of the typical model position on most issues. Across all model--issue combinations, roughly half of the expressed positions are more liberal than the average strong Democrat and only about $3\%$ are more conservative than the average strong Republican. This leftward tilt is uneven across issues. It is concentrated on issues such as healthcare, government spending, climate, and welfare, where a large majority of models sit to the left of strong Democrats, while on gun control, immigration, and taxes most models fall between the two partisan benchmarks. Positions to the left of strong Democrats correspond to views held by only a minority of American voters.

Panel A in Figure~\ref{fig:selected_topics} quantifies the ideological inconsistency of each survey respondent and each LLM on the CES 2022 questionnaire. We recode each question such that $+1$ is the conservative response and $-1$ is the liberal response, then calculate the population variance ($ddof = 0$) of each respondent's and each LLM's 46 answers. The distribution of respondents' ideological inconsistency is represented by the blue density distribution. In line with foundational works in political behavior \cite{converse2006nature}, most respondents are highly inconsistent, with the bulk of the distribution above $0.75$ and the median voter at around $0.79$. Even strong partisans---those who identify as Strong Republicans or Strong Democrats---average a variance of $0.61$, and the most ideologically consistent respondents (variance below $0.3$) make up only a small left tail.

The red density distribution shows the ideological inconsistency of LLMs. The mean LLM variance is $0.60$, essentially indistinguishable from the strong-partisan group, with the modal LLM sitting at or below the strong-partisan mean. In short, LLMs appear approximately as ideologically consistent as the most well-informed, most strongly partisan US voters. As discussed in our theoretical framework, this parallel is descriptive rather than mechanistic: the sources of consistency in LLMs (training data correlations, alignment procedures) differ fundamentally from those in humans (cognitive dissonance, social identity), but the behavioral similarity is relevant for anticipating downstream effects on users.

Panel B represents this same result using a different measure of ideological inconsistency. We fit a 2PL IRT model via \texttt{mirt}~\cite{chalmers2012mirt} to recover each respondent's and each LLM's latent ideological position $\theta$ from the same 46 CES items, then plot the inverse of the posterior standard error ($1/\mathrm{SE}$) of those $\theta$ estimates as a measure of inconsistency that runs in the same direction as Panel A (higher value $\rightarrow$ more inconsistent). Strong partisans land near $1/\mathrm{SE} = 4.5$, weak partisans near $5.4$, and the median voter near $5.6$; this ordering reverses the raw SE because the latent-trait test information is highest at the dense middle of the item-difficulty distribution and lowest at the extremes. LLMs span roughly the same range as voters on this metric, with a long left tail of well-located models and a small cluster on the right that resemble the apolitical median voter.

\begin{figure}[htbp]
\begin{centering}
\includegraphics[width=1\linewidth]{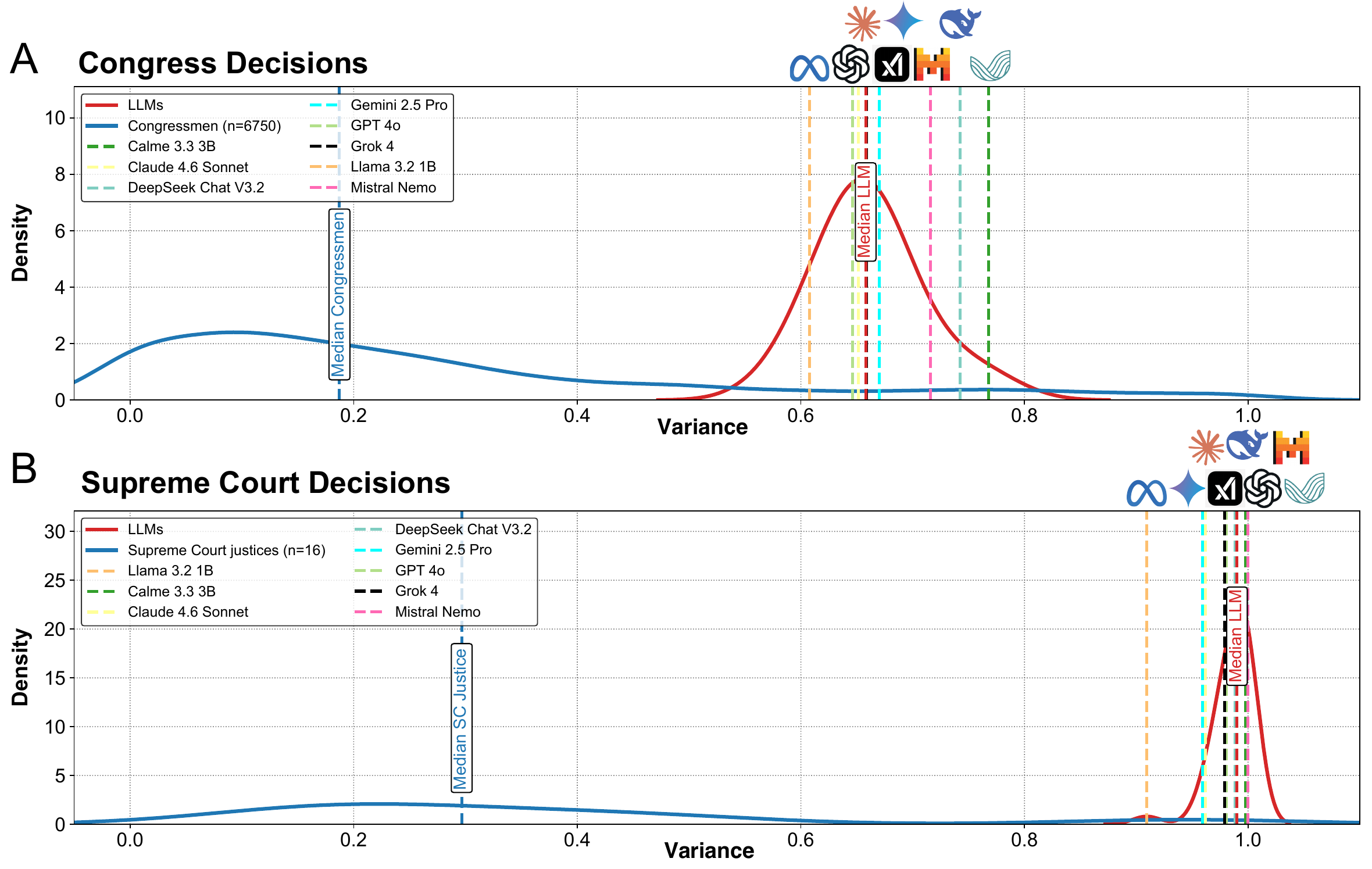}
\par\end{centering}
\caption{\textbf{Within-party / within-coalition inconsistency on bills and SCOTUS cases.} \textbf{A.} For 1{,}327 members of the 110th--119th Congresses, variance of voted-with-own-party-modal coded as $\pm 1$ (one row per (member, Congress); pooled Overall rows are also included). \textbf{B.} For the 16 Supreme Court justices serving any portion of the 2006--2026 window, variance of voted-with-own-party-modal across 1{,}749 cases. In both panels the LLM red density measures variance of matching the Republican party's modal vote; the blue density shows the corresponding human distribution. Eight named LLMs are marked as dashed verticals colored by family.}
\label{fig:bills_scotus_variance}
\end{figure}

Compare these results to those in Figure~\ref{fig:bills_scotus_variance}, which plots an analogous within-coalition inconsistency measure for legislators (110th--119th Congresses, Panel A) and Supreme Court Justices (2006--2026, Panel B), alongside LLMs. For human voters we compute, for each item, the modal vote of the voter's own party (Democratic or Republican) and code each vote as $+1$ (with party modal) or $-1$ (defection). For LLMs we instead code each vote as $+1$ if it matches the Republican modal vote and $-1$ if it matches the Democratic modal vote---so the LLM ``variance'' here measures how often it flips between siding with the two parties. Legislators (Panel A) and justices (Panel B) are highly consistent within their own party (median variance $0.19$ for Congressmen and $0.30$ for justices), while LLMs are essentially split 50/50 between matching the Republican and Democratic modals on most items (median variance $0.66$ for bills and $0.99$ for cases). This is the opposite of the CES pattern: on policy-opinion surveys, LLMs return consistent ideological positions; on legislative and judicial dockets, where the conservative-vs-liberal direction is rarely obvious from the text, their votes scatter. Supplementary Figure~18 compares LLM responses against each of the 110th to 119th Congresses.

\section*{Ideological Differences Extend to Organic Conversations}
 
{
The preceding analyses measure LLMs' expressed ideology in structured settings: roll call votes, court decisions, and survey items. These are useful for placing models on established ideological scales, but they are far removed from how most people interact with LLMs. A natural question is whether the ideological differences we observe in structured settings manifest in the kinds of open-ended conversations that characterize everyday use. If models produce substantively similar responses to organic political queries---despite differing on bills and cases---the practical implications of their measured ideology may be limited. If, however, model choice produces meaningfully different ideological framing in naturalistic conversations, this strengthens the case that users' political exposure depends on which model they happen to use.
 
To test this, we drew on a large corpus of real-world ChatGPT interactions~\cite{zhao2024wildchat} and used an LLM-based classifier to identify the political conversations within it (see Supplementary Note 3). The resulting set of 1,019 political conversations---all conducted with GPT-3.5 and GPT-4 models---spans topics including domestic policy, international relations, historical political events, and social issues. For multi-turn conversations, we extracted the first substantive political question posed by the user; in approximately 20 cases where the first message was trivial (e.g., a greeting), we used the next political turn. 
 
We replayed each user prompt through five alternative LLMs spanning five distinct AI laboratories: Mistral Large 3 (Mistral), Qwen3 32B (Alibaba), Claude Sonnet 4.6 (Anthropic), Grok 4.1 Fast (xAI), and DeepSeek V4 Flash (DeepSeek). To evaluate the ideological leaning of each response, we assembled a panel of three judge LLMs from three additional laboratories not represented among the generation models~\cite{zheng2023judging, verga2024replacing}: GPT-4o (OpenAI), Claude Opus 4.6 (Anthropic), and Mistral Magistral (Mistral). Each judge was presented with the user's original question alongside all six responses (the original GPT response plus five alternatives), labeled anonymously (A through F) with labels independently shuffled per judge to mitigate position bias. Judges performed two tasks: classifying each response as \textit{conservative}, \textit{liberal}, or \textit{neutral}, and ranking all six from most conservative to most liberal.
 
Of the 1,019 conversations, 883 received valid judgments from all three judges (GPT-4o: 906, Claude Opus: 1,014, Mistral Magistral: 998). We aggregate classifications using majority vote, requiring agreement from at least two of three judges.
 
\subsection*{Results: Model Choice Produces Meaningful Ideological Differences}
 
Table~\ref{tab:conversation_ideology} presents the main results. We summarize each model's ideological tilt by its \textit{net liberal lean}: the percentage of its responses that a majority of judges classified as liberal minus the percentage classified as conservative (in percentage points), so that higher values indicate a more liberal slant overall. The original GPT responses are largely neutral (80.8\%) and only modestly liberal on balance---12.8\% were classified as liberal against 6.5\% as conservative, a net liberal lean of $+6.3$pp. Even though OpenAI's GPT family is among the most consistently liberal models in the structured comparisons above, in the domains users actually discuss with LLMs \textit{every} alternative model leans more liberal still than the original GPT responses, with net liberal leans ranging from DeepSeek V4 Flash's relatively modest $+12.5$pp to Mistral Large 3's substantial $+23.5$pp. All differences from the original are statistically significant ($\chi^2$ tests, $p < 0.05$ for all comparisons; $p < 0.001$ for all except DeepSeek).

\begin{table}[htbp]
\centering
\small
\centering
\begin{tabular}{lcccccc}
\toprule
\textbf{Model} & \textbf{Lab} & \textbf{Con. (\%)} & \textbf{Neu. (\%)} & \textbf{Lib. (\%)} & \textbf{Net lean (pp)} & \textbf{$\chi^2$} \\
\midrule
Original (GPT-3.5/4)  & OpenAI   &  6.5 & 80.8 & 12.8 &  $+$6.3  & --- \\
\hdashline
Mistral Large 3        & Mistral  &  6.0 & 64.4 & 29.6 & $+$23.5  & 71.7$^{***}$ \\
Qwen3 32B              & Alibaba  &  2.9 & 76.2 & 20.9 & $+$18.0  & 24.1$^{***}$ \\
Claude Sonnet 4.6      & Anthropic&  2.8 & 79.6 & 17.6 & $+$14.8  & 17.3$^{***}$ \\
DeepSeek V4 Flash      & DeepSeek &  4.1 & 79.4 & 16.6 & $+$12.5  &  9.1$^{*}$ \\
Grok 4.1 Fast          & xAI      & 13.4 & 66.2 & 20.4 &  $+$7.1  & 22.0$^{***}$ \\
\bottomrule
\end{tabular}
\caption{\textbf{Ideological classification of LLM responses in organic political conversations.} Majority-vote classifications from a three-judge panel (GPT-4o, Claude Opus 4.6, Mistral Magistral). Net lean = \% liberal $-$ \% conservative. All alternative models differ significantly from the original ($^{***}p<0.001$; $^{**}p<0.01$; $^{*}p<0.05$).}
\begin{tablenotes}
\footnotesize
\item \textit{Note:} $n = 883$ conversations with all three judges valid. Labels shuffled per judge. Inter-judge reliability: Cohen's $\kappa = 0.38$ (average across judge pairs). Unanimous agreement: 51.6--63.5\%.
\end{tablenotes}
\label{tab:conversation_ideology}
\end{table}
 
Notably, the ideological differences between models are not uniform across topics. Mistral Large 3 and Qwen3 32B are the most liberal on criminal justice and economic systems, while Claude Sonnet 4.6 and DeepSeek V4 Flash more closely resemble the original GPT responses in their neutrality profile. Topics exhibiting the largest divergence between the original and alternatives include criminal justice reform, racial justice, death penalty abolition, economic inequality, and free education---areas where the original GPT response was typically classified as neutral but a majority of alternatives were classified as liberal. These conversational patterns mirror our CES results at the aggregate level---a predominantly liberal lean and wide cross-model variation---indicating that the skew we document on structured survey items is not a survey-format artifact but extends to the conversations users actually have, even if the agreement is directional rather than issue-by-issue.
 
Grok 4.1 Fast presents a distinctive ideological profile that merits separate discussion. Its aggregate net lean ($+$7.1pp) is closest to the original GPT among all alternative models, but this apparent moderation masks a fundamentally different pattern. Grok has both the highest conservative classification rate (13.4\%) \textit{and} a high liberal rate (20.4\%), giving it the lowest neutral rate of any model (66.2\%). In this respect Grok mirrors the broader inconsistency we document above through response variance, in that its moderate net lean masks strongly opposing positions across conversations rather than genuine centrism, much as LLMs' moderate average ideology masks strong, offsetting positions across issues.
 
This polarization is topic-dependent. On Nazi alternate-history prompts ($n = 66$), Grok is classified as conservative in 59.4\% of cases, far exceeding any other model. On war and military topics ($n = 90$), Grok is the only model that leans conservative (net $-$5.7pp), while all others lean liberal. On surveillance and privacy topics, Grok similarly leans conservative ($-$12.5pp). Conversely, on criminal justice and economic systems, Grok leans liberal at rates comparable to or exceeding the most liberal models. This bidirectional pattern---conservative on security and national strength, liberal on social justice---is unique to Grok and is consistent with a libertarian-populist ideological profile rather than the neutral-by-default posture adopted by most other models.
 
Within the original responses, GPT-4 (net $+$2.1pp) was significantly more ideologically balanced than GPT-3.5 (net $+$8.6pp). GPT-4 had both a higher conservative classification rate (10.0\% vs.\ 8.2\% for GPT-3.5 March) and a lower liberal rate (12.1\% vs.\ 16.8\%), suggesting that the model update moved OpenAI's flagship product toward greater neutrality. The June 2023 GPT-3.5 update showed a different pattern: it achieved the highest neutral rate (86.7\%) but nearly eliminated conservative responses (1.9\%), resulting in a liberal lean comparable to the March version.
 
A striking feature of these results is the asymmetry of the ideological landscape. There are far more topic areas in which all alternative models converge on liberal framing than on conservative framing. Only Nazi alternate-history topics produce a consistent conservative lean across models, and even there it is driven primarily by Grok and Mistral. In contrast, topics like criminal justice reform, racial equality, economic inequality, and environmental policy produce near-unanimous liberal framing across all five alternatives. This asymmetry implies that the liberal lean we and others have documented in structured settings is not merely an artifact of survey-style prompting; it is a robust feature of LLM behavior that extends to naturalistic conversation.
 
These findings establish that model choice produces meaningful, statistically significant differences in the ideological framing of responses to everyday political queries. Users who interacted with GPT-3.5 or GPT-4 received substantially more neutral responses than they would have received from any of the five alternative models we tested. This motivates our next question: if LLMs express systematic ideological positions in the conversations people actually have with them, do these positions influence users' own political attitudes?
}

\section*{LLMs Exert Persuasive Influence on Political Attitudes}

Do the ideological patterns expressed by LLMs influence the attitudes of their users? If LLMs express systematic ideological positions and users consult LLMs in forming their own political opinions, we might expect that those expressed positions would affect voters' attitudes. On the other hand, scholars of political behavior have long noted that behaviors like vote choice are remarkably difficult to influence, though perhaps those voters who consult LLMs about politics are those most likely to be swayed.

To test whether LLMs can influence users' expressed attitudes, we need to know which direction we expect those attitudes to shift when exposed to LLMs. We have shown above that a model's \textit{average} ideological disposition is not necessarily the same as its topic area disposition: an experiment that exposes users to an LLM and expects their attitudes to move in a liberal direction about climate change may fail to find the results they expect if that model's expressed positions are conservative about climate change despite being liberal overall. To estimate the persuasive effects of LLMs, we require topic-specific ideological estimates as we've produced above.

{
We conducted a pre-registered\footnote{An anonymized link to our experimental pre-registration and pre-analysis plan is available at \url{https://osf.io/5kz7d/overview?view_only=07c95f7b13744bb0b2d82e94b22d1261}.} survey experiment to answer this question. We recruited U.S. participants through the survey provider Prolific~\cite{douglas2023data} into a held-out confirmatory sample, pre-registered to comprise only data collected on or after a fixed cutoff date, so that our confirmatory tests are conducted on data unseen during design and piloting. The analyzable sample comprises 1{,}523 participants contributing 6{,}072 topic-level observations, after pre-registered exclusions of 45 participants who triggered a hidden attention canary and 15 who failed satisficing checks; a per-protocol sensitivity sample described below retains 1{,}470 participants and 5{,}681 observations. Recruitment was balanced by participants' Prolific-profile party affiliation, targeting equal numbers of Democrats, Independents, and Republicans in each wave, but all analyses use each respondent's self-reported party from our own instrument, which reclassifies many profile-labeled Independents who lean toward a party. By self-report, the analyzable sample comprises 610 Democrats, 363 Independents, and 550 Republicans. This divergence between recruited and analyzed party is by design, since the pre-registration sorts every respondent by self-report.

The experiment is fully within-subjects. Every participant considered all four topics (gun control, immigration, police, and taxes), each presented with one policy proposal drawn from a pool of two. For each topic independently, a server-side randomizer assigned the respondent to one of six conditions. Four are advocacy conditions that cross the direction of the model's expressed position (conservative or liberal) with the strength of steering (explicit or base); the remaining two are a neutral condition and a no-conversation control. In the explicit conditions the model received a gentle system prompt instructing it to argue the assigned side of the shown proposal, whereas in the base conditions the same model was told only to stay on topic and behave naturally, with no steering. Strength therefore varies only the system prompt and never the model, so the base condition isolates a naturally, informationally-behaving LLM. Advocacy models were selected per topic and pole from our own ideology-scaling estimates, taking the most conservative-leaning and most liberal-leaning models on each topic. For gun control, for instance, conservative-leaning models such as Phi 4, Grok 4.3, and Claude 4.6 Sonnet were pitted against liberal-leaning models such as Gemini 2.5 Pro, GPT-4o, and Mistral Large, while immigration paired conservative-leaning GLM 5 Turbo and Claude 4.5 Haiku against liberal-leaning Llama 4 Scout and Nemotron 3 120B. The neutral condition used Claude 4.5 Sonnet, which appears in no advocacy pole, with a balanced present-both-sides prompt. Assignment was single-blind, so respondents saw only whether a chat box was present, never the condition, model, or direction, and topic order was randomized. Base advocacy was deliberately over-weighted in the randomizer because the informational (base) contrast is the hardest to power and our featured question. Respondents assigned to a chat condition discussed the proposal with the model in a timed three-minute conversation before recording a single Support or Oppose stance on the proposal, and control respondents recorded their stance directly. A hidden canary word detected respondents pasting in an external chatbot. The study was deemed exempt by the authors' Institutional Review Board (protocol number anonymized for review).
}

{
Our outcome for each topic is whether the respondent's post-conversation stance aligned with the side implied by the assigned model's measured ideological position on that topic; that measured position was never shown to the respondent. Because each respondent contributes up to four such observations, our unit of analysis is the topic-slot, and we model alignment with a mixed-effects logistic regression that includes a respondent random intercept, proposal fixed effects within topic, and a pre-registered block of covariates (party, ideology, age, gender, education, income, pre-treatment trust in and perceived neutrality of LLMs, LLM familiarity, political engagement, and news breadth). We use a respondent random intercept rather than respondent fixed effects because the outcome varies within respondent across four independently randomized topics. Our estimand is a toward-advocated contrast, the difference in alignment between the conservative-pole and liberal-pole conditions, so that an odds ratio above one denotes movement toward whichever side the model expressed. We report odds ratios alongside average marginal effects in percentage points. Our primary estimand is intention-to-treat, retaining the 391 silent chat-slots whose participants idled through the timer without messaging, and a per-protocol analysis that drops these slots is reported as a sensitivity check.

We pre-registered five hypotheses. LLMs persuade on average (H1); naturally-behaving base models persuade (H2, our featured hypothesis); explicit steering amplifies persuasion relative to base (H3); a battery of six respondent-level moderators shapes the effect (H4a--f); and the neutral condition is equivalent to control (H5, a validity check). All confirmatory tests are two-sided at the $.05$ level, with Holm correction across \{H1, H2, H3\} and three per-stratum Benjamini--Hochberg families of six for the moderators. The study follows a pre-registered futility-only sequential design, recruiting in waves of 500 toward a registered maximum of 2{,}000 participants and permitting early stopping for futility alone. Recruitment stopped under this rule at the 1{,}500-participant look, where the conditional power for the featured base contrast fell below the registered floor of $0.20$. Because futility stopping can never produce an early rejection, all confirmatory tests are run once at the achieved sample, with no adjustment of the significance threshold for the interim looks.
}

\subsection*{Results: Political Persuasion}

\begin{figure}[hbtp]
    \centering
    \includegraphics[width=\linewidth]{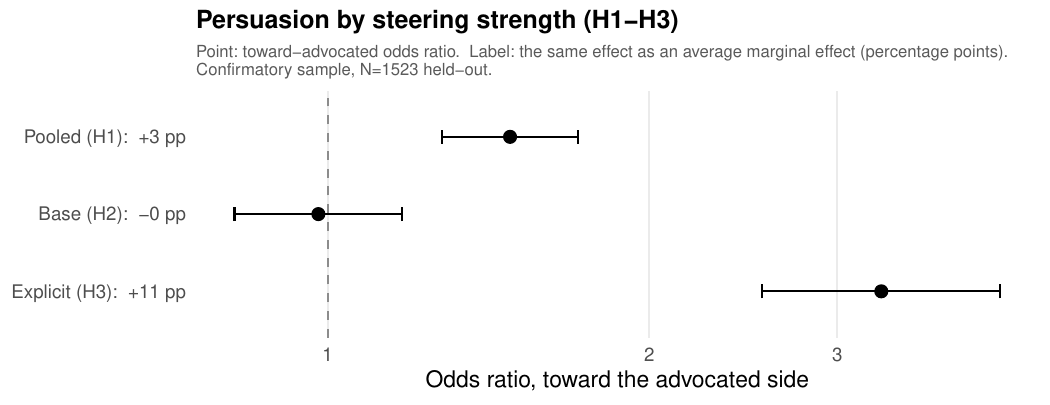}
    \caption{\textbf{Persuasion rises with the strength of steering.} Toward-advocated odds ratio (point, with 95\% confidence interval) for the pooled treatment (H1), the base/natural condition (H2), and the explicit condition (H3), from the held-out confirmatory sample; the dashed line at one marks no effect, and labels give the corresponding average marginal effect in percentage points. Odds ratios above one indicate movement toward whichever side the model expressed. The gradient from base to explicit is the paper's central persuasion result.}
    \label{fig:persuasion_strength}
\end{figure}

{
The clearest result in our data is that explicitly steering a model to argue a side sharply increases its persuasive effect. Figure~\ref{fig:persuasion_strength} traces the toward-advocated odds ratio from the no-conversation baseline through the base (natural) condition to the explicit condition, and the gradient is monotonic. The explicit-minus-base contrast is an odds ratio of $3.36$ (95\% CI $[2.46, 4.61]$; average marginal effect $+10.6$ percentage points; $p<.0001$), and a pre-registered likelihood-ratio test of the ordered Direction-by-Strength interaction (no advocacy $<$ base $<$ explicit) is significant ($\chi^2 = 66.2$, 2 df, $p<.0001$; Holm-adjusted $p<.0001$). Steering a model to argue a position, rather than letting it discuss the topic naturally, is what turns a measurable expressed ideology into a large shift in expressed attitudes.

Both explicitly steered directions move respondents toward the side they argue. Because our outcome is coded on the conservative side, a conservative-steered model should raise it and a liberal-steered model should lower it, and both do. The conservative-explicit condition raises alignment with the conservative side (OR $2.16$, 95\% CI $[1.74, 2.67]$, $p<.0001$), and the liberal-explicit condition shifts respondents toward the liberal side (OR $0.65$ on the conservative-coded outcome, 95\% CI $[0.52, 0.82]$, $p=.0002$). An odds ratio below one on the liberal arm is persuasion toward the advocated side, not backfire; the two arms differ in magnitude but not in direction.
}

\begin{table}[hbtp] \centering
  \caption{\textbf{Persuasion by steering strength (held-out confirmatory sample).} Mixed-effects logistic estimates of aligning with the assigned model's measured ideological position, coded on the conservative side. The toward-advocated contrast compares the conservative-pole and liberal-pole conditions, so an odds ratio above one denotes movement toward whichever side the model expressed; for the individual liberal arms an odds ratio below one likewise denotes movement toward the advocated (liberal) side. AME is the average marginal effect in percentage points. Estimates are from $1{,}523$ respondents contributing $6{,}072$ topic-slots; tests are two-sided, with Holm correction across H1--H3. The explicit contrast is large and significant, while the base (informational) contrast is null, with a confidence interval that excludes the pre-registered smallest effect of interest.}
  \label{tab:survey1}
\begin{tabular}{@{\extracolsep{5pt}}lccc}
\\[-1.8ex]\hline
\hline \\[-1.8ex]
Contrast / condition & OR [95\% CI] & AME (pp) & $p$ \\
\hline \\[-1.8ex]
\multicolumn{4}{l}{\textit{Primary contrasts (H1--H3)}} \\
H1: Pooled persuasion & $1.48\ [1.28, 1.72]$ & $+3.5$ & $<.0001$ \\
H2: Base / informational & $0.98\ [0.82, 1.17]$ & $-0.2$ & $.829$ \\
H3: Explicit $-$ base & $3.36\ [2.46, 4.61]$ & $+10.6$ & $<.0001$ \\
H3: Direction$\times$Strength (LRT, 2 df) & $\chi^2=66.2$ & -- & $<.0001$ \\
\hline \\[-1.8ex]
\multicolumn{4}{l}{\textit{Each arm vs.\ neutral + control baseline}} \\
Conservative $\cdot$ base & $0.97\ [0.81, 1.17]$ & -- & $.75$ \\
Liberal $\cdot$ base & $0.99\ [0.82, 1.19]$ & -- & $.92$ \\
Conservative $\cdot$ explicit & $2.16\ [1.74, 2.67]$ & -- & $<.0001$ \\
Liberal $\cdot$ explicit & $0.65\ [0.52, 0.82]$ & -- & $.0002$ \\
\hline
\hline \\[-1.8ex]
\end{tabular}
\end{table}

{
Pooling across conditions, LLMs persuade on average. A respondent who discussed a topic with any of our models was more likely to end up aligned with that model's measured position than a control respondent who did not converse (H1 pooled OR $1.48$, 95\% CI $[1.28, 1.72]$, AME $+3.5$ percentage points, $p<.0001$; Holm-adjusted $p<.0001$; Table~\ref{tab:survey1}).
}

How meaningful is this average effect? We benchmark it against the size of effects commonly observed from persuasive advertising in political campaigns, and find that LLMs' persuasive effectiveness is at least as large. For example, \cite{coppock2020small} find that campaign advertisements move candidate favorability by an average of $0.049$ points on a five-point scale, and move vote choice by $0.7$ percentage points, an effect not statistically distinguishable from zero. \cite{hewitt2024experiments} find that the average advertisement moves immediately measured vote choice by $2.3$ percentage points in 2018 downballot races, $1.2$ points in 2020 downballot races, and $0.8$ points in the 2020 presidential race. Our pooled effect of $3.5$ percentage points is therefore larger than any of these benchmarks rather than merely comparable to them. We note that these campaign studies typically measure vote choice or candidate favorability, whereas our study measures policy alignment, so the two sets of estimates involve somewhat different outcome types. We therefore treat the comparison as a benchmark of magnitude rather than a claim that the outcomes are equivalent.

\begin{figure}[hbtp]
    \centering
    \includegraphics[width=\linewidth]{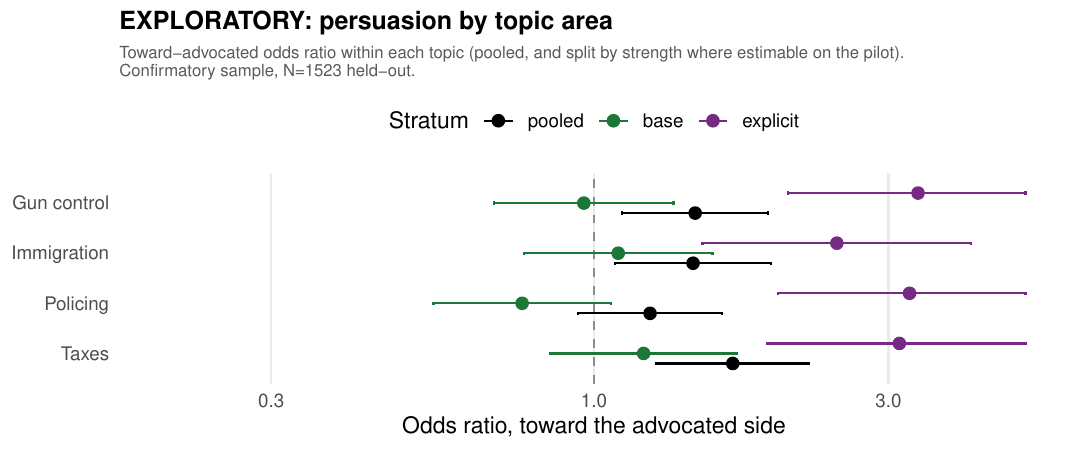}
    \caption{\textbf{Persuasion by topic area (exploratory).} Toward-advocated odds ratio within each of the four topics, pooled and split by steering strength where estimable, from the held-out confirmatory sample. Odds ratios above one indicate movement toward the side the model expressed. These by-topic tests are exploratory and uncorrected.}
    \label{fig:persuasion_topic}
\end{figure}

{
The featured question of our design is narrower. Does a model that is not steered to argue a side, but simply discusses the topic naturally, move attitudes on its own? At our final sample we detect no such effect, and the data bound how large any undetected effect could be. The base (informational) contrast is null (H2 OR $0.98$, 95\% CI $[0.82, 1.17]$, AME $-0.2$ percentage points, $p=.829$), and neither base arm differs from baseline on its own (conservative-base OR $0.97$; liberal-base OR $0.99$). The confidence interval excludes our pre-registered smallest effect of interest, a per-arm odds ratio of $1.25$, corresponding to a contrast odds ratio of roughly $1.56$. Under the registered futility rule, the conditional power of continuing to the maximum sample fell to $0.000$ under the observed trend, below the $0.20$ floor, and recruitment stopped at the 1{,}500-participant look. If naturally-behaving models persuade at all, the effect is smaller than the smallest effect this study was designed to treat as meaningful.

These conclusions are stable across specifications. The pooled, base, and explicit-contrast estimates change little under a per-protocol analysis that drops the silent chat-slots (H1 OR $1.58$; H2 OR $1.02$; H3 contrast OR $3.66$), or when the baseline is restricted to the neutral or the control condition alone, and they are robust to presentation-order and drop-one-topic refits (Supplementary Tables~15 and~18). In an exploratory breakdown by topic (Figure~\ref{fig:persuasion_topic}; Supplementary Table~16), persuasion toward the advocated side is strongest on taxes (OR $1.68$, $p=.0004$) and present on gun control (OR $1.46$, $p=.007$) and immigration (OR $1.45$, $p=.013$), while policing is directionally similar but not statistically significant (OR $1.23$, $p=.13$).
}

\subsection*{Results: Heterogeneous Effects}
\begin{figure}[hbtp]
    \centering
    \includegraphics[width=\linewidth]{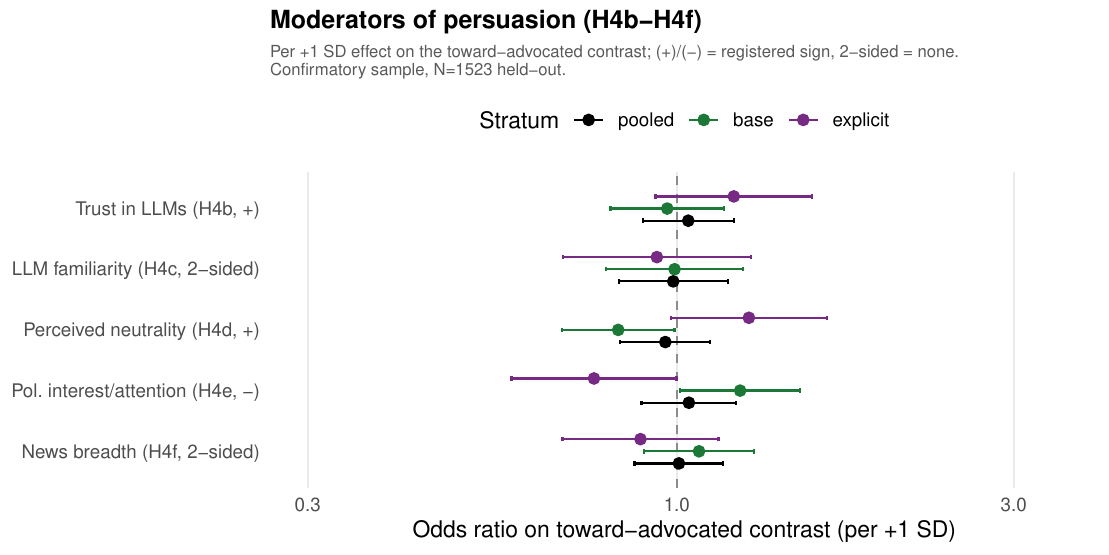}
    \caption{\textbf{No moderator survives correction.} Effect of a $+1$ SD change in each pre-registered receptivity or resistance moderator on the toward-advocated contrast (odds ratio, with 95\% confidence interval), estimated separately in the pooled, base, and explicit strata from the held-out confirmatory sample. The dashed line at one marks no moderation. After per-stratum Benjamini--Hochberg correction, none of these five moderators is significant in any stratum.}
    \label{fig:persuasion_moderators}
\end{figure}

{
LLMs are still a new technology, and a natural question is whether their persuasive power is concentrated among particular respondents, growing as familiarity spreads or confined to the least informed. We pre-registered six moderators of the toward-advocated effect. Five capture respondents' receptivity or resistance (trust in LLMs, LLM familiarity, perceived neutrality of LLMs, political interest, and news breadth), and the sixth, out-partisanship, is taken up in the next subsection. Each moderator is tested in three strata (pooled, base, and explicit) under three per-stratum Benjamini--Hochberg families of six. After correction, none of these five moderators is statistically significant in any stratum (Supplementary Table~14). Figure~\ref{fig:persuasion_moderators} shows the five receptivity and resistance moderators.

The pattern of point estimates is nonetheless informative. Consistent with our expectations and with a large literature on resistance to persuasion~\cite{cobb1997changing, coppock2023persuasion}, LLM familiarity, news breadth, and trust in LLMs show no moderation, so the effect is neither confined to the least engaged respondents nor visibly growing with experience using LLMs. The only nominal signals, for perceived neutrality and political interest, flip sign between strata (perceived neutrality OR $0.83$ per SD in the base stratum, $p=.041$, against $1.26$ in the explicit stratum, $p=.070$; political interest OR $1.23$ in base, $p=.038$, against $0.76$ in explicit, $p=.049$), and none survives correction. We read them as noise rather than findings.
}

\subsection*{Results: Heterogeneity by Party Identification and Direction of Persuasion}

\begin{figure}[hbtp]
    \centering
    \includegraphics[width=\linewidth]{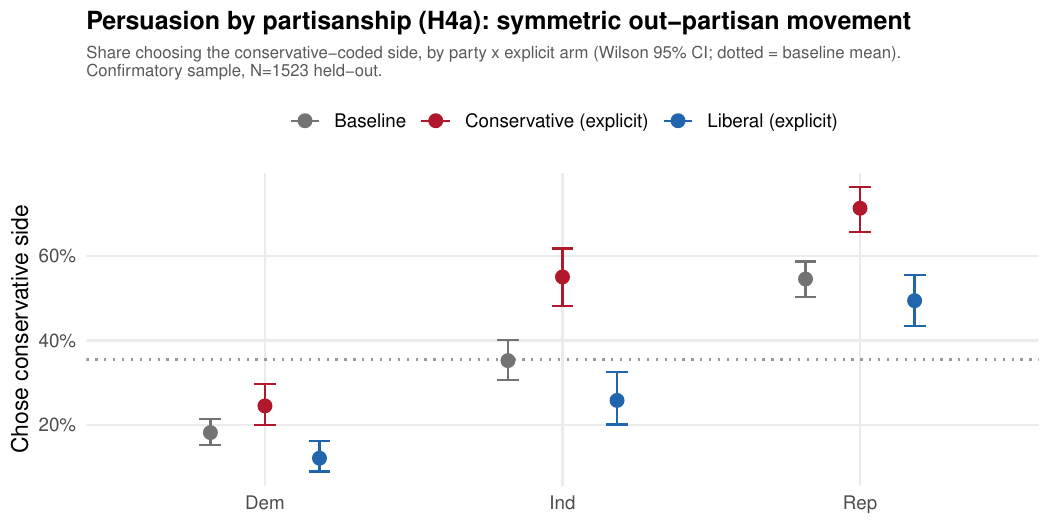}
    \caption{\textbf{Persuasion by party and steering direction.} Share choosing the conservative-coded side, by respondent party (Democrat, Independent, Republican) and explicit arm (conservative, liberal), with the no-advocacy baseline mean marked by the dotted line (Wilson 95\% confidence intervals), from the held-out confirmatory sample. The pre-registered out-partisanship moderator survives correction in the pooled stratum only, and in the direction opposite the registered prediction (out-partisans move less).}
    \label{fig:persuasion_partisanship}
\end{figure}

{
We also examine whether persuasion depends on respondents' partisan identity and on the direction the model argues. Because our design randomizes direction within each topic, some respondents encountered a model expressing conservative positions (for example, Grok 4.3 or Claude 4.6 Sonnet on policing) and others a model expressing liberal positions (for example, GPT-4o or GLM 5 Turbo on policing), so we can ask whether movement toward the advocated side depends on whose side is being argued relative to the respondent's own party. As reported above, both explicitly steered directions persuade toward the side they argue, so the effect is not driven by a single ideological direction.

Our pre-registered moderator for this question is out-partisanship, a per-slot indicator for whether the model argued against the respondent's own party, and we predicted symmetric movement (out-partisans and co-partisans moved to a similar degree). The estimates instead run the other way. Out-partisans move less than co-partisans, not more, in every stratum (pooled OR $0.51$, 95\% CI $[0.31, 0.83]$, AME $-6.1$ percentage points, raw $p=.006$, Benjamini--Hochberg $p=.037$; base OR $0.55$, $[0.31, 0.97]$, Benjamini--Hochberg $p=.082$; explicit OR $0.48$, $[0.24, 0.92]$, Benjamini--Hochberg $p=.140$). This is the opposite of what our discovery pilot suggested, exactly the kind of divergence a held-out confirmatory sample exists to catch. The pooled estimate is the one moderator in the study to survive correction, but its sign contradicts both the registered prediction and the pilot, and it does not survive within either stratum separately, so we disclose it as a reversal to be replicated rather than an established moderator. Figure~\ref{fig:persuasion_partisanship} shows the underlying party-by-arm rates.
}

\subsection*{Results: Mechanism---Persuasive Influence, Not Sycophancy}

\begin{figure}[hbtp]
    \centering
    \includegraphics[width=0.9\linewidth]{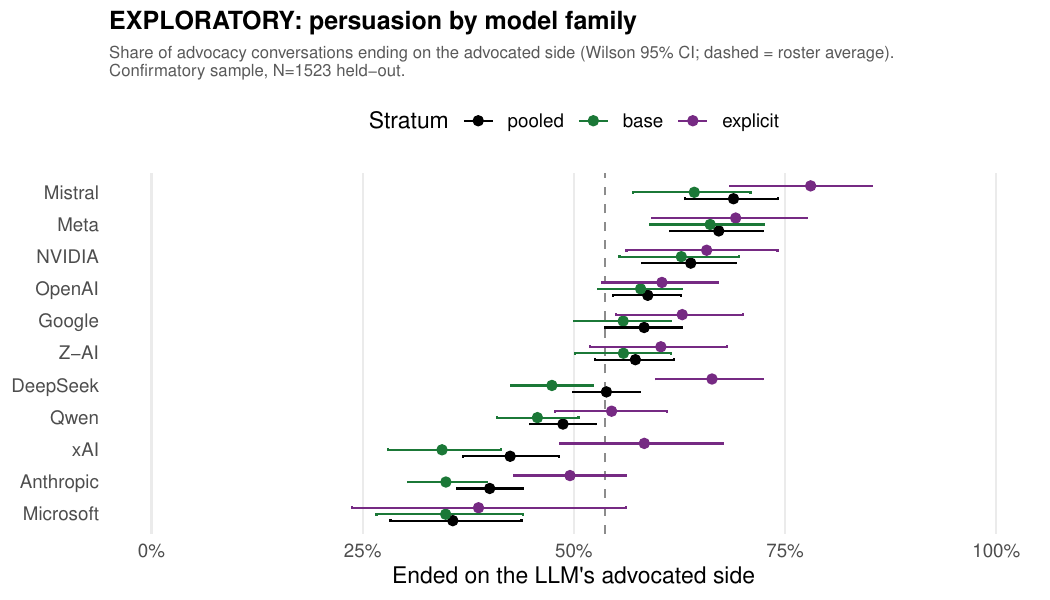}
    \caption{\textbf{Conversations end on the advocated side more often under explicit steering.} Share of advocacy conversations that ended on the model's advocated side, by model family and steering strength, with the roster average marked by the dashed line (Wilson 95\% confidence intervals), from the held-out confirmatory sample. Explicit steering raises the ended-on-advocated share relative to natural (base) conversations, confirming that the manipulation took hold.}
    \label{fig:persuasion_family}
\end{figure}

{
An important alternative interpretation is sycophancy, the tendency of an LLM to detect and mirror a user's existing view, producing the appearance of persuasion without any genuine attitude change. Our design guards against this reading more directly than a single-treatment study could. Respondents never state their prior policy position to the model before the conversation, so there is no stance for it to mirror, and each model's assigned direction was fixed in advance by our scaling analysis rather than by the content of any conversation.

The strength manipulation is the clearest evidence against a pure sycophancy account. The direction a model argued was set by the experimenter and not chosen to flatter the respondent, and yet moving from natural (base) to explicitly steered conversations turns a null effect into a shift of more than ten percentage points. A model that merely agreed with users would not become far more persuasive when instructed to argue a fixed side. Consistent with this, the neutral present-both-sides condition is statistically indistinguishable from the no-conversation control (OR $1.03$, 90\% CI $[0.84, 1.27]$, which includes one), so a balanced conversation does not by itself move respondents. A behavioral check confirms the steering manipulation took hold, since explicitly steered conversations ended on the model's advocated side more often than natural ones (60\% versus 50\%; adjusted OR $1.52$, $p<.0001$; Figure~\ref{fig:persuasion_family} and Supplementary Table~17).

We use the term ``persuasive influence'' rather than ``persuasion'' to emphasize that respondents never opted into persuasive messaging. Even in the explicit conditions the steering was invisible to the respondent, who saw only an ostensibly informational chat, and that framing may leave users less vigilant against influence than an openly persuasive appeal would.
}

\section*{Discussion}

Researchers, professionals, governments, and citizens around the world are urgently grappling with the impact that language models and other imminent artificial intelligence breakthroughs will have on every aspect of our lives. With governments and private corporations competing to produce ever-larger and more powerful models, the risks to private citizens grow more severe. These risks are especially acute in politics, where citizens are relatively uninformed but the rewards to office are immense. 

In this paper we have shown that existing LLMs exhibit large, distinct, and heterogeneous partisan positioning in the US, and that when a model is prompted to argue a side, it can meaningfully shift the expressed political attitudes of the users who converse with it. This adds to a small but growing literature showing that LLMs can be persuasive across domains when they are tasked with doing so~\cite{hackenburg2025scaling,velez2024confronting}. A model that is not steered to argue a side, but simply answers questions naturally, shows no detectable influence in our experiment. That null is informative rather than merely underpowered, since the confidence interval excludes the smallest effect our pre-registration treated as meaningful and recruitment stopped under the registered futility rule. If everyday, unsteered LLM use moves political attitudes at all, the effect is below that threshold.

{
Several features of our findings merit further discussion. First, regarding the nature of LLM ``ideology'': as we emphasize throughout, the ideological patterns we measure are role-conditioned behavioral regularities, not latent beliefs. Different prompting contexts, role instructions, and alignment procedures can shift the positions a model expresses. Our scaling exercises characterize LLMs' expressed positions under specific, reproducible conditions, and our sensitivity analyses demonstrate that the substantive conclusions are broadly robust to variations in prompt design. Nevertheless, the context-dependence of LLM ideology is itself a substantively important finding: it implies that model developers, through choices about training data, alignment objectives, and default system prompts, have meaningful leverage over the ideological positions their models express to users.

Second, our persuasion experiment demonstrates that the behavioral patterns we measure translate into real downstream effects on voters. The pooled effect is at least as large as those produced by professional campaign advertising~\cite{coppock2020small, hewitt2024experiments}, and it is strongest when a model is explicitly steered to argue a side. Our mechanism analyses indicate that these effects reflect genuine persuasive influence rather than sycophantic accommodation, since persuasion grows sharply with experimenter-set advocacy while a balanced conversation is inert. They are not confined to politically unsophisticated respondents, and no receptivity or resistance moderator survives correction for multiple comparisons. The one corrected moderator signal, out-partisanship in the pooled analysis, runs opposite both our registered prediction and our discovery pilot, with out-partisans moving less than co-partisans, and is worth revisiting in future work before being treated as established.

Our experiment measures attitudes immediately after a single conversation, so it speaks to whether LLMs can move expressed positions, not to how long any such movement lasts. The persuasion literature counsels caution here, since the effects of one-shot treatments such as campaign advertisements or media exposure typically decay within days or weeks~\cite{druckman2022framework, coppock2023persuasion}. We would expect a single LLM conversation to be similarly perishable. At the same time, everyday LLM use differs from a one-time exposure in ways that could change this expectation. People return to the same assistants repeatedly, and often for exactly the kind of information-seeking that our base condition models, so a small per-conversation shift could in principle be renewed or compounded through chronic interaction rather than fading, a dynamic that single-session designs cannot observe and that recent evidence on durable AI-driven attitude change makes plausible~\cite{costello2024durably}. We advance this as an expectation and an open question rather than a finding. Directly measuring the durability of LLM-induced attitude change, ideally by re-contacting participants after a delay, is an important direction for future work.

An important direction for future research concerns users' beliefs about LLM neutrality. Our null finding on LLM familiarity as a moderator suggests that simple awareness of these tools does not inoculate against their persuasive influence. However, familiarity is distinct from beliefs about whether LLMs are politically neutral or biased. Users who believe LLMs provide objective, unbiased information may be especially susceptible to their ideological influence, while those who understand that models can express systematic partisan positions may contextualize their outputs more critically. Future work could experimentally manipulate such beliefs---for example, by providing varying information about a model's ideological leanings before the interaction---to test whether awareness of LLM bias moderates susceptibility to persuasion.
}

The implications are stark. Any organization that can produce a cheaper or more widely-used model can imperceptibly, as far as their users are concerned, influence attitudes and behavior. Governments or private corporations may be able to influence elections around the world through selective information provision without users consciously selecting into partisan messaging as they would with news or social media. Even if LLMs develop reputations based on the partisanship of their content, many users may not know those reputations or may select the cheapest model regardless, allowing the wealthy to fund LLMs as political investments just as they do with newspapers \cite{grossman2022ultrarich}. No one can know how precisely generative AI models will become embedded in our daily lives, but that they will is beyond doubt. Indeed, LLMs are already being incorporated into tools with significant potential for persuasive influence, such as personalized search engines, smart assistants, writing aids, and content recommendation systems, where they can subtly shape how users frame issues, evaluate arguments, or form opinions. As such, it is critical that researchers borrow from diverse behavioral fields in studying how such models might affect their users' thoughts, attitudes, and behaviors.

\subsection*{Open-Source Tools for Integrating LLMs into Survey Research}

Recent work suggests the power of integrating LLMs with survey-experimental research~\cite{velez2024confronting,tjuatja2024llms,rothschild2024opportunities}. However, standard survey software like Qualtrics and SurveyMonkey does not easily support these use cases: existing approaches \cite{velez2024confronting} to integrating LLMs require building custom APIs or integrating JavaScript, which is unduly burdensome for most social science researchers. Alongside our paper's replication data and code, we also open-source our survey toolkit ({\footnotesize\url{https://github.com/comnetsAD/Interactive_LLM_Survey_Platform}}), allowing participants to interact with LLMs, and allowing researchers to record the LLMs' conversations and measure their behavior. 

Our web-based survey platform assigns participants randomized questions from predefined categories and asks them to vote on each question. Some participants are randomly selected to first chat with an LLM before submitting their votes. The platform supports a number of LLMs (currently GPT 4o, Mistral, and Llama, which could be expanded in the future), and the experimenter can specify which LLM is associated with each question. The experimenter can also specify the minimum amount of time required for participants to chat with the LLM before they are allowed to cast their votes. Users' chat history, votes, and metadata (such as Prolific IDs) are securely logged within the platform. The backend is implemented in Python and runs on Gunicorn, handling LLM API requests while ensuring concurrency-safe operations by assigning unique files per participant. The frontend, built with HTML, JavaScript, and CSS (Cascading Style Sheet), dynamically updates survey questions and the chat interface, featuring real-time interactions, a typing indicator, and a structured chat history that persists during the session. JavaScript asynchronously communicates with the backend via RESTful API calls to send user messages and retrieve LLM responses while enforcing a fixed interaction time limit.

The backend requires Python, Flask, Gunicorn, and API access credentials to run the platform for the selected LLMs. The system supports multi-model integration, allowing different survey questions to be processed by distinct LLMs (e.g., some questions use GPT-4 while others use Llama for example). The backend maintains memory within each session by tracking the conversation history, ensuring follow-up responses remain contextually relevant. The platform is deployed on a web server such as Apache or Nginx, which acts as a reverse proxy to Gunicorn for handling concurrent requests efficiently. The frontend consists of a static HTML page with JavaScript managing user interactions, session-based chat resets, and a structured post-interaction voting mechanism. The survey questions and experimental conditions are configurable. Participants access the platform via a unique URL, and their responses are automatically recorded for later analysis. This system provides a scalable and adaptable solution for AI-driven research in behavioral studies, human-computer interaction, and policy evaluation.

We anticipate an enormous range of use cases for this tool, as improved survey research technology always does~\cite{klecka1978random, kaufman2020implementing}. Aside from simply providing respondents with an LLM chatbot, the tool allows researchers to feed participants' responses from previous questions to LLMs, customizing them to the respondent's profile. These LLMs can provide in-survey fact-checks, persuasive arguments, emotion-inducement treatments, or any other type of customized text-based interactive experience that researchers might want to provide, meaningfully enhancing the range of research questions that survey experiments can address.

\subsection*{Ethical Considerations}
This study’s survey experiment was reviewed and deemed exempt by the authors’ Institutional Review Board (protocol number anonymized for review). All participants were recruited through Prolific, provided informed consent prior to participation, and were free to withdraw at any time. The study posed minimal risk, as respondents were only asked to discuss policy questions with a language model chatbot and no deception beyond randomization of conditions was used. We pre-registered the design and analysis plan, and all procedures adhered to professional standards for ethical human subjects research in political science.
\clearpage

\clearpage
\renewcommand{\figurename}{Supplementary Figure}
\renewcommand{\tablename}{Supplementary Table}
\setcounter{figure}{0}
\setcounter{table}{0}

\clearpage
\section*{Supplementary Note 1: Related Work}\label{sec:related:work}

Although multiple studies have focused entirely on the U.S.\ context when analyzing the political biases of LLMs~\cite{liu2021mitigating,mcgee2023chat,jenny2024exploring}, many other studies have considered different countries in their analysis. These include a recent study that considered the major political parties in Germany~\cite{batzner2024germanpartiesqa}, and another study that considered different political parties in both Germany and the Netherlands~\cite{hartmann2023political}. Other examples include a recent study~\cite{rozado2024political} that utilized various political orientation tests, including both the U.S.\ and the U.K. editions of the iSideWith Political Quiz~\cite{iside}, and another recent study~\cite{motoki2024more} that also considered political parties in the U.S. and the U.K., as well as those in Brazil. In contrast, Zhou and Zhang~\cite{zhou2024political} used two languages, English and simplified Chinese, to ask GPT the same questions about political issues in the U.S.\ and China, revealing that the English model is less critical of issues in the U.S.\ while the Chinese one is less critical of issues in China. Other studies that examine the political leaning of LLMs do not focus on any specific country~\cite{feng2023pretraining,rozado2023political,agiza2024politune}. 

Our work differs significantly from all the studies that considered the U.S.\ in their analysis~\cite{motoki2024more,zhou2024political,rozado2024political,liu2021mitigating,mcgee2023chat,jenny2024exploring}. In particular, none of these studies examined how interactions with LLMs influence people, and none compared LLMs to Supreme Court judges or to different demographics of American voters. Additionally, none of them considered members of the Congress in their analysis (apart from McGee~\cite{mcgee2023chat} who considered just four members), and none considered more than a single LLM in their analysis (apart from Rozado~\cite{rozado2024political} who considered 24 models). Lastly, none examined different topics in isolation, except for Liu et al.~\cite{liu2021mitigating}, who demonstrated that the model they analyze (GPT-2) exhibits political bias on certain topics, without determining the direction of the bias, e.g., whether it is in favor of a certain political party.

Let us now focus on the four studies that, like ours, compare the biases that LLMs exhibit toward Democrats (or liberals) vs.\ Republicans (or conservatives) in the U.S.~\cite{motoki2024more,rozado2024political,mcgee2023chat,jenny2024exploring}. Starting with the work of McGee~\cite{mcgee2023chat}, the author examines a single model (ChatGPT) by giving it the following prompt: ``\textit{Write an Irish Limerick using the word }$X$'' where $X$ is the name of one of 14 politicians. The results suggest that the Limericks for liberal politicians tend to be positive, whereas those for conservatives tend to be negative. Motoki et al.~\cite{motoki2024more} also focus on a single model (ChatGPT) and ask it to impersonate either a Democrat or a Republican. The resultant model is then asked to complete the Political Compass questionnaire~\cite{compass}, before comparing its responses to those produced by the default model. This analysis suggests that the default responses of ChatGPT are more associated with Democrats than with Republicans. Similarly, Jenny et al.~\cite{jenny2024exploring} prompt ChatGPT to rate various excerpts from U.S.\ presidential debates, revealing that Democratic candidates tend to be rated higher than Republicans. Finally, we mention the work of Rozado~\cite{rozado2024political} who probed 24 models to complete the U.S.\ edition of the iSideWith Quiz, showing that they tend to agree more with the Democrats than with Republicans. However, their finding is based on the iSideWith Quiz---a commercial platform that lacks transparency in its algorithms and data processing, is not subject to peer review, and has not undergone psychometric validation, limiting its credibility as a scientific instrument.

Outside the context of politics, our work is part of a growing body of research that probes LLMs to understand their emergent properties. One such study administered sentence completion prompts to a wide range of LLMs, and found that their responses exhibit social identity biases, particularly ingroup solidarity and outgroup hostility~\cite{hu2023generative}. In the context of racism, a recent study presented various LLMs with texts in either African-American English (AAE) or Standard-American English (SAE), and asked them to make predictions about the speakers who uttered the texts; the LLMs' responses exhibited prejudice against AAE speakers by ascribing more negative attributes to them---a clear sign of covert racism~\cite{hofmann2024ai}. A similar study probed GPT-3 for anti-Muslim bias using prompt completion, analogical reasoning, and story generation, revealing that the model frequently associates Muslims with violence~\cite{abid2021persistent}. In the context of cognitive psychology, a study subjected GPT-3 to a battery of canonical experiments from the literature to assess its decision-making, information search, deliberation, and causal reasoning abilities~\cite{binz2023using}. In the context of economic rationality, Chen et al.~\cite{chen2023emergence} instructed GPT to make various budgetary decisions, revealing that its choices align with utility maximization principles and reflect greater levels of rationality compared to human subjects. Finally, we mention the work of Hagendorff~\cite{hagendorff2024deception}, who used a series of language-based scenarios to reveal the ability of various LLMs to understand and induce deception strategies.

Our work also contributes to another body of research that estimates the causal effects of interacting with LLMs. In the realm of creativity, a recent study invited participants to write short stories, with the treatment group using GPT-4 for inspiration~\cite{doshi2024generative}. The results indicate that access to GPT-4 enhances the individual creativity of stories, but also causes the stories to become more similar to one another, thereby reducing the collective diversity of creative output. In a similar study, participants were asked to complete creative tasks, some with the assistance of ChatGPT~\cite{lee2024empirical}. The resulting ideas were then evaluated by external judges, revealing that ChatGPT assistance leads to more creative ideas. In the context of fact-checking, a randomized controlled trial examined how exposure to fact-checks generated by ChatGPT~3.5 influences people's perception of news headlines~\cite{deverna2024fact}. Despite the model's high accuracy in identifying false headlines, exposure to these fact-checks did not significantly improve participants' ability to discern between true and false headlines. In clinical decision-making, a randomized controlled trial assigned physicians to use either GPT-4 or conventional resources while answering complex management questions~\cite{goh2024large}; those who used GPT-4 scored higher on management reasoning tasks compared to the other condition. In a similar study, medical students engaged in simulated patient interactions, with one group receiving GPT-generated feedback on their performance~\cite{brugge2024large}. Those who received such feedback demonstrated significantly improved decision-making skills.

\clearpage
\section*{Supplementary Note 2: Comparing U.S. Legislators to LLMs}\label{sec:legislators}

\begin{figure}[hbtp]
    \centering
    \includegraphics[width=\linewidth]{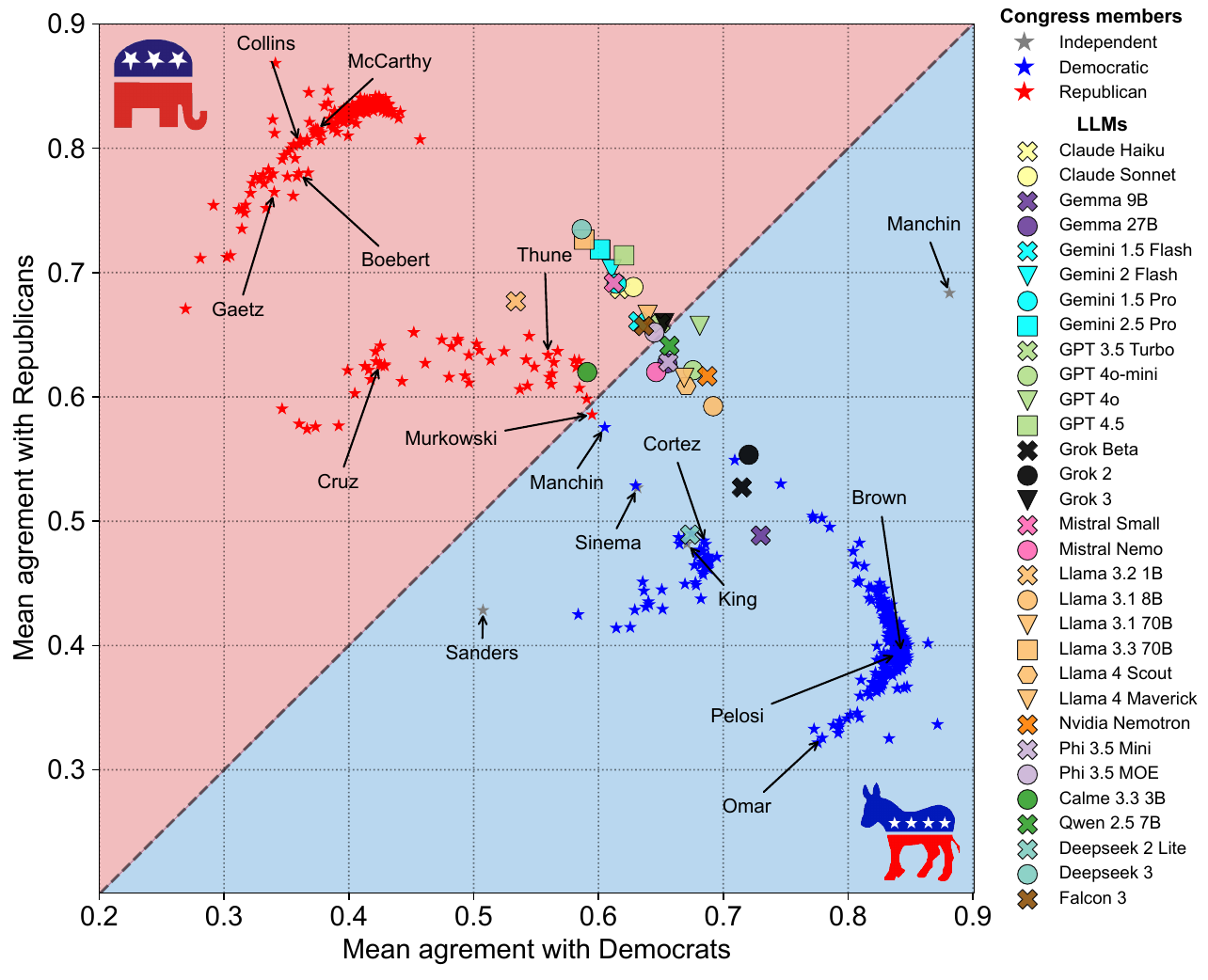}
    \caption{\textbf{Political alignment based on bill votes in the 118th Congress.} For any given data point, $d$, and any given bill, $d$'s alignment with Democrats is calculated as the proportion of Democrats whose vote on the bill matches that of $d$. By averaging out these alignments over all bills, we obtain the x-axis coordinate of $d$, which represents its overall alignment with Democrats. The y-axis coordinate is calculated in a similar way but in relation to Republicans instead of Democrats.}
    \label{fig:bills}
\end{figure}

Figure~1 of the main article compared U.S. legislators' votes to the expressed preferences of our set of LLMs. In this supplementary note, we represent the same data in a nonparametric framework. More specifically, Supplementary Figure~\ref{fig:bills} also displays legislators and LLMs together, but now the coordinates represent co-voting patterns more directly. Each point's x-axis coordinate represents the average alignment of its votes with Democratic members of Congress across all bills. More specifically, for each bill, we calculate the percentage of Democratic members who voted the same way as the point, and then compute the average percentage across all bills. The y-axis coordinate follows the same process but measures alignment with Republican members instead.

Since many bills are passed with bipartisan majorities, these are not mutually exclusive. For instance, given a particular bill, an LLM may agree with 30\% of Democrats and 40\% of Republicans; these percentages need not sum to 100\%. Based on this analysis, data points that fall under the diagonal are closer to the Democrats than the Republicans, and vice versa. 

No LLM falls exactly at the diagonal, indicating that they all vote closer to a certain party. However, most fall relatively close to the diagonal compared to the members of Congress, providing further evidence that most LLMs are relatively moderate. A few LLMs stand out, with Llama~3.2 and Llama~3.3 being the most Republican-leaning, and Grok and Gemma~9B being the most Democratic-leaning. It is noteworthy that increasing a model's size can yield the opposite political alignment. For instance, Llama~3.1 is Democratic-leaning when given 8 billion parameters but Republican-leaning when given 70 billion parameters, highlighting the role that the model's size could play in its political alignment.

\section*{Supplementary Note 3: Classifying organic conversations as political}
\label{sec:classifying}

For the naturalistic-conversation analysis we used WildChat~\cite{zhao2024wildchat}, a corpus of real-world conversations between users and GPT-3.5/GPT-4. We restricted the corpus to conversations from users in the United States (via the geolocation metadata distributed with WildChat), giving 42{,}300 conversations. Each conversation was independently classified as political or non-political by GPT-5.4-mini (temperature 0, structured JSON output). The classifier was instructed to label a conversation as political if it substantially involved government policy, legislation, elections, political parties, politicians, political ideologies, governance, civil rights, international relations, political movements, regulation, or public-policy debate, and to treat casual or incidental mentions as non-political. For each conversation it judged political, the model also returned a confidence level, a primary topic and subtopics, any named political entities, and a flag for whether the user was explicitly soliciting the assistant's own position (\textit{stance-seeking}). This procedure flagged 1{,}493 conversations as political (3.5\% of the U.S. corpus). To ensure a well-defined opening query, we restricted the analysis to conversations of at most three turns, yielding the 1{,}019 political conversations used in the main text.

\clearpage
\section*{Supplementary Figures}\label{sec:figures}

\begin{figure}[hbtp]
    \centering
    \includegraphics[width=\linewidth]{figures_si/Figure_2_2D.pdf}
    \caption{\textbf{Justices' political alignment in 2 dimensions.} Two-dimensional projection of the MCMCpack results from Figure~2 in the main text.}
    \label{fig:scotus-2d}
\end{figure}

\begin{figure}[htbp]
\begin{centering}
\includegraphics[width=1\linewidth]{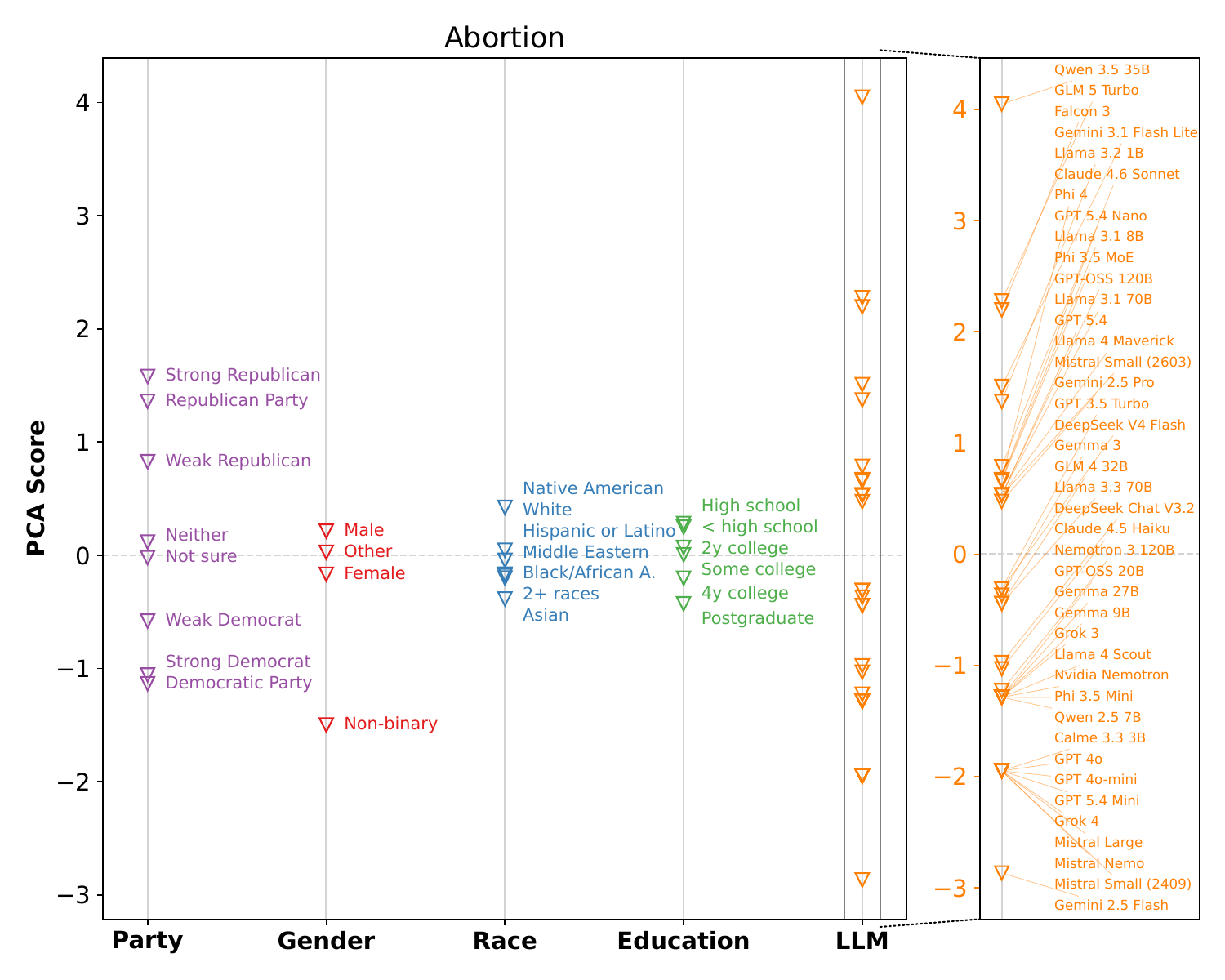}
\par\end{centering}
\caption{\textbf{Political alignment based on PCA analysis of the CES 2022, focusing on abortion.} The same as Main Figure~4, but focusing on the six questions in the CES related to abortion.}
\label{fig:CC22:Abortion}
\end{figure}

\begin{figure}[htbp]
\begin{centering}
\includegraphics[width=1\linewidth]{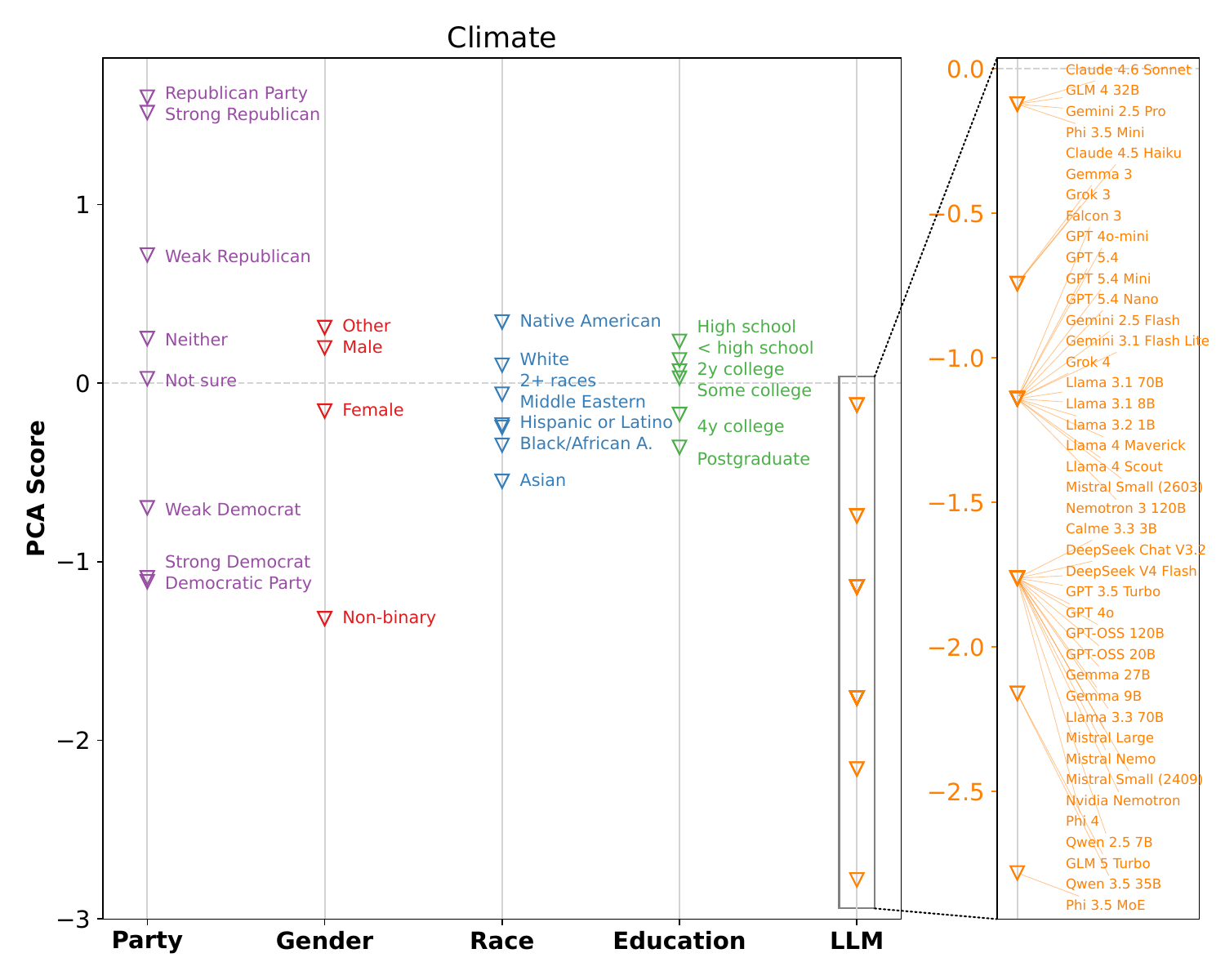}
\par\end{centering}
\caption{\textbf{Political alignment based on PCA analysis of the CES 2022, focusing on climate.} The same as Main Figure~4, but focusing on the five questions in the CES related to climate.}\label{fig:CC22:Climate}
\end{figure}

\begin{figure}[htbp]
\begin{centering}
\includegraphics[width=1\linewidth]{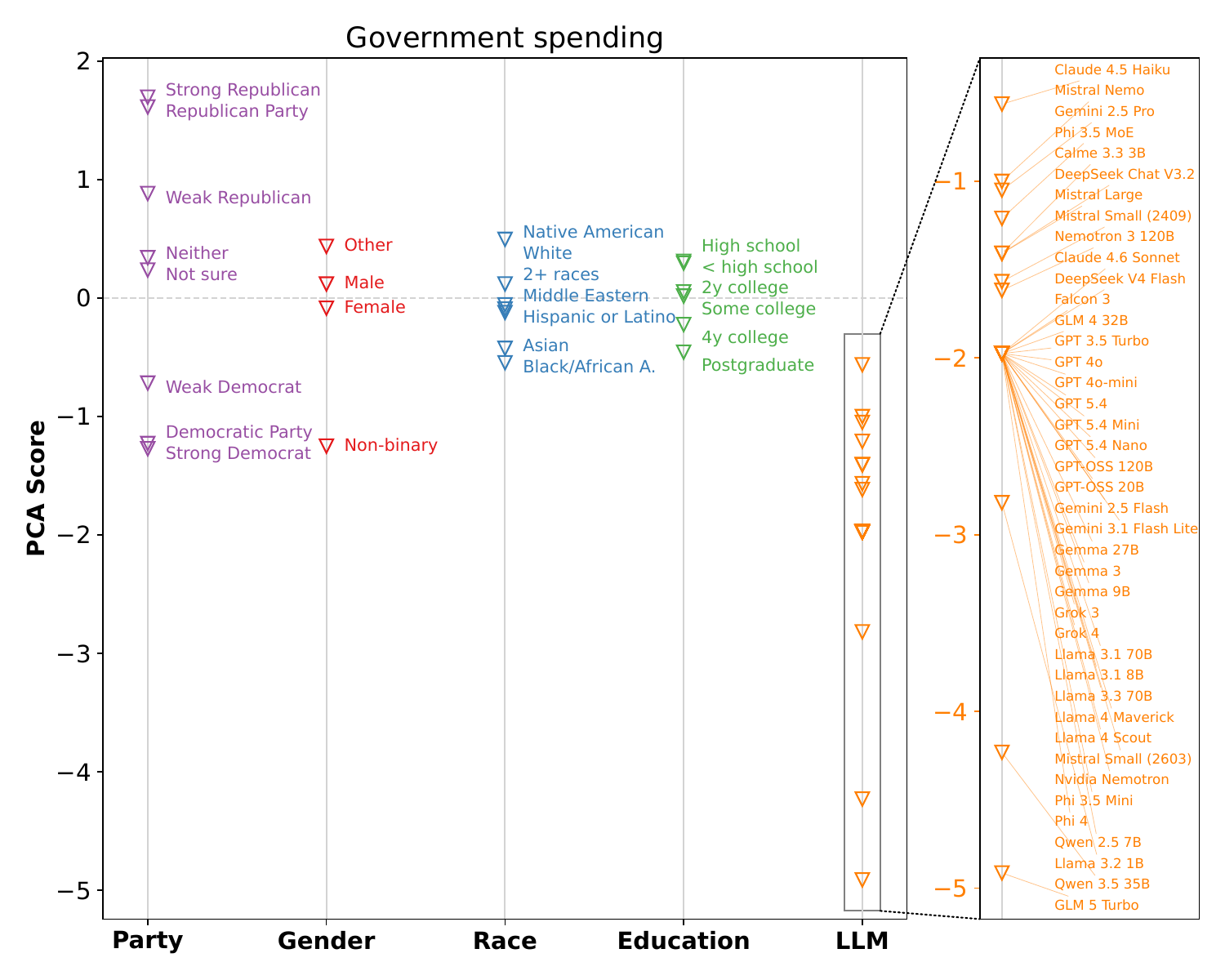}
\par\end{centering}
\caption{\textbf{Political alignment based on PCA analysis of the CES 2022, focusing on government spending.} The same as Main Figure~4, but focusing on the eight questions in the CES related to government spending.}\label{fig:CC22:Government}
\end{figure}

\begin{figure}[htbp]
\begin{centering}
\includegraphics[width=1\linewidth]{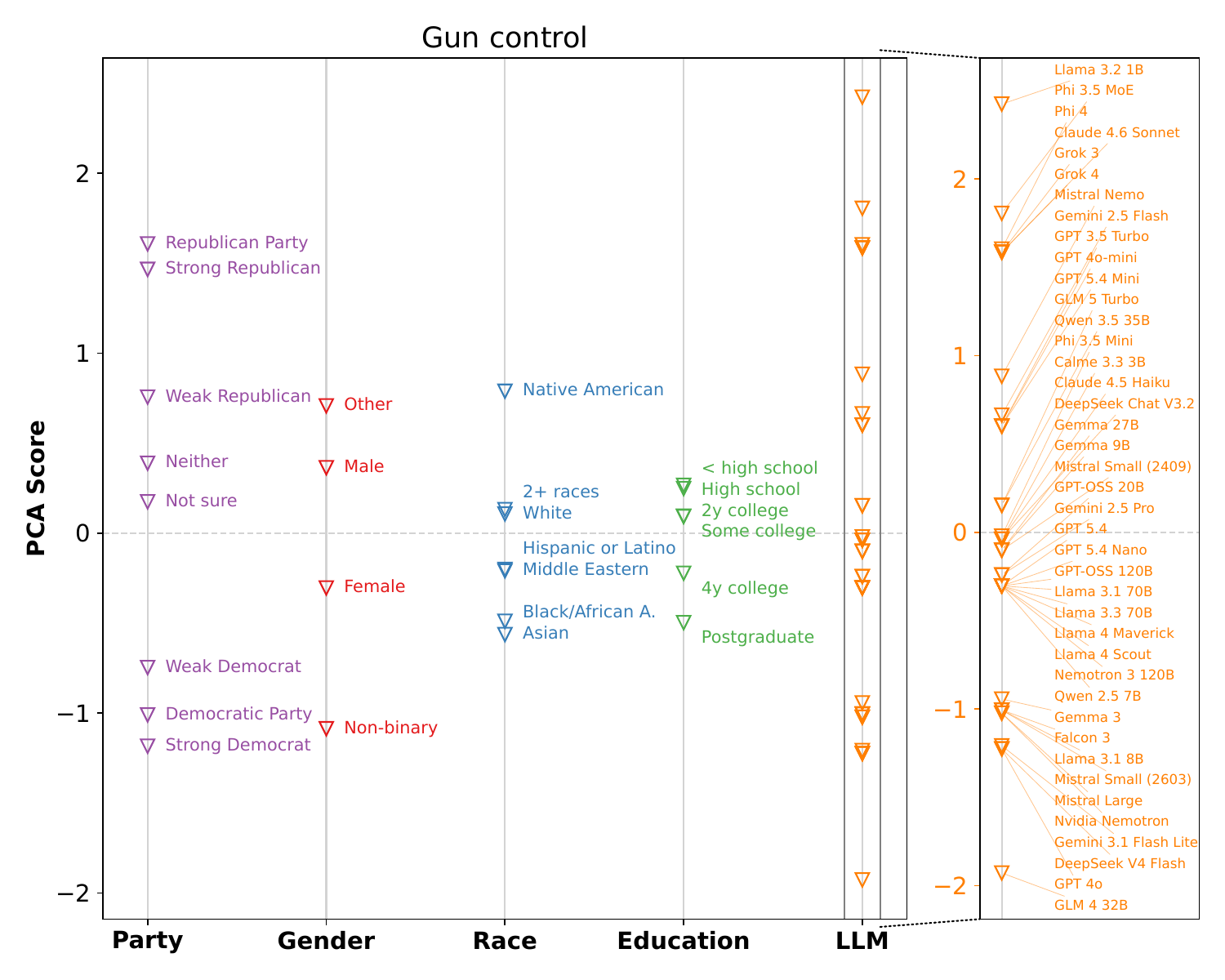}
\par\end{centering}
\caption{\textbf{Political alignment based on PCA analysis of the CES 2022, focusing on gun control.} The same as Main Figure~4, but focusing on the six questions in the CES related to gun control.}\label{fig:CC22:Gun}
\end{figure}

\begin{figure}[htbp]
\begin{centering}
\includegraphics[width=1\linewidth]{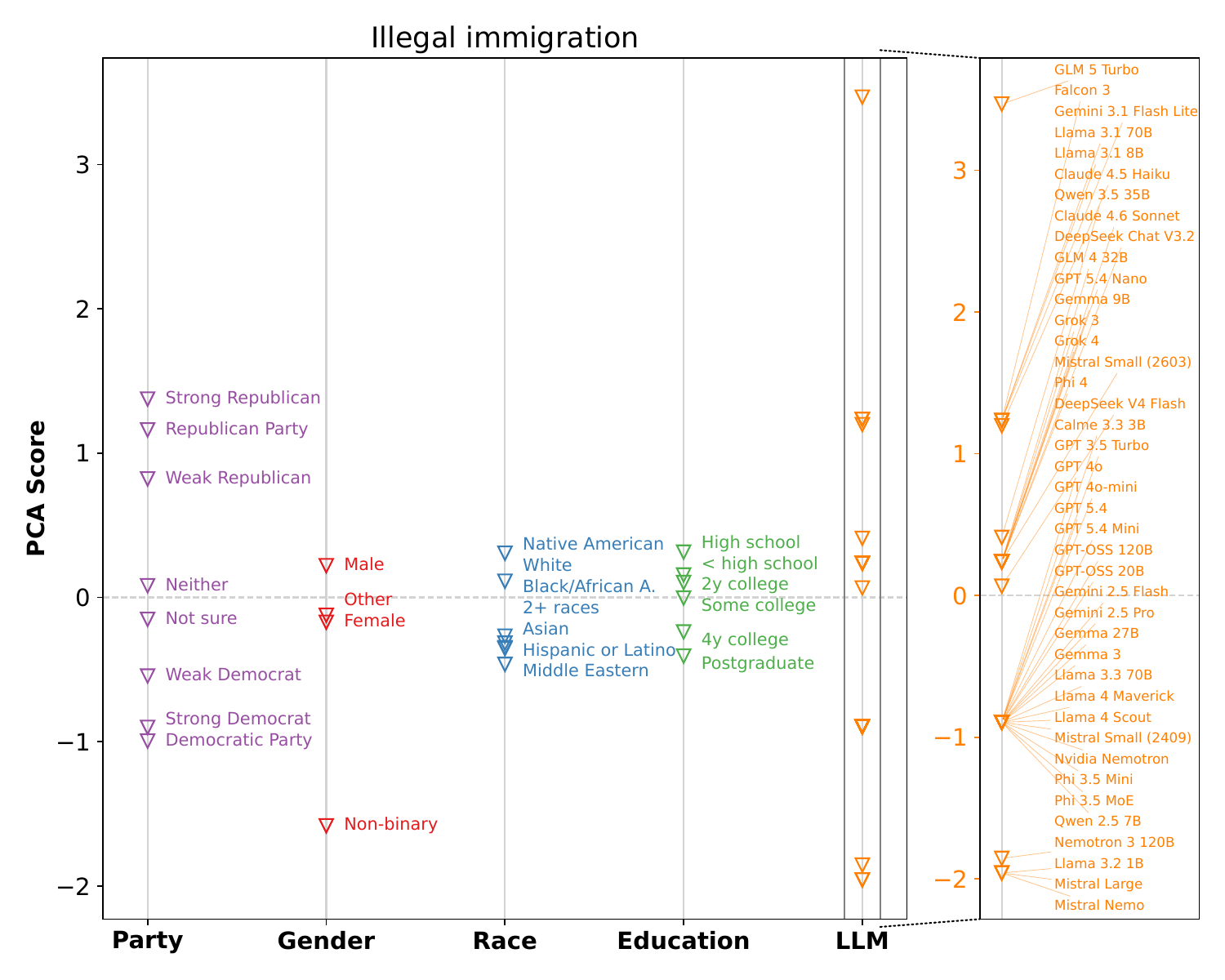}
\par\end{centering}
\caption{\textbf{Political alignment based on PCA analysis of the CES 2022, focusing on illegal immigration.} The same as Main Figure~4, but focusing on the four questions in the CES related to illegal immegration.}\label{fig:CC22:Immigration}
\end{figure}

\begin{figure}[htbp]
\begin{centering}
\includegraphics[width=1\linewidth]{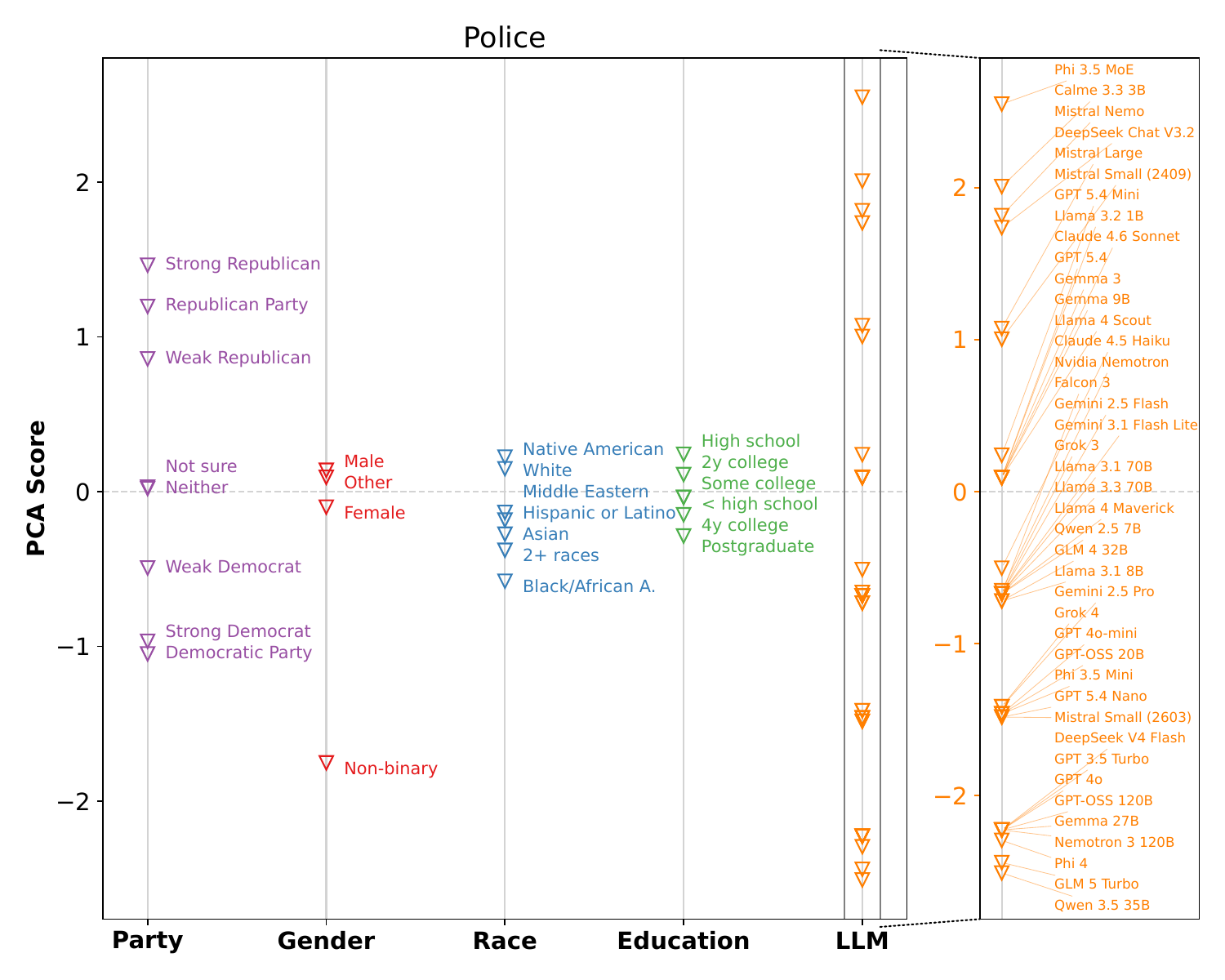}
\par\end{centering}
\caption{\textbf{Political alignment based on PCA analysis of the CES 2022, focusing on police.} The same as Main Figure~4, but focusing on the eight questions in the CES related to police.}\label{fig:CC22:Police}
\end{figure}

\begin{figure}[htbp]
\begin{centering}
\includegraphics[width=1\linewidth]{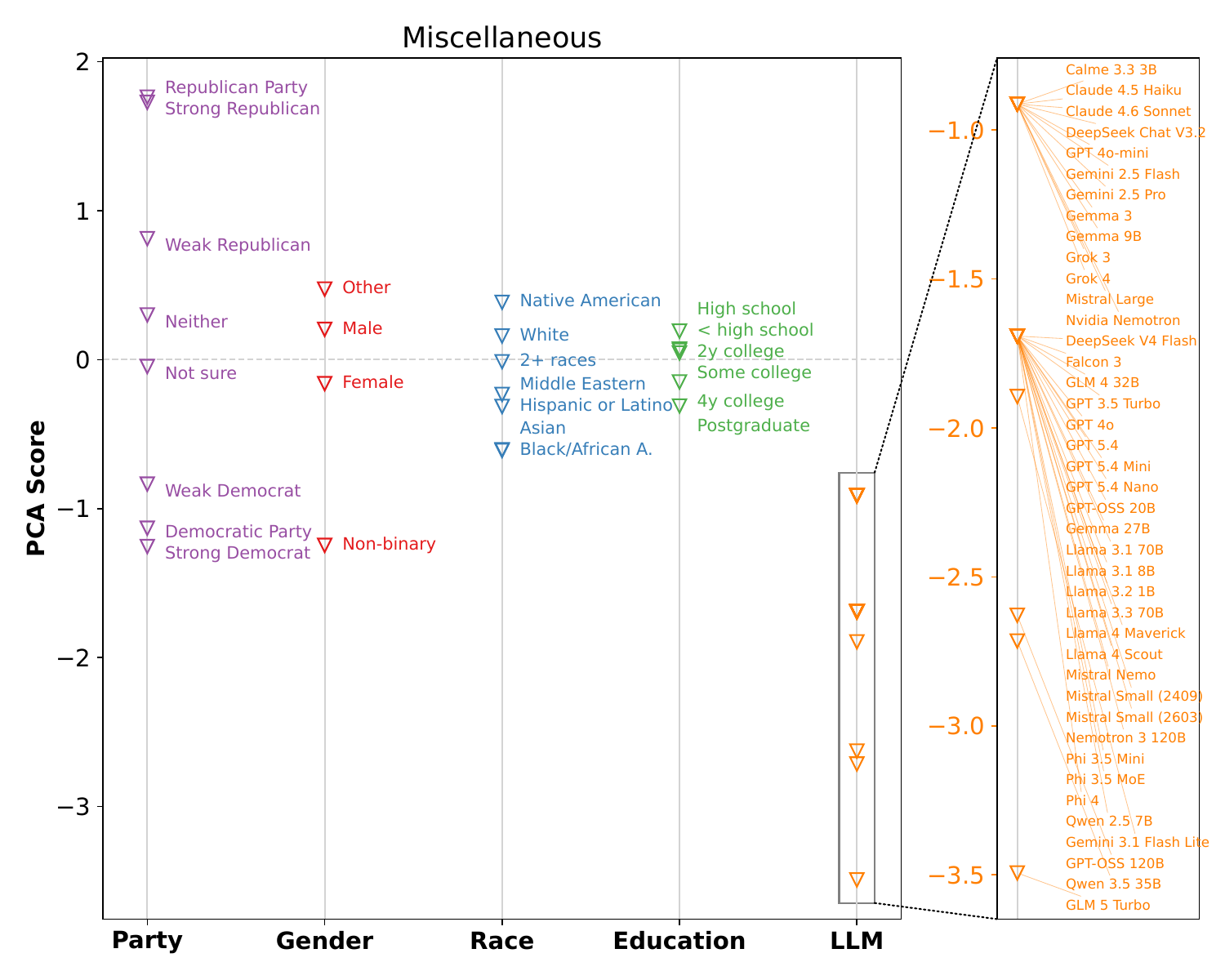}
\par\end{centering}
\caption{\textbf{Political alignment based on PCA analysis of the CES 2022, focusing on miscellaneous issues.} The same as Main Figure~4, but focusing on the five questions in the CES related to miscellaneous issues.}\label{fig:CC22:Miscellaneous}
\end{figure}

\begin{figure}[htbp]
\begin{centering}
\includegraphics[width=1\linewidth]{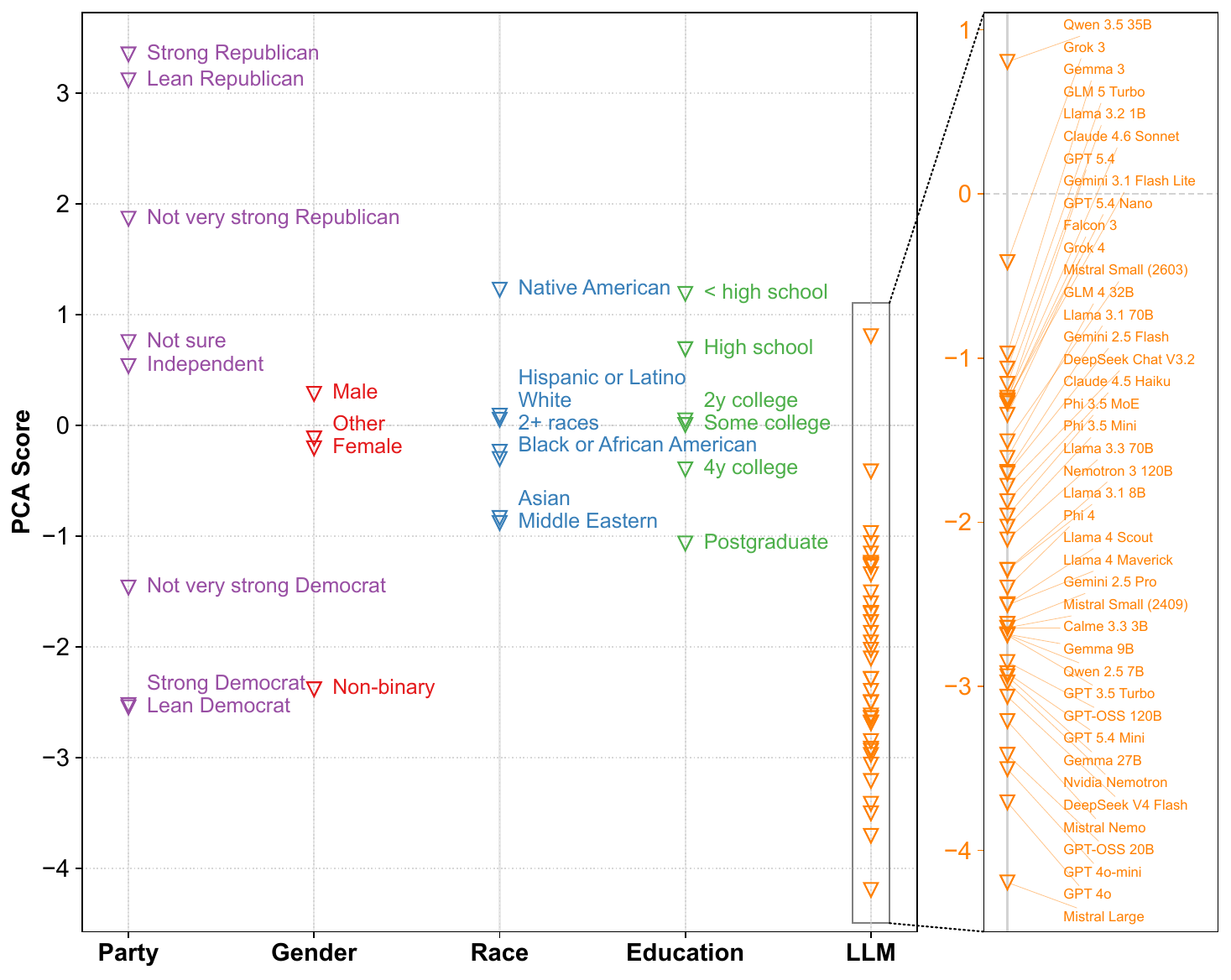}
\par\end{centering}
\caption{\textbf{Political alignment based on PCA estimation of the CES 2024.} We identified 40 questions about eight policy issues: abortion, climate change, government spending, gun control, healthcare, immigration, taxes, and miscellaneous. We compile the answers provided by the 60,000 CES (Cooperative Election Study) participants as well as those provided by the 41 LLMs. We use Principle Component Analysis (PCA) to map all answers into a single dimension. We then average the PCA scores for each LLM and each demographic category.}
\label{fig:CC24_Total}
\end{figure}

\begin{figure}[htbp]
\begin{centering}
\includegraphics[width=1\linewidth]{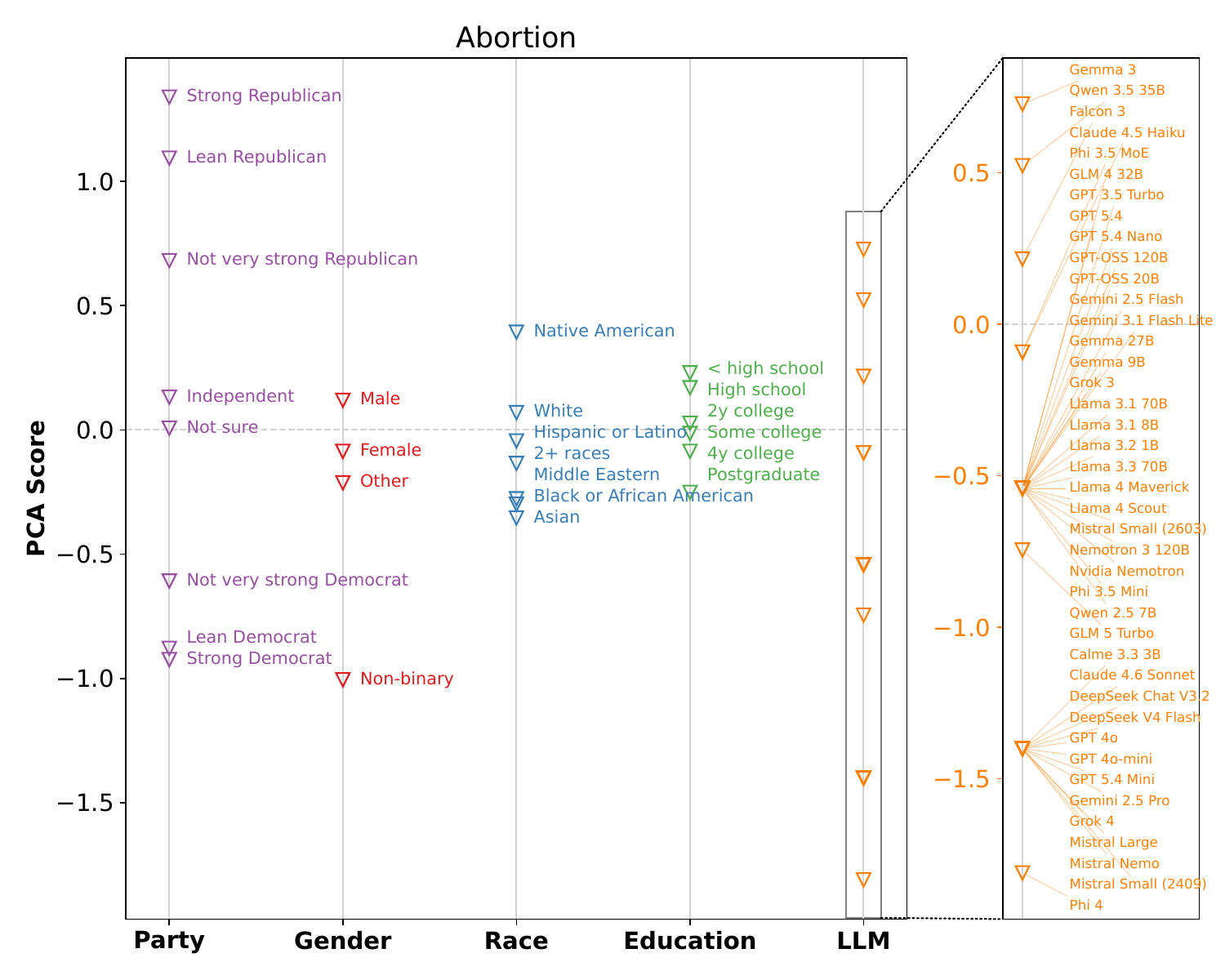}
\par\end{centering}
\caption{\textbf{Political alignment based on PCA analysis of the CES 2024, focusing on abortion.} The same as Supplementary Figure~\ref{fig:CC24_Total}, but focusing on the six questions in the CES related to abortion.}
\label{fig:CC24:Abortion}
\end{figure}

\begin{figure}[htbp]
\begin{centering}
\includegraphics[width=1\linewidth]{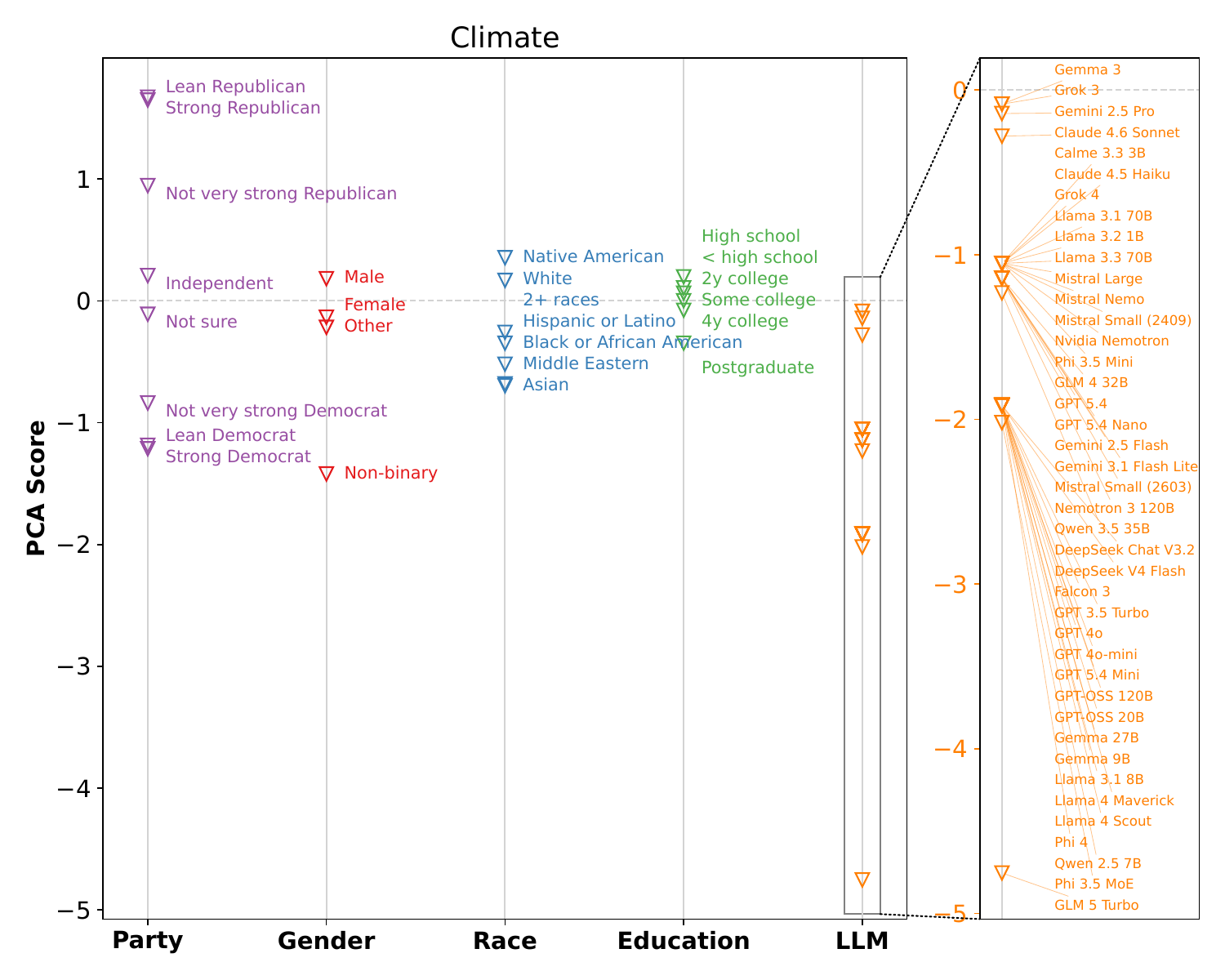}
\par\end{centering}
\caption{\textbf{Political alignment based on PCA analysis of the CES 2024, focusing on climate.} The same as Supplementary Figure~\ref{fig:CC24_Total}, but focusing on the five questions in the CES related to climate.}\label{fig:CC24:Climate}
\end{figure}

\begin{figure}[htbp]
\begin{centering}
\includegraphics[width=1\linewidth]{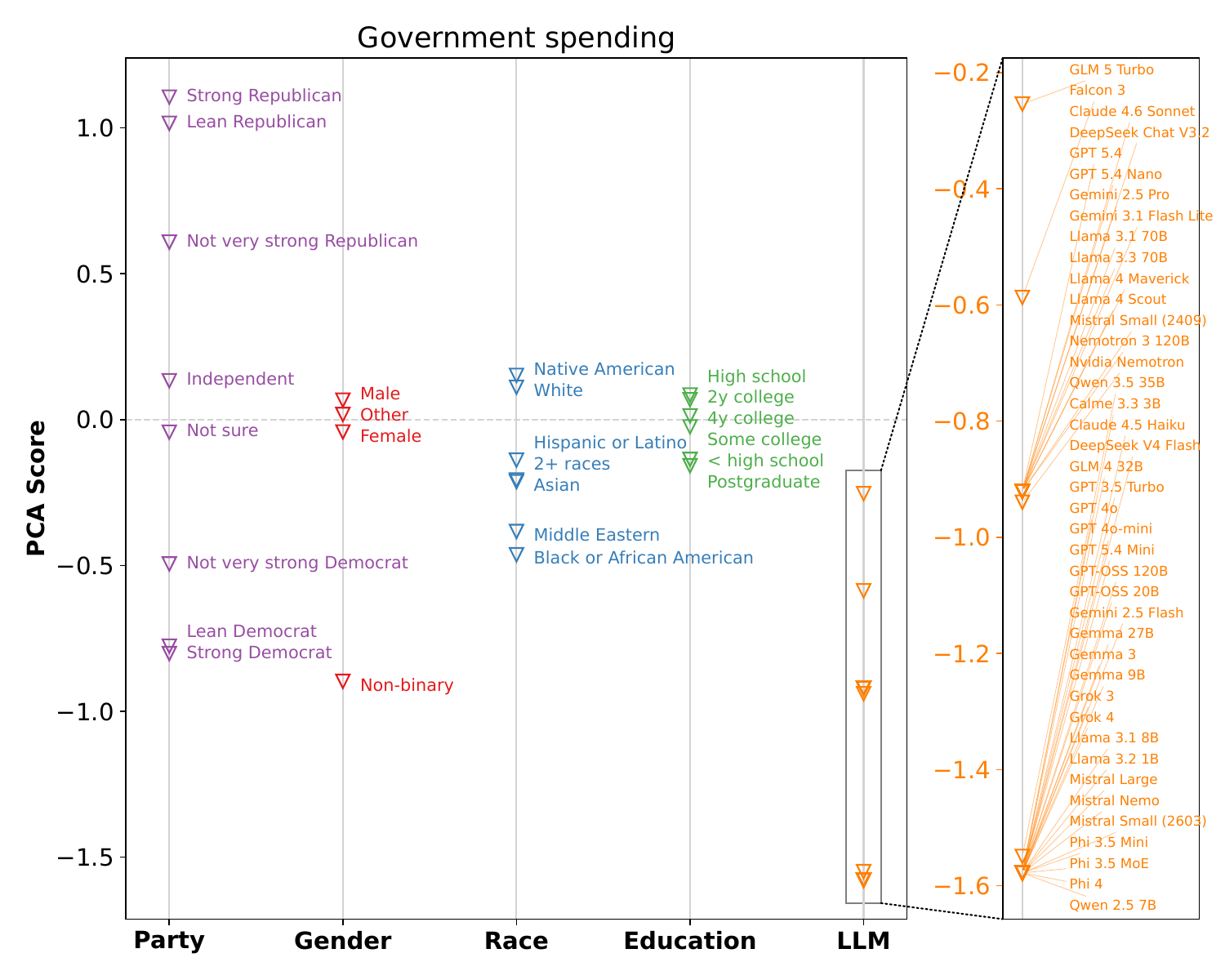}
\par\end{centering}
\caption{\textbf{Political alignment based on PCA analysis of the CES 2024, focusing on government spending.} The same as Supplementary Figure~\ref{fig:CC24_Total}, but focusing on the eight questions in the CES related to government spending.}\label{fig:CC24:Government}
\end{figure}

\begin{figure}[htbp]
\begin{centering}
\includegraphics[width=1\linewidth]{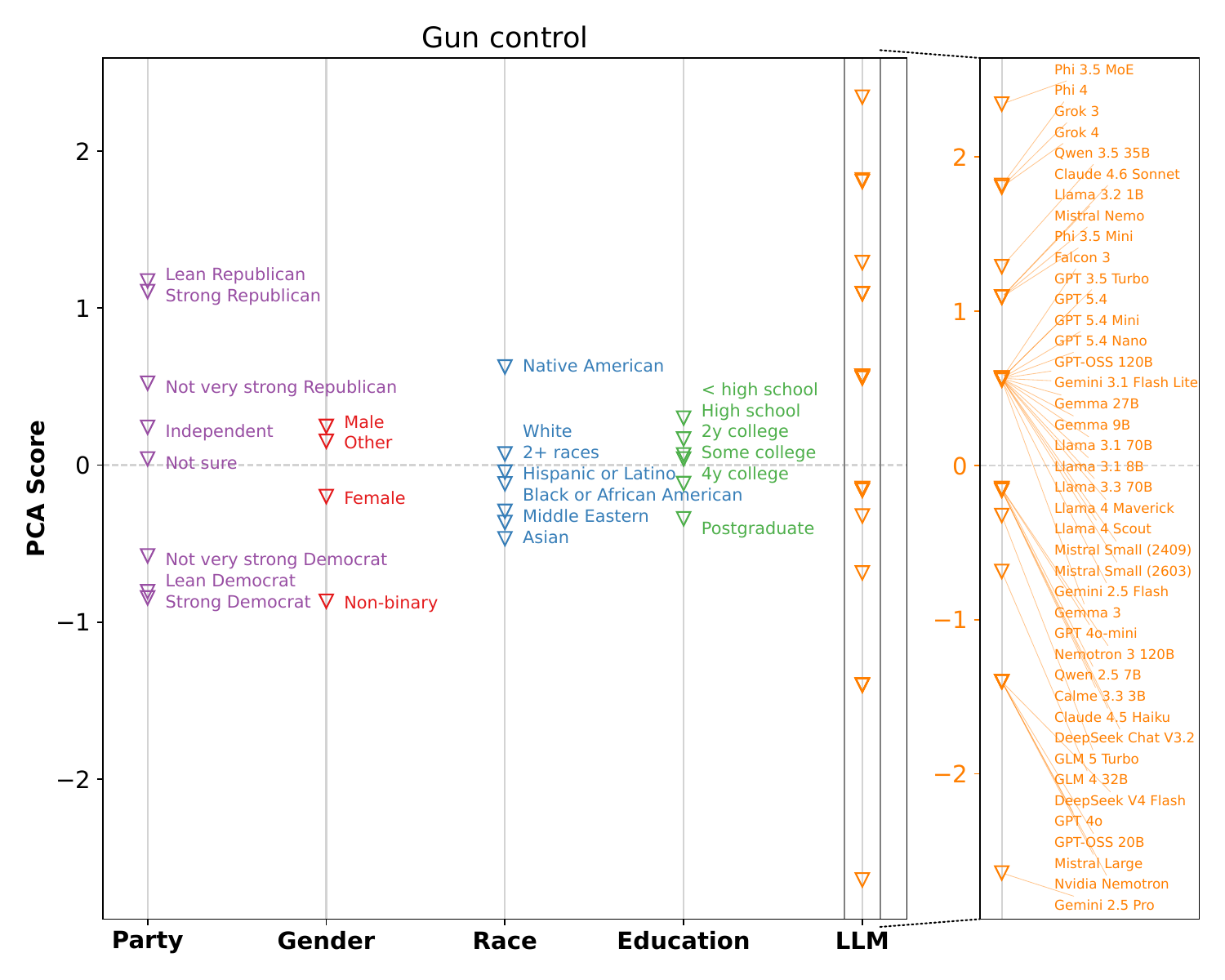}
\par\end{centering}
\caption{\textbf{Political alignment based on PCA analysis of the CES 2024, focusing on gun control.} The same as Supplementary Figure~\ref{fig:CC24_Total}, but focusing on the six questions in the CES related to gun control.}\label{fig:CC24:Gun}
\end{figure}

\begin{figure}[htbp]
\begin{centering}
\includegraphics[width=1\linewidth]{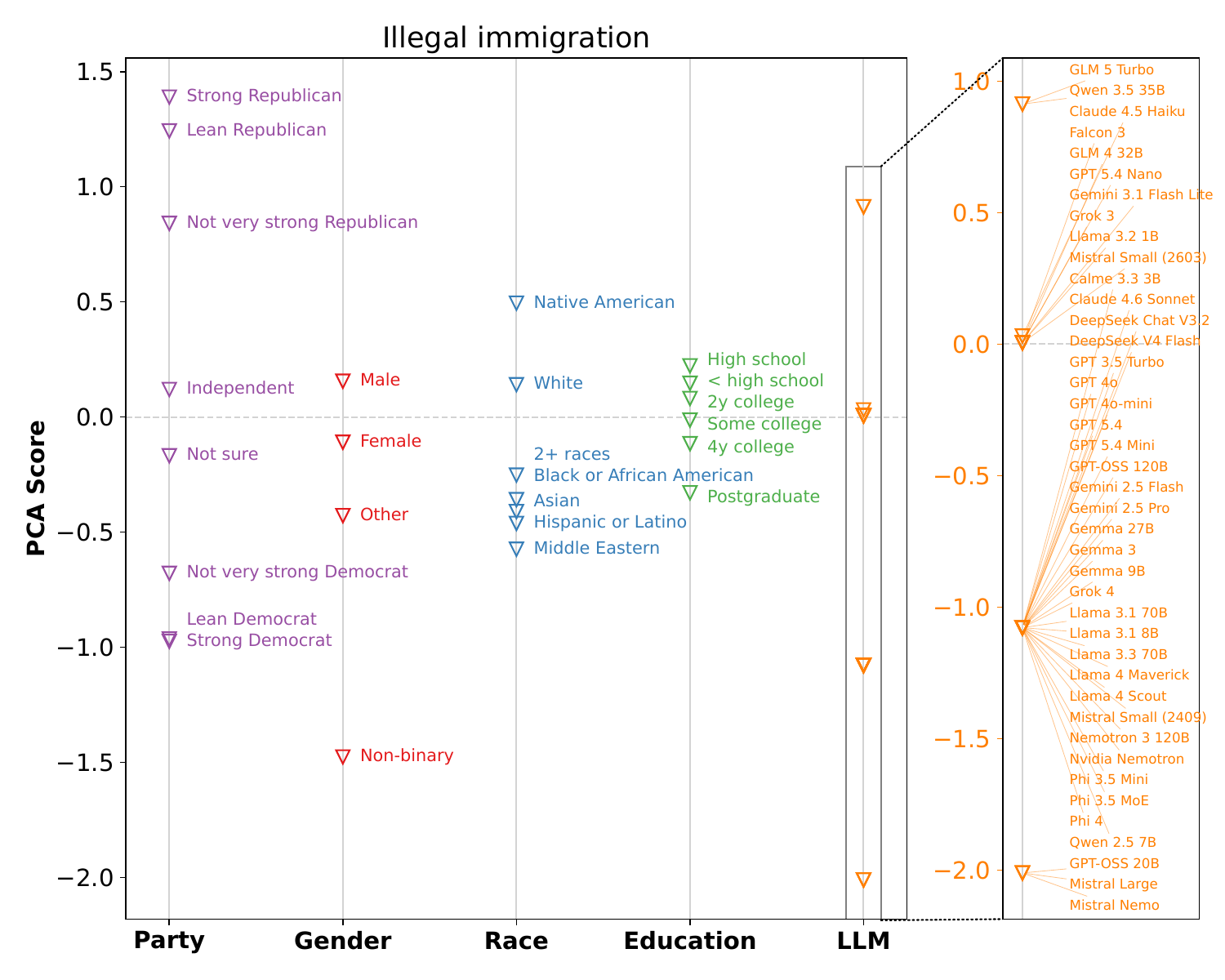}
\par\end{centering}
\caption{\textbf{Political alignment based on PCA analysis of the CES 2022, focusing on illegal immigration.} The same as Supplementary Figure~\ref{fig:CC24_Total}, but focusing on the four questions in the CES related to illegal immegration.}\label{fig:CC24:Immigration}
\end{figure}

\begin{figure}[htbp]
\begin{centering}
\includegraphics[width=1\linewidth]{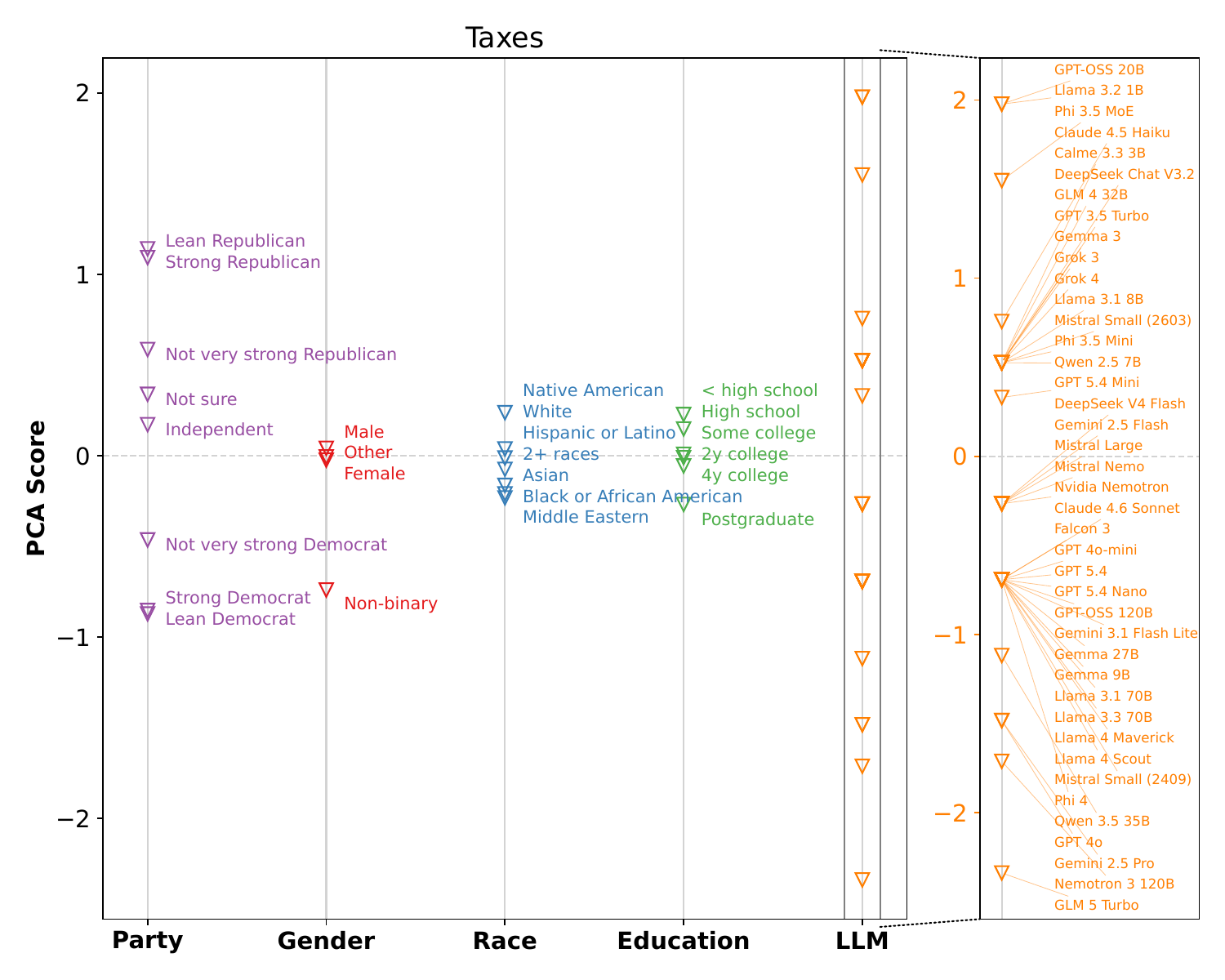}
\par\end{centering}
\caption{\textbf{Political alignment based on PCA analysis of the CES 2024, focusing on taxes.} The same as Supplementary Figure~\ref{fig:CC24_Total}, but focusing on the eight questions in the CES related to taxes.}\label{fig:CC24:Taxes}
\end{figure}

\begin{figure}[htbp]
\begin{centering}
\includegraphics[width=1\linewidth]{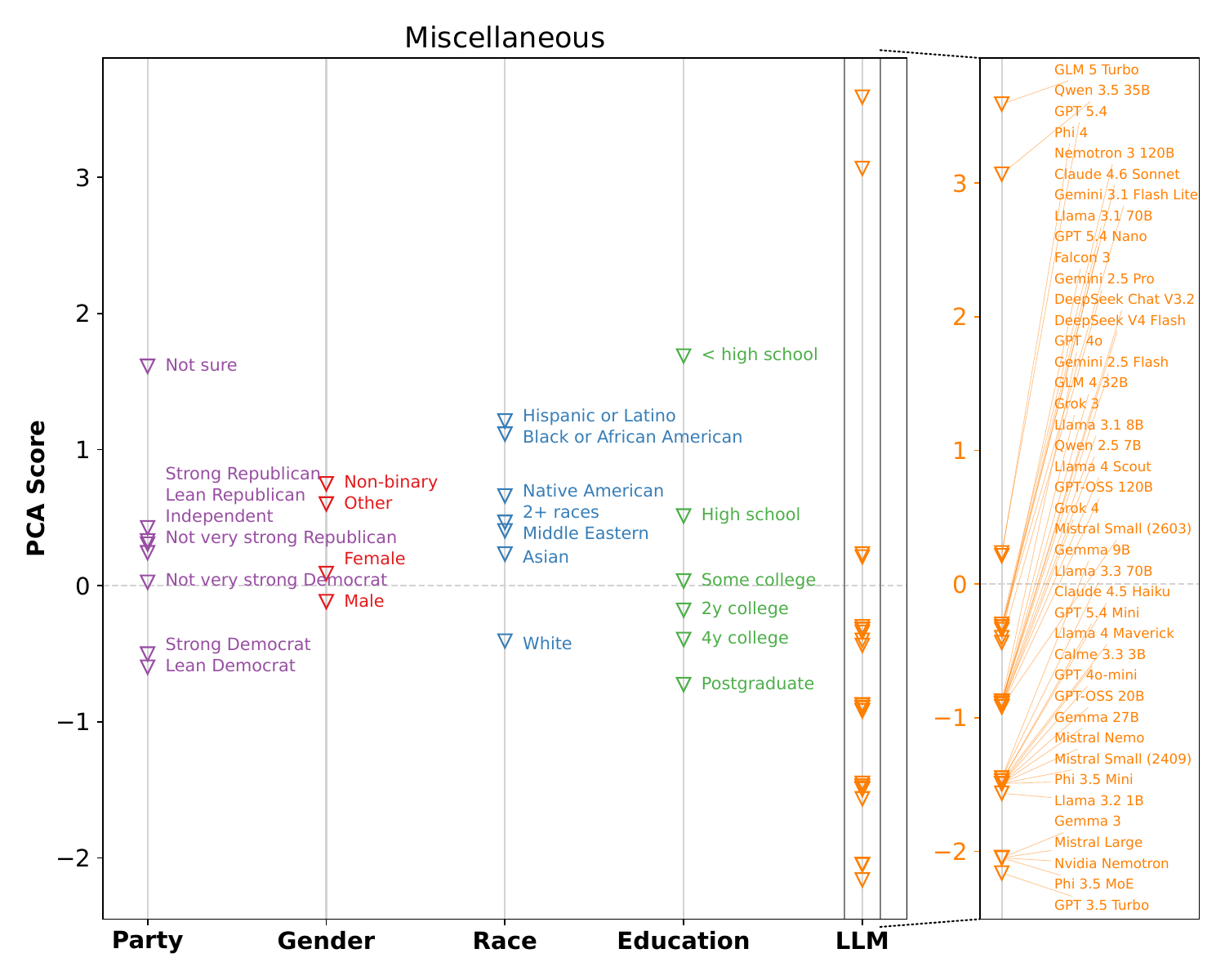}
\par\end{centering}
\caption{\textbf{Political alignment based on PCA analysis of the CES 2024, focusing on miscellaneous issues.} The same as Supplementary Figure~\ref{fig:CC24_Total}, but focusing on the five questions in the CES related to miscellaneous issues.}\label{fig:CC24:M iscellaneous}
\end{figure}

\begin{figure}[htbp]
\begin{centering}
\includegraphics[width=1\linewidth]{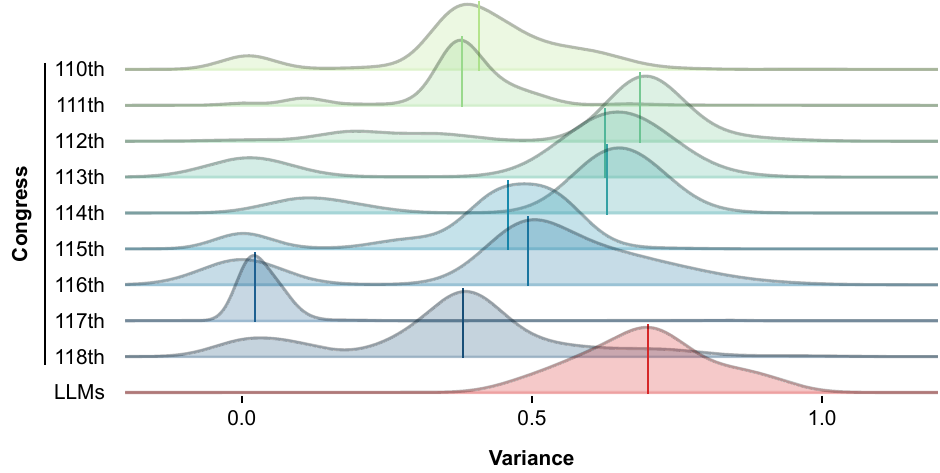}
\par\end{centering}
\caption{Ideological variance for all LLMs and different Congressional groups.}
\label{fig:congress_LLM_variance}
\end{figure}

\begin{figure}[htbp]
\begin{centering}
\includegraphics[width=1\linewidth]{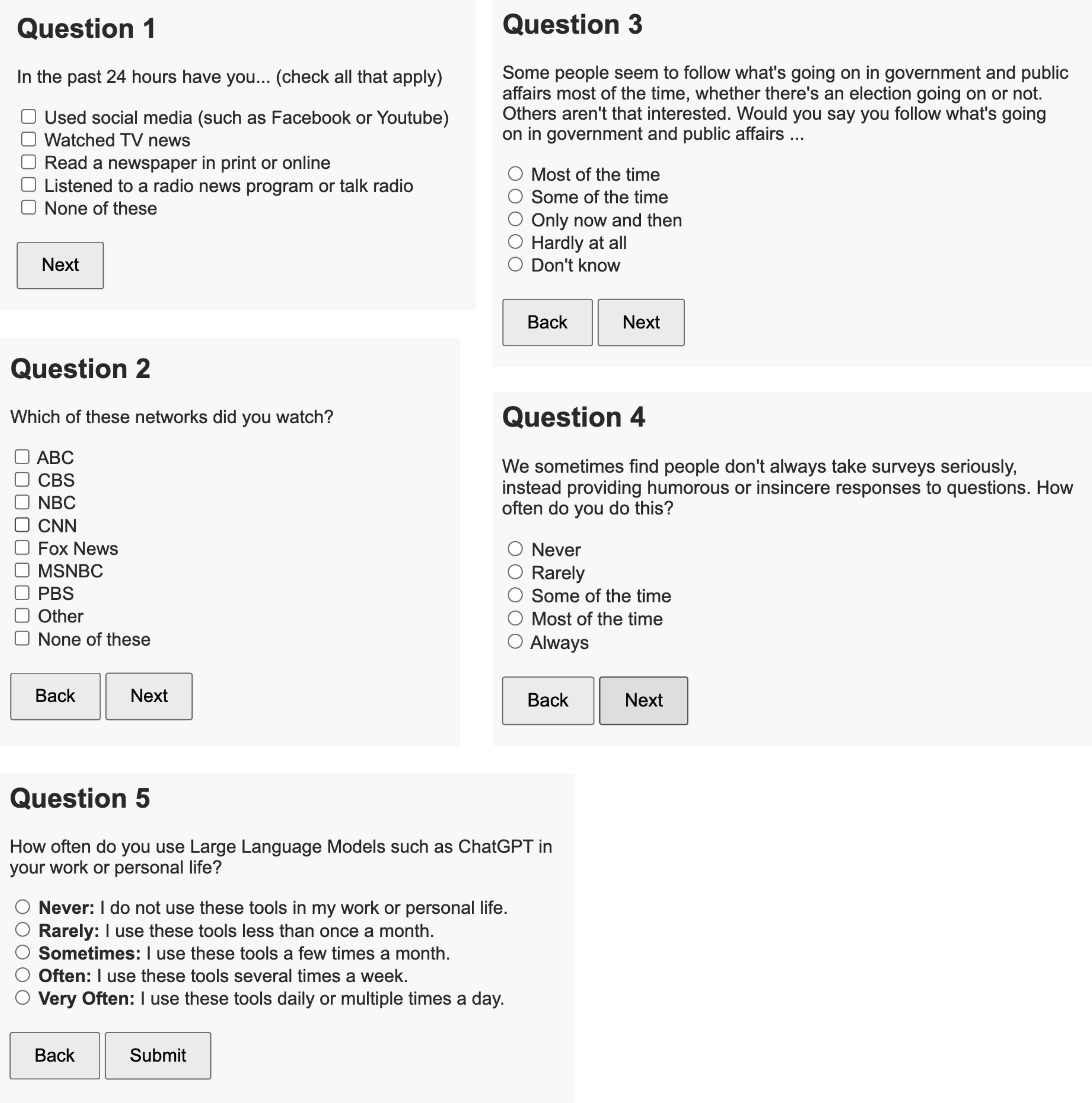}
\par\end{centering}
\caption{\textbf{Pre-treatment questions.} Screenshots of the five pre-treatment questions and attention checks asked to the survey participants.}
\label{fig:questions}
\end{figure}

\begin{figure}[htbp]
\begin{centering}
\includegraphics[width=0.8\linewidth]{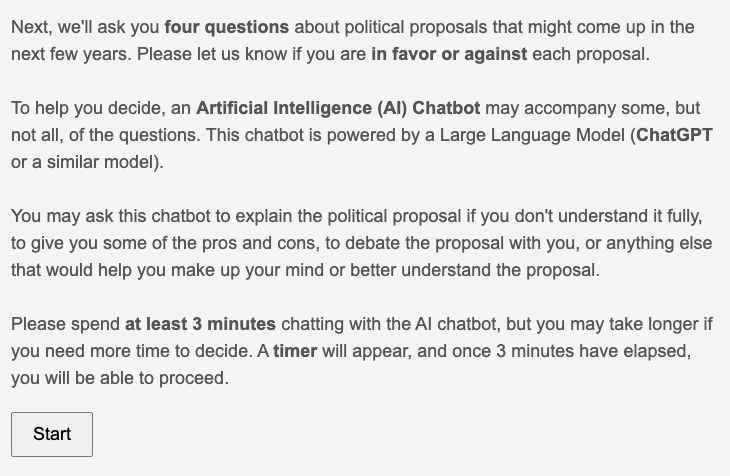}
\par\end{centering}
\caption{\textbf{Experiment description.} A screenshot of the experiment description provided to the participants before administering the treatment.}
\label{fig:consent}
\end{figure}

\begin{figure}[htbp]
\begin{centering}
\includegraphics[width=1\linewidth]{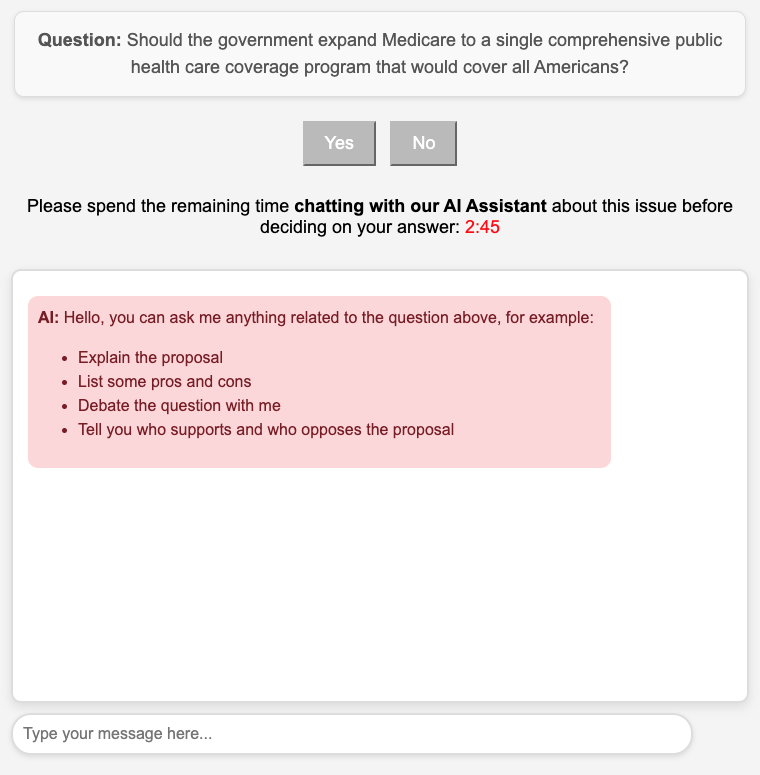}
\par\end{centering}
\caption{\textbf{Treatment example}. A screenshot of a sample treatment whereby participants discuss a political issue with an LLM-powered chatbot.}
\label{fig:survey_chatbot}
\end{figure}

\begin{figure}[htbp]
    \centering
    \includegraphics[width=\linewidth]{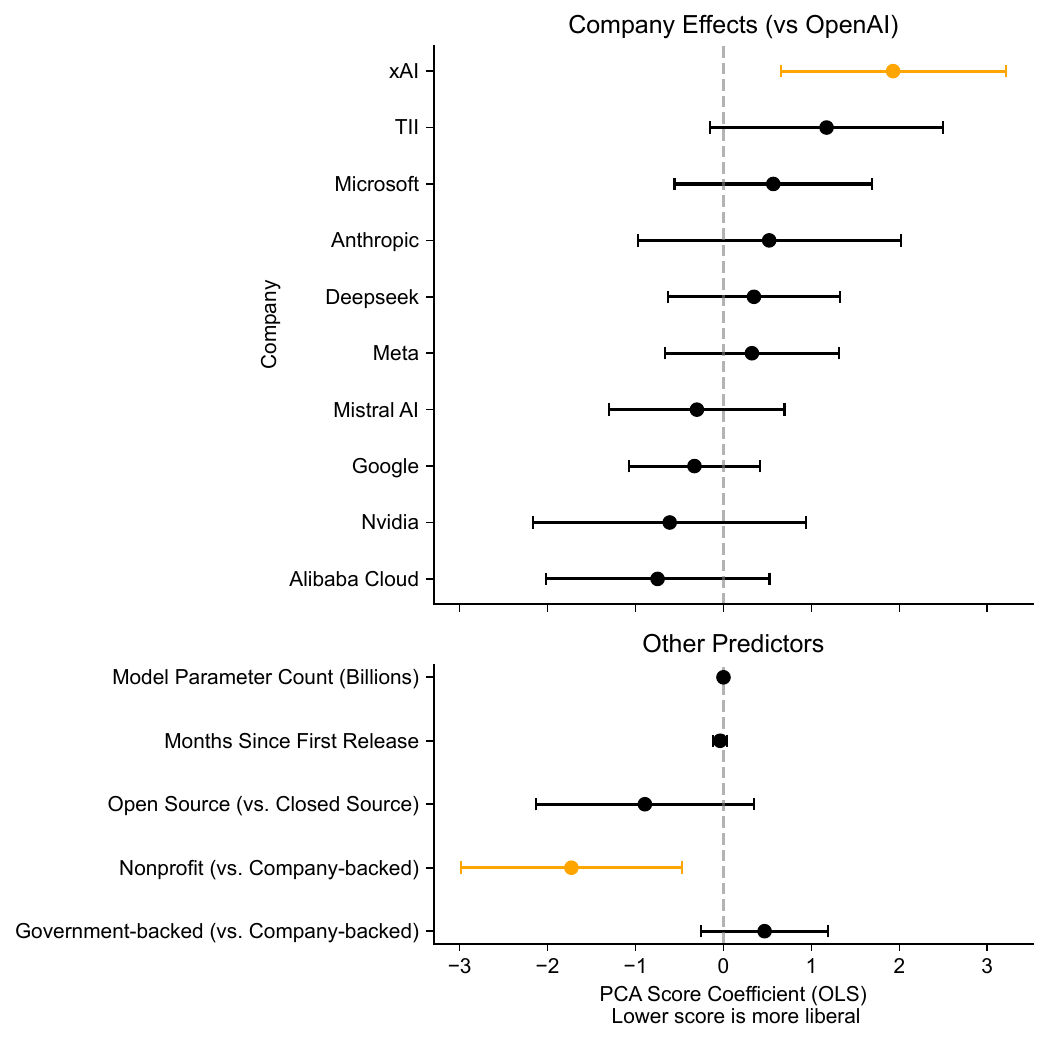}
    \caption{LLM Regression}
    \label{fig:llm_regression}
\end{figure}

\begin{figure}[htbp]
    \centering
    \includegraphics[width=\linewidth]{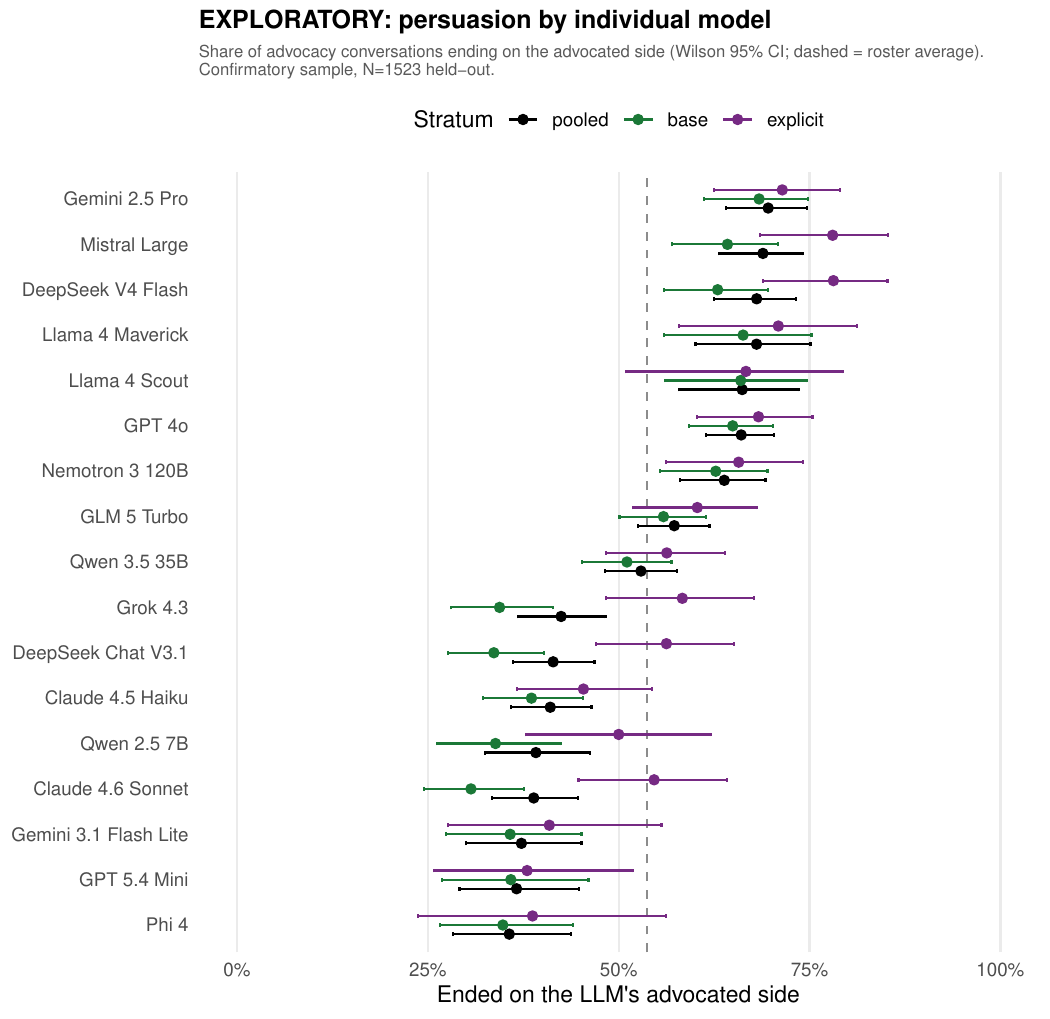}
    \caption{\textbf{Ended on the advocated side, by individual model (exploratory).} Descriptive share of advocacy conversations ending on the model's advocated side, pooled and by steering strength, with the roster average marked by the dashed line (Wilson 95\% CI).}
    \label{fig:persuasion_by_model}
\end{figure}

\begin{figure}[htbp]
    \centering
    \includegraphics[width=\linewidth]{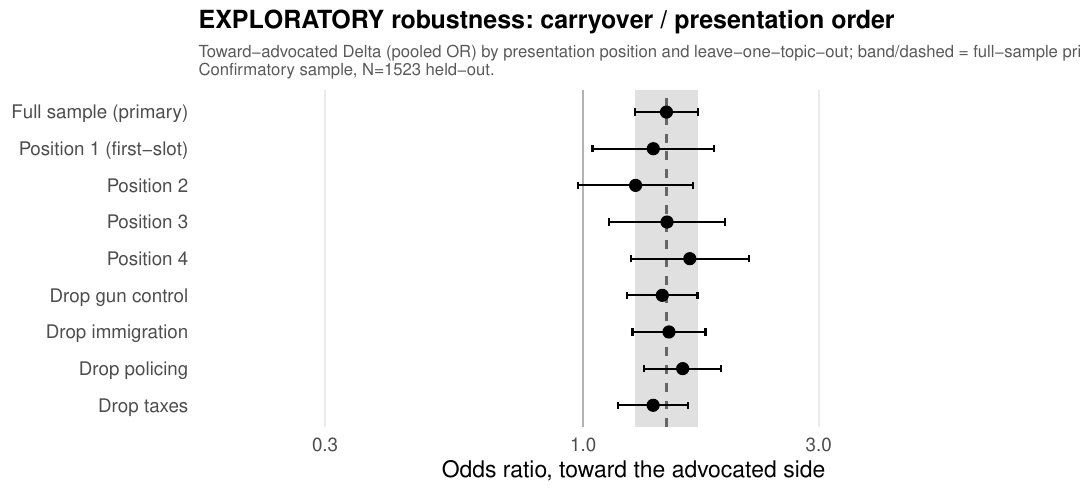}
    \caption{\textbf{Carryover and presentation-order robustness (exploratory).} Toward-advocated pooled odds ratio under first-slot-only and leave-one-topic-out refits; the shaded band and dashed line mark the full-sample primary estimate.}
    \label{fig:persuasion_carryover}
\end{figure}

\clearpage
\section*{Supplementary Tables}\label{sec:tables}

\begin{table}[htb]
\centering
\begin{tabular}{lccccc}
\hline
 & \textbf{CES 2022} & \textbf{CES 2024} & \textbf{U.S. Supreme Court} & \textbf{118th Congress} \\ 
\textbf{LLM Model}		                                                 	& \textbf{Questions} & \textbf{Questions} & \textbf{Case Votes} & \textbf{Bill Votes} \\ \hline
\textbf{Calme 3.3 3B}		                                      & 0.977              &0.991                                  & 0.986                      & 0.971          \\ 
\textbf{Claude Haiku}	& 1.0      &1.0                                              & 1.0                        & 1.0            \\ 
\textbf{Claude Sonnet} 			& 1.0       &1.0                        & 1.0                        & 0.977          \\ 
\textbf{Deepseek 2 Lite}		                                & 0.949             &0.978                                   & 0.984                      & 0.984          \\   
\textbf{Deepseek 3}		                                  & 0.949             &0.956                                   & 0.991                      & 0.713          \\  

\textbf{Falcon 3}		                                      			& 1.0        &1.0                                        & 1.0                      & 1.0          \\ 
\textbf{Gemini 1.5 Flash}		                                          	& 1.0        &1.0                                            & 1.0                        & 1.0            \\ 
\textbf{Gemini 1.5 Pro}		                                           	 	& 1.0       &1.0                                             & 0.994                      & 1.0            \\ 

\textbf{Gemini 2.0 Flash}		                                          	& 1.0          &0.977                                         & 0.981                        & 0.988            \\ 
\textbf{Gemini 2.5 Pro}		                                          	& 0.978           &0.984                                         & 0.971                        & 0.981            \\ 
\textbf{Gemma 9B}	& 1.0             &1.0                                       & 1.0                        & 1.0            \\ 
\textbf{Gemma 27B}	& 1.0           &1.0                                         & 1.0                        & 1.0            \\ 
\textbf{GPT 3.5 Turbo}		                                        		& 0.967       &0.95                                          & 0.967                      & 0.924          \\ 
\textbf{GPT 4o-mini}		                                         			& 0.96 	  &0.95                                              & 0.980                      & 1.0            \\ 
\textbf{GPT 4o}		                                                		& 0.953         &0.93                                        & 0.980                      & 0.915          \\ 

\textbf{GPT 4.5}		                                        		& 0.972          &1.0                                       & 0.947                     & 0.868         \\ 
\textbf{Grok}		                                                  		& 0.861      &0.731                                         & 0.943                      & 0.965          \\ 
\textbf{Grok 2}		                                             	& 0.964        &0.981                                 & 0.951                      & 0.96         \\ 
\textbf{Grok 3}		                                             	& 0.953         &0.970                                & 0.962                     & 0.954        \\ 

\textbf{Llama 3.1 70B}		                                				& 0.975       &0.934                                         & 0.994                      & 0.967          \\ 

\textbf{Llama 3.3 70B}		                                				& 1.0       &1.0                                             & 1.0                        & 0.992         \\ 
\textbf{Llama 3.2 1B}		                                 				& 0.882       &0.787                                         & 0.972                      & 0.626          \\ 
\textbf{Llama 3.1 8B}		                                 				& 1.0        &0.992                                          & 0.945                      & 0.822          \\ 

\textbf{Llama 4 Scout 17B-16E}		                             & 1.0             &0.967                                   & 0.992                     & 0.957          \\ 
\textbf{Llama 4 Maverick 17B-128E}		                       & 0.982               &0.982                                 & 0.968                      & 0.884          \\ 
\textbf{Mistral Nemo}	& 0.96       &0.986                                          & 0.968                      & 0.846          \\ 
\textbf{Mistral Small} 	& 1.0          &1.0                                          & 1.0                        & 0.795          \\ 
\textbf{Nvidia Nemotron}		                                & 1.0            &1.0                                        & 1.0                        & 1.0            \\ 
\textbf{Phi 3.5 Mini}		                                    & 1.0               &1.0                                     & 0.993                      & 0.988          \\ 
\textbf{Phi 3.5 MoE}		                                 & 0.991             &0.957                                     & 1.0                        & 0.979          \\ 
\textbf{Qwen 2.5 7B}		                                      & 0.976           &0.991                                     & 0.992                      & 0.951          \\ 

\hline 
\end{tabular}
\caption{Fleiss’ Kappa inter-rater agreement across same answer order.}
\end{table}

\begin{table}[htbp!]
\centering
\begin{tabular}{lcccc}
\hline
\textbf{LLM Model}                 & \textbf{\begin{tabular}[c]{@{}c@{}}CES 2022\\ Questions\end{tabular}} & \textbf{\begin{tabular}[c]{@{}c@{}}CES 2024\\ Questions\end{tabular}} & \textbf{\begin{tabular}[c]{@{}c@{}}U.S. Supreme Court\\ Case Votes\end{tabular}} & \textbf{\begin{tabular}[c]{@{}c@{}}118th Congress\\ Bill Votes\end{tabular}} \\ \hline
\textbf{Calme 3.3 3B}              & 0.86217                                                               & 0.89277                                                               & 0.91082                                                                          & 0.81412                                                                      \\
\textbf{Claude Haiku}              & 0.80513                                                               & 0.78772                                                               & 0.97203                                                                          & 0.86761                                                                      \\
\textbf{Claude Sonnet}             & 0.7641                                                                & 0.94148                                                               & 0.94868                                                                          & 0.76324                                                                      \\
\textbf{Deepseek 3}                & 0.88819                                                               & 0.89331                                                               & 0.88026                                                                          & 0.58649                                                                      \\
\textbf{Deepseek 2 Lite}           & 0.78918                                                               & 0.85256                                                               & 0.7459                                                                           & 0.23023                                                                      \\
\textbf{Falcon 3}                  & 0.69076                                                               & 0.80652                                                               & 0.92909                                                                          & 0.65003                                                                      \\
\textbf{Gemini 1.5 Flash}          & 0.96643                                                               & 0.94484                                                               & 0.82853                                                                          & 0.81683                                                                      \\
\textbf{Gemini 2.0 Flash}          & 1.0                                                                   & 0.91602                                                               & 0.92004                                                                          & 0.84446                                                                      \\
\textbf{Gemini 1.5 Pro}            & 0.84069                                                               & 0.96679                                                               & 0.98979                                                                          & 0.89722                                                                      \\
\textbf{Gemini 2.5 Pro}            & 0.77394                                                               & 0.8348                                                                & 0.80747                                                                          & 0.65358                                                                      \\
\textbf{Gemma 27B}                 & 0.88406                                                               & 0.87288                                                               & 0.87324                                                                          & 0.91031                                                                      \\
\textbf{Gemma 9B}                  & 0.96757                                                               & 0.85164                                                               & 0.77035                                                                          & 0.8528                                                                       \\
\textbf{GPT 3.5 Turbo}             & 0.95226                                                               & 0.95226                                                               & 0.91471                                                                          & 0.5237                                                                       \\
\textbf{GPT 4.5}                   & 0.79393                                                               & 0.96686                                                               & 0.8907                                                                           & 0.81249                                                                      \\
\textbf{GPT 4o}                    & 0.97287                                                               & 0.92005                                                               & 0.88155                                                                          & 0.78283                                                                      \\
\textbf{GPT 4o-mini}               & 0.92484                                                               & 0.92205                                                               & 0.85495                                                                          & 0.93216                                                                      \\
\textbf{Grok}                      & 0.88945                                                               & 0.82341                                                               & 0.83595                                                                          & 0.83104                                                                      \\
\textbf{Grok 2}                    & 0.97586                                                               & 1.0                                                                   & 0.9024                                                                           & 0.78417                                                                      \\
\textbf{Grok 3}                    & 0.90852                                                               & 0.88808                                                               & 0.89883                                                                          & 0.89657                                                                      \\
\textbf{Llama 3.1 70B}             & 0.89522                                                               & 0.82859                                                               & 0.96025                                                                          & 0.92542                                                                      \\
\textbf{Llama 3.1 8B}              & 0.88406                                                               & 0.92088                                                               & 0.80868                                                                          & 0.73817                                                                      \\
\textbf{Llama 3.2 1B}              & -0.01284                                                              & 0.17446                                                               & 0.96102                                                                          & -0.09595                                                                     \\
\textbf{Llama 3.3 70B}             & 0.95578                                                               & 0.96882                                                               & 0.83795                                                                          & 0.85348                                                                      \\
\textbf{Llama 4 Maverick 17B-128E} & 0.90948                                                               & 0.81011                                                               & 0.89409                                                                          & 0.83044                                                                      \\
\textbf{Llama 4 Scout 17B-16E}     & 0.90791                                                               & 0.88103                                                               & 0.89494                                                                          & 0.66187                                                                      \\
\textbf{Mistral Nemo}              & 0.53284                                                               & 0.54685                                                               & 0.73025                                                                          & 0.76298                                                                      \\
\textbf{Mistral Small}             & 0.60173                                                               & 0.66807                                                               & 0.77972                                                                          & 0.72604                                                                      \\
\textbf{Nvidia Nemotron}           & 0.96757                                                               & 0.97295                                                               & 0.91862                                                                          & 0.9228                                                                       \\
\textbf{Phi 3.5 Mini}              & 0.72331                                                               & 0.91926                                                               & 0.79294                                                                          & 0.80195                                                                      \\
\textbf{Phi 3.5 MoE}               & 0.7641                                                                & 0.91955                                                               & 0.81055                                                                          & 0.66341                                                                      \\
\textbf{Qwen 2.5 7B}               & 0.84069                                                               & 0.91333                                                               & 0.86603                                                                          & 0.91315                                                               \\ \hline
\end{tabular}
\caption{Fleiss Kappa inter-rater agreement when shuffling answer order.}
\end{table}

\begin{table}[htbp!]
\centering
\small
\begin{tabular}{lccc|c|cc}
\hline
\textbf{LLM Model} & \textbf{Imagine} & \textbf{State} & \textbf{Indicate} & \textbf{\begin{tabular}[c]{@{}c@{}}Rewording\\ Average\end{tabular}} & \textbf{\begin{tabular}[c]{@{}c@{}}No\\ Pretend\end{tabular}} & \textbf{As LLM} \\ \hline
\textbf{Calme 3.3 3B} & 0.895 & 0.771 & 0.768 & 0.811 & 0.811 & 0.874 \\
\textbf{Claude 4.5 Haiku} & 0.771 & 0.812 & 0.730 & 0.771 & 0.707 & 0.527 \\
\textbf{Claude 4.6 Sonnet} & 0.830 & 0.743 & 0.712 & 0.762 & 0.637 & 0.630 \\
\textbf{DeepSeek Chat V3.2} & 0.637 & 0.754 & 0.779 & 0.723 & 0.712 & 0.781 \\
\textbf{DeepSeek V4 Flash} & 0.536 & 0.728 & 0.773 & 0.679 & 0.665 & 0.740 \\
\textbf{Falcon 3} & 0.596 & 0.546 & 0.429 & 0.524 & 0.619 & 0.399 \\
\textbf{GLM 4 32B} & 0.644 & 0.710 & 0.437 & 0.597 & 0.474 & 0.454 \\
\textbf{GPT 3.5 Turbo} & 0.737 & 0.820 & 0.868 & 0.808 & 0.762 & 0.828 \\
\textbf{GPT 4o} & 0.952 & 0.976 & 0.951 & 0.960 & 0.883 & 0.817 \\
\textbf{GPT 4o-mini} & 0.898 & 0.926 & 0.874 & 0.899 & 0.926 & 0.833 \\
\textbf{GPT 5.4} & 0.721 & 0.778 & 0.897 & 0.799 & 0.571 & 0.545 \\
\textbf{GPT 5.4 Mini} & 0.796 & 0.761 & 0.813 & 0.790 & 0.796 & 0.705 \\
\textbf{GPT 5.4 Nano} & 0.495 & 0.708 & 0.618 & 0.607 & 0.562 & 0.754 \\
\textbf{GPT-OSS 120B} & 0.760 & 0.626 & 0.689 & 0.692 & 0.490 & 0.307 \\
\textbf{GPT-OSS 20B} & 0.718 & 0.772 & 0.767 & 0.752 & 0.486 & 0.467 \\
\textbf{Gemma 27B} & 0.802 & 0.877 & 0.451 & 0.710 & 0.770 & 0.503 \\
\textbf{Gemma 3} & 0.809 & 0.851 & 0.765 & 0.808 & 0.690 & 0.501 \\
\textbf{Gemma 9B} & 0.821 & 0.806 & 0.527 & 0.718 & 0.778 & 0.484 \\
\textbf{Grok 3} & 0.881 & 0.729 & 0.790 & 0.800 & 0.755 & 0.634 \\
\textbf{Grok 4} & 0.836 & 0.738 & 0.836 & 0.803 & 0.746 & 0.748 \\
\textbf{Llama 3.1 70B} & 0.657 & 0.616 & 0.597 & 0.623 & 0.712 & 0.315 \\
\textbf{Llama 3.1 8B} & 0.768 & 0.875 & 0.847 & 0.830 & 0.826 & 0.633 \\
\textbf{Llama 3.2 1B} & 0.456 & 0.461 & 0.481 & 0.466 & 0.556 & 0.506 \\
\textbf{Llama 3.3 70B} & 0.893 & 0.820 & 0.840 & 0.851 & 0.820 & 0.595 \\
\textbf{Llama 4 Maverick 17B-128E} & 0.940 & 0.911 & 0.970 & 0.940 & 0.750 & 0.602 \\
\textbf{Llama 4 Scout 17B-16E} & 0.825 & 0.857 & 0.884 & 0.855 & 0.805 & 0.781 \\
\textbf{Mistral Large} & 0.979 & 0.958 & 0.916 & 0.951 & 0.897 & 0.706 \\
\textbf{Mistral Nemo} & 0.634 & 0.654 & 0.713 & 0.667 & 0.572 & 0.733 \\
\textbf{Mistral Small (2409)} & 0.805 & 0.839 & 0.828 & 0.824 & 0.822 & 0.784 \\
\textbf{Mistral Small (2603)} & 0.641 & 0.654 & 0.651 & 0.649 & 0.644 & 0.583 \\
\textbf{Nvidia Nemotron} & 0.913 & 0.957 & 0.893 & 0.921 & 0.644 & 0.572 \\
\textbf{Phi 3.5 Mini} & 0.845 & 0.799 & 0.707 & 0.784 & 0.863 & 0.492 \\
\textbf{Qwen 2.5 7B} & 0.836 & 0.896 & 0.832 & 0.855 & 0.921 & 0.664 \\
\hline
\textbf{Average} & \textbf{0.767} & \textbf{0.780} & \textbf{0.746} & \textbf{0.765} & \textbf{0.717} & \textbf{0.621} \\
\hline
\end{tabular}
\caption{Robustness of LLM survey responses to prompt framing (Cohen's $\kappa$ vs.\ baseline). Models with coverage below 0.5 in any variation are excluded.}
\label{tab:robust_summary}
\end{table}

\begin{table}[htbp!]
\centering
\scriptsize
\begin{tabular}{lccc}
\hline
\textbf{LLM Model} & \textbf{\begin{tabular}[c]{@{}c@{}}Overall\\ $\kappa$\end{tabular}} & \textbf{\begin{tabular}[c]{@{}c@{}}Binary\\ $\kappa$\end{tabular}} & \textbf{Coverage} \\ \hline
\textbf{Calme 3.3 3B} & 0.811 & 0.798 & 0.978 \\
\textbf{Claude 4.5 Haiku} & 0.707 & 0.688 & 0.978 \\
\textbf{Claude 4.6 Sonnet} & 0.637 & 0.607 & 0.978 \\
\textbf{DeepSeek Chat V3.2} & 0.712 & 0.691 & 0.978 \\
\textbf{DeepSeek V4 Flash} & 0.665 & 0.672 & 0.957 \\
\textbf{Falcon 3} & 0.619 & 0.561 & 0.978 \\
\textbf{GLM 4 32B} & 0.474 & 0.554 & 0.870 \\
\textbf{GLM 5 Turbo} & 0.000 & -- & 0.000 \\
\textbf{GPT 3.5 Turbo} & 0.762 & 0.743 & 0.978 \\
\textbf{GPT 4o} & 0.883 & 0.874 & 0.978 \\
\textbf{GPT 4o-mini} & 0.926 & 0.920 & 0.978 \\
\textbf{GPT 5.4} & 0.571 & 0.519 & 0.978 \\
\textbf{GPT 5.4 Mini} & 0.796 & 0.778 & 0.978 \\
\textbf{GPT 5.4 Nano} & 0.562 & 0.492 & 0.978 \\
\textbf{GPT-OSS 120B} & 0.490 & 0.870 & 0.761 \\
\textbf{GPT-OSS 20B} & 0.486 & 0.770 & 0.750 \\
\textbf{Gemini 2.5 Flash} & 0.161 & 0.621 & 0.359 \\
\textbf{Gemini 2.5 Pro} & 0.013 & 1.000 & 0.022 \\
\textbf{Gemini 3 Flash} & -- & -- & 0.000 \\
\textbf{Gemini 3.1 Flash Lite} & 0.053 & 1.000 & 0.185 \\
\textbf{Gemini 3.1 Pro} & 0.000 & -- & 0.000 \\
\textbf{Gemma 27B} & 0.770 & 0.861 & 0.924 \\
\textbf{Gemma 3} & 0.690 & 0.718 & 0.946 \\
\textbf{Gemma 9B} & 0.778 & 0.802 & 0.957 \\
\textbf{Grok 3} & 0.755 & 0.738 & 0.978 \\
\textbf{Grok 4} & 0.746 & 0.744 & 0.967 \\
\textbf{Llama 3.1 70B} & 0.712 & 0.682 & 0.978 \\
\textbf{Llama 3.1 8B} & 0.826 & 0.808 & 0.978 \\
\textbf{Llama 3.2 1B} & 0.556 & 0.577 & 0.935 \\
\textbf{Llama 3.3 70B} & 0.820 & 0.804 & 0.978 \\
\textbf{Llama 4 Maverick 17B-128E} & 0.750 & 0.726 & 0.978 \\
\textbf{Llama 4 Scout 17B-16E} & 0.805 & 0.786 & 0.978 \\
\textbf{Mistral Large} & 0.897 & 0.910 & 0.967 \\
\textbf{Mistral Nemo} & 0.572 & 0.545 & 0.978 \\
\textbf{Mistral Small (2409)} & 0.822 & 0.810 & 0.978 \\
\textbf{Mistral Small (2603)} & 0.644 & 0.652 & 0.957 \\
\textbf{Nemotron 3 120B} & 0.233 & 0.575 & 0.500 \\
\textbf{Nvidia Nemotron} & 0.644 & 0.663 & 0.946 \\
\textbf{Phi 3.5 Mini} & 0.863 & 0.873 & 0.967 \\
\textbf{Phi 3.5 MoE} & 0.690 & 0.805 & 0.891 \\
\textbf{Phi 4} & 0.397 & 0.853 & 0.576 \\
\textbf{Qwen 2.5 7B} & 0.921 & 0.914 & 0.978 \\
\hline
\textbf{Average} & \textbf{0.615} & \textbf{0.744} & \textbf{0.809} \\
\hline
\end{tabular}
\caption{No-pretend prompt framing: per-model Cohen's $\kappa$ vs.\ baseline. Overall $\kappa$ treats refusals and non-binary responses as a third class; binary $\kappa$ is restricted to the Support/Oppose subset; coverage is the fraction of items retained for the binary calculation.}
\label{tab:robust_nopretend}
\end{table}

\begin{table}[htbp!]
\centering
\scriptsize
\begin{tabular}{lccc}
\hline
\textbf{LLM Model} & \textbf{\begin{tabular}[c]{@{}c@{}}Overall\\ $\kappa$\end{tabular}} & \textbf{\begin{tabular}[c]{@{}c@{}}Binary\\ $\kappa$\end{tabular}} & \textbf{Coverage} \\ \hline
\textbf{Calme 3.3 3B} & 0.874 & 0.866 & 0.978 \\
\textbf{Claude 4.5 Haiku} & 0.527 & 0.799 & 0.761 \\
\textbf{Claude 4.6 Sonnet} & 0.630 & 0.602 & 0.978 \\
\textbf{DeepSeek Chat V3.2} & 0.781 & 0.766 & 0.978 \\
\textbf{DeepSeek V4 Flash} & 0.740 & 0.740 & 0.967 \\
\textbf{Falcon 3} & 0.399 & 0.344 & 0.946 \\
\textbf{GLM 4 32B} & 0.454 & 0.891 & 0.717 \\
\textbf{GLM 5 Turbo} & 0.108 & 0.000 & 0.043 \\
\textbf{GPT 3.5 Turbo} & 0.828 & 0.813 & 0.978 \\
\textbf{GPT 4o} & 0.817 & 0.867 & 0.946 \\
\textbf{GPT 4o-mini} & 0.833 & 0.841 & 0.967 \\
\textbf{GPT 5.4} & 0.545 & 0.494 & 0.978 \\
\textbf{GPT 5.4 Mini} & 0.705 & 0.683 & 0.978 \\
\textbf{GPT 5.4 Nano} & 0.754 & 0.713 & 0.978 \\
\textbf{GPT-OSS 120B} & 0.307 & 0.837 & 0.609 \\
\textbf{GPT-OSS 20B} & 0.467 & 0.740 & 0.739 \\
\textbf{Gemini 2.5 Flash} & 0.065 & -- & 0.152 \\
\textbf{Gemini 2.5 Pro} & -- & -- & -- \\
\textbf{Gemini 3 Flash} & -- & -- & 0.000 \\
\textbf{Gemini 3.1 Flash Lite} & 0.000 & -- & 0.000 \\
\textbf{Gemini 3.1 Pro} & 0.000 & -- & 0.000 \\
\textbf{Gemma 27B} & 0.503 & 0.734 & 0.772 \\
\textbf{Gemma 3} & 0.501 & 0.840 & 0.717 \\
\textbf{Gemma 9B} & 0.484 & 0.708 & 0.772 \\
\textbf{Grok 3} & 0.634 & 0.609 & 0.978 \\
\textbf{Grok 4} & 0.748 & 0.749 & 0.967 \\
\textbf{Llama 3.1 70B} & 0.315 & 0.494 & 0.750 \\
\textbf{Llama 3.1 8B} & 0.633 & 0.603 & 0.978 \\
\textbf{Llama 3.2 1B} & 0.506 & 0.493 & 0.935 \\
\textbf{Llama 3.3 70B} & 0.595 & 0.565 & 0.978 \\
\textbf{Llama 4 Maverick 17B-128E} & 0.602 & 0.586 & 0.967 \\
\textbf{Llama 4 Scout 17B-16E} & 0.781 & 0.760 & 0.978 \\
\textbf{Mistral Large} & 0.706 & 0.947 & 0.826 \\
\textbf{Mistral Nemo} & 0.733 & 0.716 & 0.978 \\
\textbf{Mistral Small (2409)} & 0.784 & 0.769 & 0.978 \\
\textbf{Mistral Small (2603)} & 0.583 & 0.609 & 0.935 \\
\textbf{Nemotron 3 120B} & 0.149 & 0.567 & 0.315 \\
\textbf{Nvidia Nemotron} & 0.572 & 0.881 & 0.761 \\
\textbf{Phi 3.5 Mini} & 0.492 & 1.000 & 0.674 \\
\textbf{Phi 3.5 MoE} & 0.134 & 0.000 & 0.304 \\
\textbf{Phi 4} & 0.105 & 0.519 & 0.272 \\
\textbf{Qwen 2.5 7B} & 0.664 & 0.772 & 0.902 \\
\hline
\textbf{Average} & \textbf{0.526} & \textbf{0.673} & \textbf{0.742} \\
\hline
\end{tabular}
\caption{As-LLM prompt framing: per-model Cohen's $\kappa$ vs.\ baseline. Columns as in Table~\ref{tab:robust_nopretend}.}
\label{tab:robust_asllm}
\end{table}

\begin{table}[htbp!]
\centering
\scriptsize
\begin{tabular}{lccc}
\hline
\textbf{LLM Model} & \textbf{\begin{tabular}[c]{@{}c@{}}Overall\\ $\kappa$\end{tabular}} & \textbf{\begin{tabular}[c]{@{}c@{}}Binary\\ $\kappa$\end{tabular}} & \textbf{Coverage} \\ \hline
\textbf{Calme 3.3 3B} & 0.811 & 0.799 & 0.978 \\
\textbf{Claude 4.5 Haiku} & 0.771 & 0.756 & 0.978 \\
\textbf{Claude 4.6 Sonnet} & 0.762 & 0.763 & 0.967 \\
\textbf{DeepSeek Chat V3.2} & 0.723 & 0.703 & 0.978 \\
\textbf{DeepSeek V4 Flash} & 0.679 & 0.657 & 0.974 \\
\textbf{Falcon 3} & 0.524 & 0.416 & 0.978 \\
\textbf{GLM 4 32B} & 0.597 & 0.550 & 0.978 \\
\textbf{GLM 5 Turbo} & 0.339 & 0.880 & 0.188 \\
\textbf{GPT 3.5 Turbo} & 0.808 & 0.790 & 0.978 \\
\textbf{GPT 4o} & 0.960 & 0.956 & 0.978 \\
\textbf{GPT 4o-mini} & 0.899 & 0.891 & 0.978 \\
\textbf{GPT 5.4} & 0.799 & 0.769 & 0.978 \\
\textbf{GPT 5.4 Mini} & 0.790 & 0.773 & 0.978 \\
\textbf{GPT 5.4 Nano} & 0.607 & 0.550 & 0.978 \\
\textbf{GPT-OSS 120B} & 0.692 & 0.851 & 0.895 \\
\textbf{GPT-OSS 20B} & 0.752 & 0.760 & 0.964 \\
\textbf{Gemini 2.5 Flash} & 0.523 & 0.698 & 0.772 \\
\textbf{Gemini 2.5 Pro} & 0.372 & 0.383 & 0.870 \\
\textbf{Gemini 3 Flash} & 0.000 & -- & 0.000 \\
\textbf{Gemini 3.1 Flash Lite} & 0.328 & 0.786 & 0.580 \\
\textbf{Gemini 3.1 Pro} & 0.000 & -- & 0.000 \\
\textbf{Gemma 27B} & 0.710 & 0.868 & 0.870 \\
\textbf{Gemma 3} & 0.808 & 0.795 & 0.978 \\
\textbf{Gemma 9B} & 0.718 & 0.785 & 0.913 \\
\textbf{Grok 3} & 0.800 & 0.784 & 0.978 \\
\textbf{Grok 4} & 0.803 & 0.794 & 0.974 \\
\textbf{Llama 3.1 70B} & 0.623 & 0.590 & 0.978 \\
\textbf{Llama 3.1 8B} & 0.830 & 0.812 & 0.978 \\
\textbf{Llama 3.2 1B} & 0.466 & 0.584 & 0.848 \\
\textbf{Llama 3.3 70B} & 0.851 & 0.837 & 0.978 \\
\textbf{Llama 4 Maverick 17B-128E} & 0.940 & 0.934 & 0.978 \\
\textbf{Llama 4 Scout 17B-16E} & 0.855 & 0.840 & 0.978 \\
\textbf{Mistral Large} & 0.951 & 0.948 & 0.978 \\
\textbf{Mistral Nemo} & 0.667 & 0.646 & 0.978 \\
\textbf{Mistral Small (2409)} & 0.824 & 0.810 & 0.978 \\
\textbf{Mistral Small (2603)} & 0.649 & 0.611 & 0.978 \\
\textbf{Nemotron 3 120B} & 0.286 & 0.530 & 0.608 \\
\textbf{Nvidia Nemotron} & 0.921 & 0.930 & 0.971 \\
\textbf{Phi 3.5 Mini} & 0.784 & 0.804 & 0.956 \\
\textbf{Phi 3.5 MoE} & 0.651 & 0.756 & 0.891 \\
\textbf{Phi 4} & 0.386 & 0.787 & 0.630 \\
\textbf{Qwen 2.5 7B} & 0.855 & 0.840 & 0.978 \\
\hline
\textbf{Average} & \textbf{0.669} & \textbf{0.750} & \textbf{0.866} \\
\hline
\end{tabular}
\caption{Voter-rewording prompt framings (mean of Imagine, State, Indicate): per-model Cohen's $\kappa$ vs.\ baseline. Columns as in Table~\ref{tab:robust_nopretend}.}
\label{tab:robust_rewording}
\end{table}

\begin{table}[hbtp]
\centering
\small
\begin{tabular}{lp{3.5in}ll}
\textbf{Topic}       & \textbf{Question}   & \textbf{LLM}          & \textbf{LLM Answer} \\
\toprule
Gun Control & Should the government make it easier for people to obtain concealed-carry permit?                                                                      & GPT-4o       & No         \\
Gun Control & Should the government ban assault rifles?                                                                                                              & GPT-4o       & Yes        \\
Immigration & Should the government reduce legal immigration by 50 percent over the next 10 years by eliminating the visa lottery and ending family-based migration? & Llama 3.2 1B & Yes        \\
Immigration & Should the government increase spending on border security by \$25 billion?                                                                            & Llama 3.2 1B & Yes        \\
Healthcare  & Should the government expand Medicare to a single comprehensive public health care coverage program that would cover all Americans?                    & GPT-4o       & Yes        \\
Healthcare  & Should the government allow states to import prescription drugs from other countries?                                                                  & GPT-4o       & Yes        \\
Police      & Should the government end the Department of Defense program that sends surplus military weapons and equipment to police departments?                   & Mistral Nemo & No         \\
Police      & Should the government create a national registry of police who have been investigated or disciplined for misconduct?                                   & Mistral Nemo & No        \\ \hline
\end{tabular}
\label{tab:survey_questions}
\caption{Survey questions and their corresponding LLM answers.}
\end{table}

\begin{table}[!htbp]
\centering
\begin{tabular}{lcc}
Topic               & CES 2022 Question & CES 2024 Question \\ \hline
Gun Control         & CC22\_330         & CC24\_321        \\
Illegal Immigration & CC22\_331         & CC24\_323        \\
Abortion            & CC22\_332        & CC24\_324       \\
Climate             & CC22\_333         & CC24\_326        \\
Government Spending & CC22\_350         & CC24\_328        \\
Miscellaneous       & CC22\_355         & CC24\_444         \\
Health care & CC22\_327 & $~\sim$ \\
Police              & CC22\_334         & $\sim$            \\
Taxes               & $\sim$            & CC24\_341     \\ \hline   
\end{tabular}
\caption{The questions taken from each of the 2022 and 2024 CES surveys on each of the studied topics. Each of these questions contains multiple sub-questions. Please refer to~\cite{harvardCooperativeElection} for question texts and the full CES questionnaires.}
\end{table}

\begin{table}[htbp!]
\centering
{\tiny
\begin{tabular}{lll}
\textbf{LLM Model} & \textbf{Repository/API-based} & Date\\ \hline
\textbf{claude-4-5-haiku}~\cite{claude_haiku} & \url{API-based}  & 15-30 April \\ 

\textbf{@vertexai/anthropic.claude-4-6-sonnet}~\cite{claude_sonnet} & \url{API-based} & 15-30 April
\\ 
\textbf{deepseek-chat(deepseek\_chat\_v3\_2)}~\cite{liu2025deepseek} & \url{API-based} & 15-30 April \\ 
\textbf{deepseek/deepseek\_v4\_flash}~\cite{deepseek_flash1,deepseek_flash2} & \url{API-based} & 15-30 April \\  
\textbf{@vertexai/gemini-2.5-flash}~\cite{comanici2025gemini} & \url{API-based} & 15-30 April \\ 
\textbf{@vertexai/gemini-2.5-pro}~\cite{comanici2025gemini} & \url{API-based} & 15-30 April \\ 
\textbf{@vertexai/gemini-3-flash-preview}~\cite{gemini_3_flash} & \url{API-based} & 15-30 April \\ 
\textbf{@vertexai/gemini-3.1-flash-lite-preview}~\cite{gemini_3.1_flash_lite} & \url{API-based} & 15-30 April \\ 
\textbf{@vertexai/gemini-3.1-pro-preview}~\cite{gemini_3.1_pro} & \url{API-based} & 15-30 April \\ 
\textbf{gpt-3.5-turbo-0125}~\cite{GPT-3.5-turbo} & \url{API-based} & 15-30 April \\ 
\textbf{gpt-4o-mini-2024-07-18}~\cite{GPT-4o_mini} & \url{API-based} & 15-30 April \\ 
\textbf{gpt-4o-2024-08-06}~\cite{openai2024gpt4ocard}  & \url{API-based} & 15-30 April \\ 
\textbf{gpt-5.4-nano-2026-03-17}~\cite{gpt_nano_mini}  & \url{API-based} & 15-30 April \\ 
\textbf{gpt-5.4-mini-2026-03-17}~\cite{gpt_nano_mini}  & \url{API-based} & 15-30 April \\ 
\textbf{gpt-5.4-2026-03-05}~\cite{gpt_5.4}  & \url{API-based} & 15-30 April \\ 
\textbf{gpt-oss-20b}~\cite{gpt-oss}  & \url{API-based} & 15-30 April \\ 
\textbf{gpt-oss-120b}~\cite{gpt-oss}  & \url{API-based} & 15-30 April \\ 
\textbf{Grok-3-beta}~\cite{grok3} & \url{API-based} & 15-30 April \\ 
\textbf{Grok-4}~\cite{grok4} & \url{API-based} & 15-30 April \\ 

\textbf{meta-llama/Llama-3.1-8B-Instruct}~\cite{grattafiori2024llama,Llama-3.1-8B-Instruct} & \url{https://huggingface.co/meta-llama/Llama-3.1-8B-Instruct} & 1-15 May \\ 
\textbf{meta-llama/Llama-3.1-70B-Instruct}~\cite{grattafiori2024llama,Llama-3.1-70B-Instruct} & \url{https://huggingface.co/meta-llama/Llama-3.1-70B-Instruct} & 1-15 May \\ 
\textbf{meta-llama/Llama-3.2-1B-Instruct}~\cite{grattafiori2024llama,Llama-3.2-1B-Instruct} & \url{https://huggingface.co/meta-llama/Llama-3.2-1B-Instruct} & 1-15 May \\ 
\textbf{meta-llama/Llama-3.3-70B-Instruct}~\cite{grattafiori2024llama,Llama-3.3-70B-Instruct} & \url{https://huggingface.co/meta-llama/Llama-3.3-70B-Instruct} & 1-15 May \\ 
\textbf{meta-llama/Llama-4-Scout-17B-16E-Instruct}~\cite{Llama4,Llama-4-Scout-17B-16E-Instruct} & \url{https://huggingface.co/meta-llama/Llama-4-Scout-17B-16E-Instruct} & 1-15 May \\ 
\textbf{meta-llama/llama-4-maverick}~\cite{Llama4} & \url{API-based} & 15-30 April \\

\textbf{google/gemma-2-9b-it}~\cite{gemmateam2024gemma2improvingopen,gemma-2-9b-it} & \url{https://huggingface.co/google/gemma-2-9b-it} & 1-15 May \\ 
\textbf{google/gemma-2-27b-it}~\cite{gemmateam2024gemma2improvingopen,gemma-2-27b-it} & \url{https://huggingface.co/google/gemma-2-27b-it} & 1-15 May \\ 
\textbf{google/gemma-3-12b-it}~\cite{gemmateam2025gemma3technicalreport,gemma-3-12b-it} & \url{https://huggingface.co/google/gemma-3-12b-it} & 1-15 May \\ 
\textbf{MaziyarPanahi/calme-3.3-instruct-3b}~\cite{calme-3.3-instruct-3b} & \url{https://huggingface.co/MaziyarPanahi/calme-3.3-instruct-3b} & 1-15 May \\ 

\textbf{mistralai/Mistral-Nemo-Instruct-2407}~\cite{Mistral-Nemo,Mistral-Nemo-Instruct} & \url{https://huggingface.co/mistralai/Mistral-Nemo-Instruct-2407} & 1-15 May \\

\textbf{mistralai/Mistral-Small-Instruct-2409}~\cite{Mistral-Small-Instruct-2409} & \url{https://huggingface.co/mistralai/Mistral-Small-Instruct-2409} & 1-15 May\\ 

\textbf{mistralai/mistral-small-2603}~\cite{Mistral-4} & \url{API-based} & 15-30 April \\ 

\textbf{mistralai/mistral-large-2411}~\cite{Mistral-large} & \url{API-based} & 15-30 April \\ 

\textbf{nvidia/Llama-3.1-Nemotron-70B-Instruct}~\cite{wang2024helpsteer2preferencecomplementingratingspreferences,nvidia_-Nemotron} & \url{https://huggingface.co/nvidia/Llama-3.1-Nemotron-70B-Instruct} & 1-15 May \\ 

\textbf{nvidia/nemotron-3-super-120b-a12b
}~\cite{chandiramani2026nemotron} & \url{API-based} & 15-30 April \\ 

\textbf{microsoft/Phi-3.5-mini-instruct}~\cite{abdin2024phi3technicalreporthighly,Phi-3.5-mini-instruct} & \url{https://huggingface.co/microsoft/Phi-3.5-mini-instruct} & 1-15 May \\ 
\textbf{microsoft/Phi-3.5-MoE-instruct}~\cite{abdin2024phi3technicalreporthighly,Phi-3.5-MoE-instruct} & \url{https://huggingface.co/microsoft/Phi-3.5-MoE-instruct} & 1-15 May \\ 

\textbf{microsoft/phi-4}~\cite{aabdin2024phi} & \url{API-based} & 15-30 April \\ 

\textbf{Qwen/Qwen2.5-72B-Instruct}~\cite{qwen2025qwen25technicalreport,qwen2.5} & \url{https://huggingface.co/Qwen/Qwen2.5-72B-Instruct} & 1-15 May \\ 

\textbf{qwen/qwen3.5-35b-a3b}~\cite{Qwen3.6} & \url{API-based} & 15-30 April \\ 

\textbf{Falcon 3}~\cite{almazrouei2023falcon,Falcon3} & \url{https://huggingface.co/tiiuae/Falcon3-7B-Instruct} & 1-15 May \\ 

\textbf{z-ai/glm-4-32b}~\cite{glm2024chatglm} & \url{API-based} & 15-30 April \\ 

\textbf{z-ai/glm-5-turbo}~\cite{hong2026glm,GLM-5-Turbo} & \url{API-based} & 15-30 April \\ 

\hline
\end{tabular}
}
\caption{The LLMs used in our study.}
\label{tab:llm_sources}
\end{table}

\begin{table}[htbp!]
\centering
\begin{tabular}{lc}
\textbf{Dataset}             & \textbf{Source}                                               \\ \hline
Congress voting history      & GovInfo Bulk Data Repository~\cite{govinfoGovinfoBulkdata}                                  \\
Supreme Court voting history & Supreme Court database code book~\cite{spaeth2014supreme}                              \\
CES 2022                     & Dataverse~\cite{DVN/PR4L8P_2023} \\
CES 2024                     & Dataverse~\cite{DVN/X11EP6_2025} \\ \hline
\end{tabular}
\caption{Sources for voting and survey data used in our study.}
\label{tab:human_sources}
\end{table}

\begin{table}[htbp!]
\centering
\scriptsize
\setlength{\tabcolsep}{4pt}
\renewcommand{\arraystretch}{1.05}
\begin{tabular}{l*{8}{c}}
\hline
\textbf{Model} & \rotatebox{90}{Healthcare} & \rotatebox{90}{Gun control} & \rotatebox{90}{Immigration} & \rotatebox{90}{Abortion} & \rotatebox{90}{Climate} & \rotatebox{90}{Police} & \rotatebox{90}{Spending} & \rotatebox{90}{Exec orders} \\
\hline
\textbf{Median voter} & 0.39 & 0.33 & 0.44 & 0.26 & 0.14 & 0.43 & 0.25 & 0.17 \\
\hline
Calme 3.3 3B & -0.07 & 0.43 & 0.00 & -0.34 & -0.26 & 1.22 & -0.05 & 0.11 \\
Claude 4.5 Haiku & 0.49 & 0.43 & 0.92 & 0.01 & 0.13 & 0.19 & 0.24 & 0.11 \\
Claude 4.6 Sonnet & -0.21 & 1.05 & 0.50 & 0.70 & 0.37 & 0.44 & -0.24 & 0.11 \\
DeepSeek Chat V3.2 & -0.21 & 0.43 & 0.50 & 0.03 & -0.26 & 1.11 & -0.05 & 0.11 \\
DeepSeek V4 Flash & -0.21 & -0.02 & 0.42 & 0.28 & -0.26 & -0.52 & -0.24 & -0.15 \\
Falcon 3 & 0.39 & 0.07 & 0.94 & 1.23 & -0.02 & 0.12 & -0.24 & -0.15 \\
GLM 4 32B & -0.21 & -0.28 & 0.50 & 0.23 & 0.37 & 0.10 & -0.24 & -0.15 \\
GLM 5 Turbo & -0.90 & 0.50 & 1.92 & 1.26 & -0.41 & -0.61 & -1.23 & -0.75 \\
GPT 3.5 Turbo & -0.21 & 0.67 & 0.00 & 0.28 & -0.26 & -0.52 & -0.24 & -0.15 \\
GPT 4o & -0.21 & -0.02 & 0.00 & -0.34 & -0.26 & -0.52 & -0.24 & -0.15 \\
GPT 4o-mini & -0.21 & 0.67 & 0.00 & -0.34 & -0.02 & -0.20 & -0.24 & 0.11 \\
GPT 5.4 & -0.21 & 0.33 & 0.00 & 0.60 & -0.02 & 0.44 & -0.24 & -0.15 \\
GPT 5.4 Mini & -0.21 & 0.67 & 0.00 & -0.34 & -0.02 & 0.50 & -0.24 & -0.15 \\
GPT 5.4 Nano & -0.21 & 0.33 & 0.50 & 0.65 & -0.02 & -0.21 & -0.24 & -0.15 \\
GPT-OSS 120B & -0.34 & 0.33 & 0.00 & 0.60 & -0.26 & -0.52 & -0.24 & -0.46 \\
GPT-OSS 20B & -0.21 & 0.36 & 0.00 & -0.09 & -0.26 & -0.20 & -0.24 & -0.15 \\
Gemini 2.5 Flash & -0.21 & 0.70 & 0.00 & -0.69 & -0.02 & 0.12 & -0.24 & 0.11 \\
Gemini 2.5 Pro & -0.21 & 0.36 & 0.00 & 0.58 & 0.37 & -0.18 & 0.07 & 0.11 \\
Gemini 3.1 Flash Lite & -0.21 & -0.01 & 0.94 & 0.97 & -0.02 & 0.12 & -0.24 & -0.21 \\
Gemma 27B & -0.21 & 0.41 & 0.00 & -0.09 & -0.26 & -0.52 & -0.24 & -0.15 \\
Gemma 3 & -0.21 & 0.09 & 0.00 & 0.26 & 0.13 & 0.44 & -0.24 & 0.11 \\
Gemma 9B & -0.21 & 0.41 & 0.50 & -0.09 & -0.26 & 0.44 & -0.24 & 0.11 \\
Grok 3 & -0.21 & 1.05 & 0.50 & -0.09 & 0.13 & 0.12 & -0.24 & 0.11 \\
Grok 4 & -0.21 & 1.05 & 0.50 & -0.34 & -0.02 & -0.18 & -0.24 & 0.11 \\
Llama 3.1 70B & -0.21 & 0.33 & 0.94 & 0.60 & -0.02 & 0.12 & -0.24 & -0.15 \\
Llama 3.1 8B & -0.21 & 0.07 & 0.94 & 0.65 & -0.02 & 0.10 & -0.24 & -0.15 \\
Llama 3.3 70B & -0.21 & 0.33 & 0.00 & 0.23 & -0.26 & 0.12 & -0.24 & -0.15 \\
Llama 4 Maverick & -0.21 & 0.33 & 0.00 & 0.60 & -0.02 & 0.12 & -0.24 & -0.15 \\
Llama 4 Scout & -0.21 & 0.33 & 0.00 & -0.09 & -0.02 & 0.44 & -0.24 & -0.15 \\
Mistral Large & -0.21 & 0.06 & -0.46 & -0.34 & -0.26 & 0.84 & -0.05 & 0.11 \\
Mistral Small (2409) & -0.21 & 0.41 & 0.00 & -0.34 & -0.26 & 0.81 & -0.10 & -0.15 \\
Mistral Small (2603) & -0.21 & 0.07 & 0.50 & 0.60 & -0.02 & -0.21 & -0.24 & -0.15 \\
Nemotron 3 120B & -0.21 & 0.33 & -0.42 & -0.06 & -0.02 & -0.52 & -0.12 & -0.15 \\
Nvidia Nemotron & -0.21 & 0.06 & 0.00 & -0.09 & -0.26 & 0.13 & -0.24 & 0.11 \\
Phi 3.5 Mini & -0.21 & 0.44 & 0.00 & -0.09 & 0.37 & -0.20 & -0.24 & -0.15 \\
Phi 3.5 MoE & -0.21 & 1.13 & 0.00 & 0.65 & -0.65 & 1.45 & 0.02 & -0.15 \\
Phi 4 & 0.29 & 1.05 & 0.50 & 0.66 & -0.26 & -0.55 & -0.24 & -0.15 \\
Qwen 2.5 7B & -0.21 & 0.33 & 0.00 & -0.09 & -0.26 & 0.12 & -0.24 & -0.15 \\
Qwen 3.5 35B & 0.29 & 0.50 & 0.58 & 1.94 & -0.41 & -0.63 & -1.00 & -0.49 \\
\hline
\end{tabular}
\caption{\textbf{Each LLM's expressed position by issue, CES 2022.} For every issue, each model's responses are projected onto a one-dimensional PCA scale fit on the CES 2022 respondents and normalized so that $0$ is the average strong Democrat and $1$ the average strong Republican; values below $0$ are more liberal than strong Democrats and values above $1$ more conservative than strong Republicans. A model's distance from a benchmark on a given issue is its difference from that reference down the column: from the strong Democrat ($0$), the strong Republican ($1$), or the median CES voter (top row). Four models (Gemini~3.1~Pro, Gemini~3~Flash, Llama~3.2~1B, and Mistral~Nemo) are excluded for excessive refusals, instability, or response-order bias.}
\label{tab:llm_topic_distances}
\end{table}

\begin{table}[htbp!]
\centering
\scriptsize
\setlength{\tabcolsep}{4pt}
\renewcommand{\arraystretch}{1.05}
\begin{tabular}{l*{9}{c}}
\hline
\textbf{Model} & \rotatebox{90}{Gun control + police} & \rotatebox{90}{Immigration} & \rotatebox{90}{Abortion} & \rotatebox{90}{Climate} & \rotatebox{90}{Welfare + housing} & \rotatebox{90}{Mixed misc} & \rotatebox{90}{Taxes} & \rotatebox{90}{Cultural issues} & \rotatebox{90}{SCOTUS topics} \\
\hline
\textbf{Median voter} & 0.08 & 0.41 & 0.17 & 0.33 & 0.27 & 0.21 & 0.08 & 0.17 & 0.98 \\
\hline
Calme 3.3 3B & 0.35 & -0.04 & -0.21 & 0.06 & -0.41 & 0.31 & 0.71 & -1.07 & -1.51 \\
Claude 4.5 Haiku & 0.35 & 0.43 & 0.37 & 0.06 & -0.41 & 0.21 & 0.83 & -1.05 & -1.51 \\
Claude 4.6 Sonnet & 0.99 & -0.04 & -0.21 & 0.33 & -0.06 & 0.21 & 0.08 & 0.19 & 0.98 \\
DeepSeek Chat V3.2 & 0.35 & -0.04 & -0.21 & -0.24 & -0.06 & 0.78 & 0.71 & -0.40 & 0.98 \\
DeepSeek V4 Flash & -0.28 & -0.04 & -0.21 & -0.24 & -0.41 & 0.21 & 0.30 & -0.40 & 0.98 \\
Falcon 3 & 0.72 & 0.41 & 0.50 & -0.24 & 0.11 & 0.21 & 0.08 & 0.11 & -0.22 \\
GLM 4 32B & 0.08 & 0.41 & 0.17 & 0.03 & -0.41 & 0.21 & 0.71 & -0.43 & -0.32 \\
GLM 5 Turbo & 0.27 & 0.80 & 0.08 & -1.24 & 0.29 & 0.11 & -0.76 & 4.41 & 2.27 \\
GPT 3.5 Turbo & 0.72 & -0.04 & 0.17 & -0.24 & -0.41 & 0.21 & 0.71 & -1.79 & -1.51 \\
GPT 4o & -0.28 & -0.04 & -0.21 & -0.24 & -0.41 & 0.21 & -0.32 & -0.40 & -0.22 \\
GPT 4o-mini & 0.36 & -0.04 & -0.21 & -0.24 & -0.41 & 0.21 & 0.08 & -1.07 & -1.51 \\
GPT 5.4 & 0.72 & -0.04 & 0.17 & 0.03 & -0.06 & 0.21 & 0.08 & 0.79 & 0.98 \\
GPT 5.4 Mini & 0.72 & -0.04 & -0.21 & -0.24 & -0.41 & 0.21 & 0.61 & -1.05 & -1.51 \\
GPT 5.4 Nano & 0.72 & 0.41 & 0.17 & 0.03 & -0.06 & 0.21 & 0.08 & 0.17 & -0.22 \\
GPT-OSS 120B & 0.72 & -0.04 & 0.17 & -0.24 & -0.41 & -0.36 & 0.08 & -0.45 & -0.32 \\
GPT-OSS 20B & -0.28 & -0.44 & 0.17 & -0.24 & -0.41 & -0.25 & 1.45 & -1.07 & -1.51 \\
Gemini 2.5 Flash & 0.72 & -0.04 & 0.17 & 0.03 & -0.41 & 0.21 & 0.30 & -0.40 & 0.98 \\
Gemini 2.5 Pro & -0.92 & -0.04 & -0.21 & 0.37 & -0.06 & 0.21 & -0.32 & 0.07 & 0.98 \\
Gemini 3.1 Flash Lite & 0.72 & 0.41 & 0.17 & 0.03 & -0.06 & 0.21 & 0.08 & 0.19 & -0.22 \\
Gemma 27B & 0.72 & -0.04 & 0.17 & -0.24 & -0.41 & 0.30 & 0.08 & -1.07 & -1.51 \\
Gemma 3 & 0.72 & -0.04 & 0.73 & 0.39 & -0.41 & 0.21 & 0.71 & -1.67 & 0.98 \\
Gemma 9B & 0.72 & -0.04 & 0.17 & -0.24 & -0.41 & 0.30 & 0.08 & -1.02 & -0.22 \\
Grok 3 & 1.36 & 0.41 & 0.17 & 0.39 & -0.41 & 0.21 & 0.71 & -0.43 & -0.22 \\
Grok 4 & 1.36 & -0.04 & -0.21 & 0.06 & -0.41 & 0.21 & 0.71 & -0.45 & 0.98 \\
Llama 3.1 70B & 0.72 & -0.04 & 0.17 & 0.06 & -0.06 & 0.21 & 0.08 & 0.19 & -0.22 \\
Llama 3.1 8B & 0.72 & -0.04 & 0.17 & -0.24 & -0.41 & 0.21 & 0.71 & -0.43 & -1.51 \\
Llama 3.3 70B & 0.72 & -0.04 & 0.17 & 0.06 & -0.06 & 0.21 & 0.08 & -1.02 & -0.22 \\
Llama 4 Maverick & 0.72 & -0.04 & 0.17 & -0.24 & -0.06 & 0.21 & 0.08 & -1.05 & -0.22 \\
Llama 4 Scout & 0.72 & -0.04 & 0.17 & -0.24 & -0.06 & 0.21 & 0.08 & -0.43 & -1.51 \\
Mistral Large & -0.28 & -0.44 & -0.21 & 0.06 & -0.41 & -0.25 & 0.30 & -1.67 & -1.51 \\
Mistral Small (2409) & 0.72 & -0.04 & -0.21 & 0.06 & -0.06 & 0.21 & 0.08 & -1.07 & -1.51 \\
Mistral Small (2603) & 0.72 & 0.41 & 0.17 & 0.03 & -0.41 & 0.21 & 0.71 & -0.45 & -1.51 \\
Nemotron 3 120B & 0.36 & -0.04 & 0.17 & 0.03 & -0.06 & 0.21 & -0.44 & 0.22 & -0.22 \\
Nvidia Nemotron & -0.28 & -0.04 & 0.17 & 0.06 & -0.07 & 0.21 & 0.30 & -1.67 & -0.22 \\
Phi 3.5 Mini & 0.99 & -0.04 & 0.17 & 0.06 & -0.41 & 0.21 & 0.71 & -1.07 & -1.51 \\
Phi 3.5 MoE & 1.63 & -0.04 & 0.37 & -0.28 & -0.41 & -0.17 & 1.23 & -1.67 & -0.32 \\
Phi 4 & 1.36 & -0.04 & -0.39 & -0.24 & -0.41 & 0.30 & 0.08 & 0.77 & -1.51 \\
Qwen 2.5 7B & 0.36 & -0.04 & 0.17 & -0.24 & -0.41 & -0.25 & 0.71 & -0.43 & -0.32 \\
Qwen 3.5 35B & 1.09 & 0.80 & 0.64 & -0.01 & -0.39 & -0.52 & -0.14 & 3.85 & 2.27 \\
\hline
\end{tabular}
\caption{\textbf{Each LLM's expressed position by issue, CES 2024.} For every issue, each model's responses are projected onto a one-dimensional PCA scale fit on the CES 2024 respondents and normalized so that $0$ is the average strong Democrat and $1$ the average strong Republican; values below $0$ are more liberal than strong Democrats and values above $1$ more conservative than strong Republicans. A model's distance from a benchmark on a given issue is its difference from that reference down the column: from the strong Democrat ($0$), the strong Republican ($1$), or the median CES voter (top row). Four models (Gemini~3.1~Pro, Gemini~3~Flash, Llama~3.2~1B, and Mistral~Nemo) are excluded for excessive refusals, instability, or response-order bias.}
\label{tab:llm_topic_distances_24}
\end{table}

\begin{table}[htbp!]
\centering
\begin{tabular}{lcccc}
\toprule
 & \multicolumn{2}{c}{Response variance (items)} & \multicolumn{2}{c}{Conservative answers, \%} \\
\cmidrule(lr){2-3}\cmidrule(lr){4-5}
Model & Base & Instruct & Base & Instruct \\
\midrule
Falcon 3      & 0.97 (41) & 0.67 (42) & 37 & 20 \\
Llama 3.1 70B & 0.77 (46) & 0.57 (46) & 26 & 17 \\
Qwen 2.5 7B   & 1.00 (45) & 0.63 (46) & 46 & 20 \\
\bottomrule
\end{tabular}
\caption{\textbf{Ideological consistency of base vs.\ instruction-tuned variants, CES 2022 (exploratory).} Population variance of conservative-coded $\{-1,+1\}$ answers across the 46 CES 2022 items (the same consistency measure applied to voters and deployed LLMs in the main text), computed from each variant's modal answer per item over ten runs; the number of items with a valid answer is in parentheses, alongside the share of answered items given the conservative response. Voter benchmarks on this measure are 0.61 for strong partisans, 0.75 for weak partisans, and 0.79 for the median voter, and the deployed 43-model roster averages 0.60. Base models are at or above median-voter inconsistency, near the theoretical maximum of 1 (an even conservative--liberal split), while every instruction-tuned variant falls to the strong-partisan and deployed-roster level. Mistral Nemo was not run on the CES 2022 items and is omitted. This comparison is exploratory and was not pre-registered.}
\label{tab:base_instruct_variance}
\end{table}

\clearpage
\section*{Supplementary Materials: Persuasion Experiment (v2)}\label{sec:persuasion_v2}

This section reports supporting analyses for the persuasion experiment described in the main text. All estimates are from the pre-registered, held-out confirmatory sample (intention-to-treat: 1{,}523 participants, 6{,}072 topic-slots; per-protocol: 1{,}470 participants, 5{,}681 slots), using a mixed-effects logistic model of the toward-advocated outcome (respondent random intercept, proposal fixed within topic, pre-registered covariate block). All tests are two-sided at $\alpha=.05$. Recruitment stopped under the pre-registered futility rule at the 1{,}500-participant look, so these are the study's final confirmatory results.

\begin{table}[htbp!]
\centering
{\footnotesize
\begin{tabular}{lccc}
\toprule
 & \multicolumn{3}{c}{OR [95\% CI] on toward-advocated contrast (BH $p$)} \\
\cmidrule(lr){2-4}
Moderator ($+1$ SD) & Pooled & Base & Explicit \\
\midrule
H4a Out-partisanship & 0.51 [0.31, 0.83] (.04) & 0.55 [0.31, 0.97] (.08) & 0.48 [0.24, 0.92] (.14) \\
H4b Trust in LLMs        & 1.04 [0.90, 1.20] (.93) & 0.97 [0.81, 1.17] (.89) & 1.20 [0.93, 1.55] (.23) \\
H4c LLM familiarity      & 0.99 [0.83, 1.18] (.93) & 0.99 [0.79, 1.24] (.95) & 0.94 [0.69, 1.27] (.68) \\
H4d Perceived neutrality & 0.96 [0.83, 1.11] (.93) & 0.83 [0.69, 0.99] (.08) & 1.26 [0.98, 1.63] (.14) \\
H4e Pol. interest        & 1.04 [0.89, 1.21] (.93) & 1.23 [1.01, 1.49] (.08) & 0.76 [0.58, 1.00] (.14) \\
H4f News breadth         & 1.01 [0.87, 1.16] (.93) & 1.08 [0.90, 1.29] (.64) & 0.89 [0.69, 1.15] (.43) \\
\bottomrule
\end{tabular}
}
\caption{\textbf{Moderators of persuasion (H4a--f), by stratum.} Effect of a $+1$ SD change in each moderator on the toward-advocated contrast, estimated separately in the pooled, base, and explicit strata; Benjamini--Hochberg $p$ in parentheses (three per-stratum families of six). None of the five receptivity or resistance moderators (H4b--f) is significant after correction in any stratum. Out-partisanship (H4a) is a per-slot indicator for whether the model argued against the respondent's party; its estimates run opposite the registered symmetric prediction (out-partisans move less) and reverse the discovery pilot, surviving correction in the pooled stratum only.}
\label{tab:persuasion_moderators}
\end{table}

\begin{table}[htbp!]
\centering
{\footnotesize
\begin{tabular}{lcccc}
\toprule
Contrast & Primary (ITT) & Per-protocol & Control-only base. & Neutral-only base. \\
\midrule
H1 Pooled          & 1.48 [1.28, 1.72] & 1.58 [1.36, 1.84] & 1.49 [1.29, 1.72] & 1.48 [1.28, 1.71] \\
H2 Base            & 0.98 [0.82, 1.17] & 1.02 [0.84, 1.23] & 0.98 [0.82, 1.17] & 0.98 [0.82, 1.17] \\
H3 Explicit$-$base & 3.36 [2.46, 4.61] & 3.66 [2.64, 5.09] & 3.44 [2.51, 4.72] & 3.38 [2.47, 4.64] \\
\bottomrule
\end{tabular}
}
\caption{\textbf{Estimand and baseline-sensitivity refits (H1--H3).} Toward-advocated odds ratios under the primary intention-to-treat estimand, the per-protocol estimand (silent chat-slots dropped), and baselines restricted to the control-only or neutral-only conditions. The primary conclusions are stable across all four specifications.}
\label{tab:persuasion_robustness}
\end{table}

\begin{table}[htbp!]
\centering
\begin{tabular}{lcc}
\toprule
Topic & Pooled OR [95\% CI] & $p$ \\
\midrule
Gun control & 1.46 [1.11, 1.91] & .007 \\
Immigration & 1.45 [1.08, 1.93] & .013 \\
Policing    & 1.23 [0.94, 1.61] & .127 \\
Taxes       & 1.68 [1.26, 2.23] & .0004 \\
\bottomrule
\end{tabular}
\caption{\textbf{Persuasion by topic (exploratory).} Pooled toward-advocated odds ratio within each topic. These by-topic tests are exploratory and uncorrected.}
\label{tab:persuasion_topic}
\end{table}

\begin{table}[htbp!]
\centering
\begin{tabular}{lcc}
\toprule
Model family & Ended on advocated side, \% [95\% CI] & $n$ \\
\midrule
Mistral   & 69 [63, 74] & 270 \\
Meta      & 67 [61, 72] & 280 \\
NVIDIA    & 64 [58, 69] & 282 \\
OpenAI    & 59 [55, 63] & 572 \\
Google    & 58 [54, 63] & 439 \\
Z-AI      & 57 [53, 62] & 426 \\
DeepSeek  & 54 [50, 58] & 611 \\
Qwen      & 49 [45, 53] & 618 \\
xAI       & 42 [37, 48] & 285 \\
Anthropic & 40 [36, 44] & 612 \\
Microsoft & 36 [28, 44] & 143 \\
\bottomrule
\end{tabular}
\caption{\textbf{Ended on the advocated side, by model family (exploratory).} Descriptive share of advocacy conversations ending on the model's advocated side (advocacy slots, $n=4{,}538$; roster average 53.7\%; Wilson 95\% CI).}
\label{tab:persuasion_family}
\end{table}

\begin{table}[htbp!]
\centering
\input{persuasion_carryover_table_si.tex}
\caption{\textbf{Carryover and presentation-order null tests.} Order-nuisance interactions with the toward-advocated contrast (all expected null), reported as robustness rather than confirmatory.}
\label{tab:persuasion_carryover}
\end{table}

\clearpage
\bibliographystyle{naturemag}
\bibliography{scibib}

\end{document}

%% file: persuasion_carryover_table_si.tex
\begin{tabular}{lcc}
\toprule
Order feature & OR & $95\%$ CI \\
\midrule
Position $\times\,\Delta$ (per position later) & 1.08 & [0.94, 1.23] \\
Cumulative exposure $\times\,\Delta$ (per prior advocacy arm) & 1.05 & [0.91, 1.21] \\
Preceding arm: same-side (vs.\ none) & 1.08 & [0.92, 1.27] \\
Preceding arm: opposite-side & 1.07 & [0.91, 1.26] \\
Preceding arm: neutral & 1.06 & [0.84, 1.35] \\
Preceding arm: control & 1.18 & [0.93, 1.50] \\
\bottomrule
\end{tabular}